\documentclass[11pt]{article}

\pdfoutput=1
\usepackage{cite}
\usepackage[parsep]{collref}

\usepackage{simpler-wick}
\usepackage{datetime2}
\usepackage{comment}
\usepackage{xcolor}

\usepackage{caption}
\usepackage{graphicx}
\usepackage{amsmath}
\usepackage{amsthm}
\usepackage{mathtools}
\usepackage{tabularx}
\usepackage{multirow}
\usepackage{multicol}
\usepackage{enumitem}
\usepackage[utf8]{inputenc}
\usepackage{comment}
\usepackage{amsmath}
\usepackage{amsfonts}
\usepackage{braket}
\usepackage{cancel}
\usepackage{upgreek}
\usepackage{booktabs}
\usepackage{amssymb}

\usepackage[compat=1.1.0]{tikz-feynman}

\theoremstyle{definition}

\theoremstyle{remark}

\usepackage{amsmath, amssymb,amsthm}
\usepackage{stackrel}
\numberwithin{equation}{section}
\usepackage{bm,environ,mathrsfs,array,arydshln}
\usepackage{booktabs,float,slashed}
\usepackage{appendix}
\usepackage{tensor}
\usepackage[vcentermath]{youngtab}

\usepackage{simplewick}

\newcommand{\tr}{\mathrm{tr}}

\newcommand{\T}{\mathrm{T}}

\newcommand{\ads}{\mathrm{AdS}_{4}}

\newcommand{\Le}{\mathrm{L}}
\newcommand{\vol}{\mathrm{vol}}
\newcommand{\ses}{\mathrm{SE}_{7}}
\newcommand{\cyf}{\mathrm{CY}_{4}}

\newcommand{\ind}{\mathrm{ind}}
\newcommand{\B}{\mathcal{B}}

\usepackage{xcolor}

\colorlet{linkred}{red!50!black}

\usepackage[
  colorlinks=true,
  linkcolor=black,
  citecolor=linkred,
  urlcolor=linkred,
  linktoc=none
]{hyperref}

\makeatletter
\renewcommand{\eqref}[1]{%
  \textup{(}%
  \hyperref[#1]{\textcolor{linkred}{\ref*{#1}}}%
  \textup{)}%
}
\makeatother

\begin{document}

\vspace{-3cm}
\thispagestyle{empty}

\begin{flushright}
Imperial-TP-2026-SK-03
\end{flushright}

\begin{center}
\vspace{1.2cm}

{\Large\bf Monopoles on the edge

}

\vspace{0.8cm}

Stefan A. Kurlyand

\vskip 0.5cm

{\em
Abdus Salam Centre for Theoretical Physics, Imperial College London, SW7 2AZ, UK
}

\vskip 0.35cm

E-mail: \texttt{s.kurlyand23@imperial.ac.uk}

\end{center}

\vskip 1.2cm

\begin{abstract}

Quantum M2-brane instantons are studied in eleven-dimensional Freund-Rubin backgrounds. The construction uses the identification of the Calabi-Yau cone over the Sasaki-Einstein manifold $\mathrm{SE}_7$ with the geometric branch of the abelian moduli space of the dual superconformal field theory. A class of BPS cycles in $\mathrm{SE}_7$ wrapped by M2-branes is identified as links of two-dimensional holomorphic cones whose coordinate rings contain BPS monopole operators. The algebraic structure and R-charge assignment on the moduli space further allow the computation of both the classical exponents and one-loop partition functions of the corresponding M2-brane instantons. Several three-dimensional superconformal field theories are analysed explicitly, and the relation to the corresponding localisation matrix models is discussed.

\end{abstract}

\newpage
\tableofcontents

\section{Introduction and summary of results}


In the $\ads/\mathrm{CFT}_{3}$ correspondence, a class of three-dimensional superconformal gauge theories with $\mathcal{N}\geq 2$ supersymmetry appears as the low-energy description of M2-branes probing a Calabi-Yau fourfold singularity. Their partition functions on $S^3$ can be computed by supersymmetric localisation \cite{Kapustin:2009kz,Jafferis:2010un,Hama:2010av}, which reduces the path integral to a finite-dimensional matrix integral. In the large-$N$ limit, the resulting partition functions contain both perturbative contributions and exponentially suppressed corrections.

Considerable progress has been made in understanding the perturbative large-$N$ behaviour of these matrix models. In particular, with Chern-Simons levels and number of flavours held fixed, the leading free energy for large $N$ can be obtained by solving saddle-point equations for continuous eigenvalue densities \cite{Herzog:2010hf,Jafferis:2011zi}. At the same time, the Fermi-gas formulation \cite{Marino:2011eh,Marino:2012az} provides further control over the perturbative expansion for suitable classes of theories. The non-perturbative contributions to the localisation matrix model, however, are generically more difficult to extract and may involve several instanton sectors.

A well-studied example is the three-dimensional $\mathcal{N}=6$ ABJM theory with gauge group $U(N)_k\times U(N)_{-k}$ \cite{Aharony:2008ug}. Its matrix model contains two basic non-perturbative sectors, usually referred to as ``worldsheet'' and ``D2-brane'' instantons \cite{Drukker:2010nc,Drukker:2011zy}. At finite $k$, these corrections can be studied using the Fermi-gas approach \cite{Marino:2011eh}, and in the modified grand potential they are organised schematically as \cite{Hatsuda:2012dt,Calvo:2012du,Hatsuda:2013gj,Hatsuda:2013oxa}
\begin{align}\label{1.111}
    J_{\mathrm{np}}(\mu,k)
    =
    \sum_{m, l}
    f_{m,l}(k,\mu)
    \exp\Big[-\big(\frac{4m}{k}+2l\big)\mu\Big] \ .
\end{align}
Here $\mu$ is the chemical potential, $m$ and $l$ label the ``worldsheet'' and ``D2-brane'' sectors, respectively, while terms with $m,l\neq0$ describe their bound-state contributions. Individual sector coefficients can develop poles at special values of $k$, which cancel in the combined non-perturbative contribution. This pole-cancellation mechanism provides non-trivial constraints on the instanton coefficients \cite{Hatsuda:2013gj}. Detailed non-perturbative analyses have also been performed for several $\mathcal{N}=4$ theories \cite{Grassi:2014vwa,Honda:2014ica,Hatsuda:2014vsa,
Moriyama:2014gxa,Moriyama:2014nca,Moriyama:2015jsa,Hatsuda:2015lpa}. However, extending such calculations to more general
$\mathcal{N}\geq 2$ theories appears to be non-trivial.

Holography provides a complementary approach to studying the ``worldsheet'' sector. For ABJM, in the small M-theory circle limit, the relevant holographic configurations are type IIA string instantons in $\ads\times\mathbb{CP}^{3}$, wrapping $\mathbb{CP}^{1}\subset\mathbb{CP}^{3}$ \cite{Cagnazzo:2009zh}. The one-loop partition functions of type IIA string instantons were analysed for several backgrounds in \cite{Gautason:2023igo}. At finite $k$, the gravity dual is M-theory, and these string instantons lift to M2-branes, which are pointlike in $\ads$ and wrap $S^3/\mathbb{Z}_{k}\subset S^7/\mathbb{Z}_{k}$. These M2-brane configurations are BPS and, as observed in \cite{Gautason:2025per}, localise at the fixed points of the Killing vector associated with the preserved supersymmetry in $\ads$. In \cite{Beccaria:2023ujc}, the one-loop partition function around the $S^3/\mathbb{Z}_{k}$ saddle was computed in an expansion at large effective tension $\T_{2}=\frac{\Le^3}{(2\pi)^2\ell_{p}^3}\gg 1$, where $\Le$ is the $S^7$ radius. For $k>2$, the result of \cite{Beccaria:2023ujc} reproduces the leading ``worldsheet'' instanton prefactor in \eqref{1.111} (see also \cite{Tseytlin:2026dhy} for a recent two-loop analysis in this case).

A direct extension of this calculation to other $\ses$ geometries would require identifying the relevant cycles and analysing their bosonic and fermionic fluctuation operators. Even when an explicit metric is available, the cancellation between these fluctuations can be non-trivial \cite{Beccaria:2023ujc}.\footnote{In this respect, semiclassical M2-branes describing BPS Wilson loops, studied in the ABJM context in \cite{Giombi:2023vzu}, were recently considered in several $\ses$ backgrounds in \cite{Gautason:2025bft}.
In these cases, the M2-branes wrap $\mathrm{AdS}_{2}\subset\ads$, and their internal embedding is specified by the choice of a supersymmetric M-theory circle \cite{Farquet:2013cwa,Farquet:2014bda}. An explicit mode-by-mode evaluation of the fluctuation determinants can nevertheless lead to complicated intermediate expressions.} A simplification was obtained in \cite{Kurlyand:2026yke}, where the worldvolume supersymmetry was explicitly determined for M2-branes wrapping three-dimensional Sasakian links in $\ses$. Such links arise from holomorphic subcones of complex dimension two in the Calabi-Yau cone $\cyf=C(\ses)$, and the corresponding M2-branes preserve the two real Killing spinors associated with the Sasaki-Einstein structure. As derived in \cite{Kurlyand:2026yke}, supersymmetry organises the one-loop partition function in terms of the eigenvalues of the Lie derivative along the Reeb vector field and equivariant indices of Dolbeault-type operators. To apply this framework, it remains to identify suitable cycles and determine the equivariant data of their embeddings.\footnote{Note that the BPS property at the classical level does not by itself guarantee that a supersymmetric saddle is isolated or that its contribution to a specified observable is non-vanishing. For instance, in certain heterotic compactifications described by $(2,2)$ or $(2,0)$ non-linear sigma models on Calabi-Yau threefolds, worldsheet instanton contributions to the effective superpotential vanish despite the existence of holomorphic curves \cite{Distler:1986wm,Distler:1987ee}. A related question was recently considered in \cite{Lin:2026uzp}, where the one-loop partition function of supersymmetric M5-branes wrapping $(S^3\times S^3)/\mathbb{Z}_{k}\subset S^7/\mathbb{Z}_{k}$ was argued to vanish.}

A helpful intuition can be developed by taking the cone $\cyf$ over the Sasaki-Einstein manifold and identifying it with the geometric branch of the moduli space of the abelian version of the dual theory. In this case, the vacuum expectation values of the chiral operators on the geometric branch determine the allowed positions of the mobile M2-brane in $\cyf$ \cite{Aharony:2008ug, Martelli:2008si, Hanany:2008cd, Hanany:2008fj}. The superconformal R-charges determine the Reeb weights, while magnetic charges encode the action of the M-theory circle. The graded chiral ring therefore provides algebraic coordinates in which both the candidate M2-brane cycles and their normal fluctuations can be described.

In this work, I use this description to identify a class of BPS M2-brane cycles and compute the corresponding one-loop partition functions. Of special interest are the two-dimensional subcones of the geometric branch on which the vacuum expectation values of the BPS monopole operators are non-vanishing. The selected subcones are invariant under the M-theory circle action, and the corresponding three-dimensional links wrap the M-theory circle direction. The associated single M2-brane contributions are therefore naturally identified with ``worldsheet''-type matrix-model instantons.

Both the classical action and one-loop determinant can then
be determined from the graded coordinate ring and the
embedding of the chosen subcone. Specifically, for the classical action, the character formula~\cite{Martelli:2006yb} expresses the volume of the Sasakian link $\Sigma$ in terms of the Hilbert series of its cone $\mathcal{M}=C(\Sigma)$. If $R$ denotes the R-charge, the volume of the link can be expressed as
\begin{align}
    \frac{\mathrm{Vol}(\Sigma)}{\mathrm{Vol}(S^3)} =\lim_{t\to1}\, (1-t)^2h(t,\mathcal{M}) \ , \qquad h(t,\mathcal{M})=\operatorname{Tr}_{\mathbb{C}[\mathcal{M}]} t^{2R} \ , 
\end{align}
and determines the classical action. At the same time, for the one-loop prefactor, local monopole and mesonic coordinates determine the normal-bundle transition functions and their Reeb weights. Applying the equivariant-index formula of \cite{Kurlyand:2026yke} then gives the determinant without requiring an explicit Sasaki-Einstein metric or the complete fluctuation spectrum.

I apply this construction explicitly to several three-dimensional SCFTs, with the resulting single M2-brane one-loop partition functions summarised in Table~\ref{tab:summ2}. The one-loop approximation to the M2-brane partition function corresponds, on the gauge side, to the large-$N$/large chemical potential limit, with Chern-Simons levels and the number of flavours held fixed. The results are expressed in terms of
\begin{align}
    \mu_{\ast}=\sqrt{\frac{N}{C}} \ , \qquad
    C=\frac{2}{\pi^2}
    \frac{\mathrm{Vol}(\ses)}{\mathrm{Vol}(S^7)} \ ,
\end{align}
so that the effective M2-brane tension is $\T_2=2\mu_{\ast}/\pi^2$.

\begin{table}[h!]
    \centering
    \begingroup
    \small
    \setlength{\tabcolsep}{10pt}
    \renewcommand{\arraystretch}{1.15}

    \begin{tabular}{@{}lccc@{}}
        \toprule[0.6pt]
        \textit{SCFT}
        & \textit{Classical} $S$
        & \textit{One-loop} $Z_1$
        & $\mu_{\ast}$ \\
        \midrule[0.3pt]
        \addlinespace[6pt]

        $\mathcal{N}=3$ flavoured ABJM
        & $\dfrac{4k\mu_{\ast}}{(N_f+k)^2}$
        & $\dfrac{1}{4\sin^2\!\bigl(\frac{2\pi}{N_f+k}\bigr)}$
        & $\pi(N_f+k)\sqrt{\dfrac{N}{N_f+2k}}$ \\
        \addlinespace[9pt]

        Two-node $Q^{1,1,1}/\mathbb{Z}_{N_f}$
        & $\dfrac{\mu_{\ast}}{N_f}$
        & $\dfrac{1}{4\sin^2\!\bigl(\frac{\pi}{N_f}\bigr)}$
        & $2\pi\sqrt{\dfrac{N_fN}{3}}$ \\
        \addlinespace[9pt]

        Two-node $V^{5,2}/\mathbb{Z}_{k}$
        & $\dfrac{9\mu_{\ast}}{4k}$
        & $\dfrac{\sin\!\bigl(\frac{2\pi}{k}\bigr)e^{\delta S(k)}}
            {4\sin\!\bigl(\frac{\pi}{k}\bigr)
            \sin^2\!\bigl(\frac{3\pi}{2k}\bigr)}$
        & $\dfrac{8\pi}{9}\sqrt{kN}$ \\
        \addlinespace[9pt]

        ADHM
        & $\dfrac{4\mu_{\ast}}{N_f}$
        & $\dfrac{\gamma\, 2\mu_{\ast}}
            {\pi\sin\!\bigl(\frac{2\pi}{N_f}\bigr)}$
        & $\pi\sqrt{\dfrac{N_fN}{2}}$ \\
        \addlinespace[9pt]

        Flavoured $V^{5,2}/\mathbb{Z}_{N_f}$
        & $\dfrac{9\mu_{\ast}}{4N_f}$
        & $\dfrac{\gamma \, 2N_f\mu_{\ast}}
            {3\pi\cos^2\!\bigl(\frac{\pi}{2N_f}\bigr)}$
        & $\dfrac{8\pi}{9}\sqrt{N_fN}$ \\
        \addlinespace[6pt]
        \bottomrule[0.6pt]
    \end{tabular}
    \endgroup

    \caption{Above, the one-loop M2-brane instanton contributions to the partition function are $Z_1e^{-S}$, before summing over cycles and orientations. The Chern-Simons levels are assumed to be $k>2$ for flavoured ABJM and $V^{5,2}/\mathbb{Z}_{k}$ theories; the number of flavours is assumed to be $N_{f}>1$ for $Q^{1,1,1}/\mathbb{Z}_{N_{f}}$ and $V^{5,2}/\mathbb{Z}_{N_{f}}$ theories, and $N_{f}>2$ for ADHM. For $V^{5,2}/\mathbb{Z}_{k}$, $\delta S(k)=k\big(\ell(1/k)+\ell(2/k)-2\ell(3/(2k))\big) $, with $\ell(z)$ defined in \eqref{3.48aaaa}. For ADHM and $V^{5,2}/\mathbb{Z}_{N_{f}}$ theories, $\gamma$ is the numerical coefficient, to be determined from the integral over the unpaired bosonic zero modes and the possible determinant phase (see Section~\ref{m2inst}).}
    \label{tab:summ2}
\end{table}

The resulting M2-brane partition functions can be independently checked against the matrix-model results. For the ADHM theory, the classical exponent and the leading large-$\mu$ prefactor agree with the ``worldsheet''-instanton expression proposed in \cite{Hatsuda:2014vsa} (up to the factor $\gamma$, which is expected to be independent of $N_f$ and whose derivation requires explicit evaluation of the rescaled integral over the bosonic zero modes). 
For $\mathcal{N}=3$ flavoured ABJM, I analyse the large-energy periods of the complexified Fermi surface, following the Fermi-gas and tropical-curve approaches of \cite{Marino:2011eh,Santamaria:2010dm}. This analysis supports the classical exponent $S = 4k\mu_{\ast}/(N_{f}+k)^2$ obtained from the M2-brane action.

The chiral ring interpretation provides a further connection
between M2-brane instantons and the matrix model. This connection follows from the relation between asymptotic counting of chiral operators and the large-$N$ eigenvalue density \cite{Gulotta:2011si,Gulotta:2011aa}, together with the geometric interpretation of the latter in terms of the M-theory Hamiltonian \cite{Farquet:2013cwa}. In Section~\ref{sec4}, I further relate the volumes of the three-dimensional monopole links, and hence their classical instanton actions, to discontinuities of the eigenvalue
density. This relation is checked explicitly for the $\mathcal{N}=3$ flavoured ABJM, $Q^{1,1,1}/\mathbb{Z}_{N_f}$, and $V^{5,2}/\mathbb{Z}_k$ theories.

The rest of the paper is organised as follows.
Section~\ref{2} reviews the geometric branch of the
abelian theories and determines the monopole loci
and their Hilbert series. Section~\ref{m-theory}
identifies the corresponding M2-brane cycles and
computes their classical actions and one-loop
prefactors. Section~\ref{sec4} compares these results
with the matrix model and develops the relations
to Fermi-surface periods and eigenvalue-density
discontinuities. Section~\ref{sec5} contains concluding
remarks and comments on multi-instanton configurations.
Appendix~\ref{A} reviews the KK-monopole solution in
eleven-dimensional supergravity from the GIT quotient
perspective.

\section{Monopole operators on the geometric branch}\label{2}

In this section I identify the monopole loci and compute the Hilbert series of the corresponding coordinate rings for several examples. I start by reviewing the construction of the geometric branch of the moduli space for abelian three-dimensional $\mathcal{N}=2$ Chern-Simons matter theories, describing a single mobile M2-brane. 

\subsubsection*{Moduli space of abelian theories}

The construction of the moduli space of three-dimensional $\mathcal{N}=2$ theories resembles that of the moduli space of four-dimensional $\mathcal{N}=1$ theories. In particular, in three dimensions the $\mathcal{N}=2$ Poincaré superalgebra has four
real supercharges, which can be assembled into a complex two-component
spinor. 
In contrast to four dimensions, however, the Yang-Mills coupling in three dimensions is dimensionful, making the Yang-Mills kinetic term irrelevant in the IR. A weakly coupled description can be obtained by introducing Chern-Simons terms, with the inverse Chern-Simons level playing the role of an effective coupling \cite{Schwarz:2004yj}. In three dimensions, in the Wess-Zumino gauge, the vector superfield may be parametrised as
\begin{align}
 V=-2i\theta \bar{\theta} \sigma+2\theta \gamma^{i}\bar{\theta}A_{i}+\sqrt{2}i\theta^2(\bar{\theta}\bar{\chi})-\sqrt{2}i\bar{\theta}^2(\theta\chi)+\theta^2\bar{\theta}^2D \ , 
\end{align}
where $A_{i}$ is the gauge field, $\chi$ is the gaugino, $\sigma$ is a real scalar, and $D$ is a real auxiliary scalar. For the abelian gauge group $G = U(1)^n$, the gauge-field part of the Chern-Simons action is
\begin{align}
    S_{\mathrm{CS}}(A) = \sum_{i=1}^{n}\frac{k_{i}}{4\pi }\int A_{i}\wedge dA_{i} \ .
\end{align}
I will assume that the sum of the Chern-Simons levels vanishes\footnote{Without this constraint, a diagonal monopole operator carries a non-zero gauge charge under the overall diagonal $U(1)$. In the type IIA reduction, $\sum_i k_i$ is identified with the Romans mass \cite{Gaiotto:2009mv}. As reviewed below, when the Chern-Simons couplings receive quantum corrections, the conditions in \eqref{cszero} are understood as conditions on the effective couplings.}
\begin{align}\label{cszero}
     \sum_{i=1}^{n}k_{i} = 0 \ , \qquad k_{i}\in \mathbb{Z} \ .
\end{align}
The matter fields are described by chiral multiplets $X_{ij}$ transforming in the bifundamental representation of the gauge group $U(1)_{i}\times U(1)_{j}$. After eliminating the auxiliary fields in the vector and chiral multiplets, the vanishing of the bosonic potential requires \cite{Martelli:2008si, Hanany:2008cd}
\begin{align}\label{modulieq}
    \partial W(X)=0 \ , \qquad 
    \mu_i(X)=\frac{k_i}{2\pi}\sigma_i \ , \qquad \sigma_iX_{ij}-X_{ij}\sigma_j=0 \ ,
\end{align}
where, in the abelian case, the moment map for the $i$-th gauge group is
\begin{align}
    \mu_i(X)
    =
    \sum_{j=1}^{n} |X_{ij}|^2
    -\sum_{j=1}^{n} |X_{ji}|^2 \ .
\end{align}

On the geometric branch, describing a mobile M2-brane, one can take $\sigma_i=\sigma$, allowing generic vacuum expectation values for the bifundamental scalars. The remaining equations then give
\begin{align}
    \sigma_i=\sigma \ ,  \qquad \mu_{i}(X) = \frac{k_{i}}{2\pi}\sigma \ , \qquad \sum_{i=1}^{n}\mu_{i}(X) = 0 \ . \label{2.6}
\end{align}
Introducing an $(n-2)$-dimensional space, with $\mathbf{1}=(1,\dots,1)$,
\begin{align}
    \mathfrak{h}_{\mathbf{k}}
    =
    \big{\{}
       \mathbf{v}\in\mathbb{R}^{n}:
       \langle \mathbf{v}, \mathbf{1}\rangle =0, \ 
       \langle \mathbf{v}, \mathbf{k}\rangle = 0 
    \big{\}} \ , \qquad \mathbf{k} = (k_{1}, \dots, k_{n})\neq 0 \ ,
    \label{hk-algebra}
\end{align}
the equations \eqref{2.6} define $(n-2)$ independent homogeneous moment-map equations 
\begin{align}\label{2.8}
    \sum_{i=1}^{n}\mathrm{v}^{i}\mu_{i}(X) = 0  \ .
\end{align}
Together with the quotient by the corresponding $U(1)^{n-2}$ gauge transformations, these equations define the symplectic reduction subject to the F-term equations, while the remaining gauge direction, parametrised by~$\mathbf{k}$, determines $\sigma$. In addition, due to the presence of the monopole operators on the geometric branch, the moduli space should also be quotiented by the residual discrete action, as reviewed below.

One can also consider flavoured $\mathcal{N}=2$ Chern-Simons matter theories introduced in \cite{Benini:2009qs, Jafferis:2009th}. From the gauge-theory perspective, one introduces $N_f$ flavour pairs of fundamental and antifundamental chiral multiplets $(p_{\alpha}, q_{\alpha})$, each coupled to a bifundamental operator $\mathcal{O}_{\alpha}$ via the superpotential
\begin{align}
    \delta W = \sum_{\alpha=1}^{N_{f}}p_{\alpha}\mathcal{O}_{\alpha}q_{\alpha} \ ,
\end{align}
where $\mathcal{O}_{\alpha}$ may be constructed as a product of bifundamental matter fields. On the geometric branch considered here, the vacuum expectation values of the flavour fields $(p_{\alpha}, q_{\alpha})$ are set to zero and therefore do not affect this branch classically. They nevertheless modify the quantum chiral ring through the one-loop corrections to the charges of monopole operators. At this stage, note that since each superpotential term has R-charge two and is gauge invariant, it follows that
\begin{align}\label{2.10}
    R[p_{\alpha}]+R[q_{\alpha}]+R[\mathcal{O}_{\alpha}] = 2 \ , \qquad 
    Q_{i}[p_{\alpha}]+Q_{i}[q_{\alpha}]+Q_{i}[\mathcal{O}_{\alpha}] = 0 \ .
\end{align}
Introducing $N_{f}$ identical pairs gives a $U(N_{f})$ flavour symmetry.

\subsubsection*{Monopole operators}

The chiral ring of a three-dimensional $\mathcal{N}=2$ Chern-Simons-matter theory may also contain sectors generated by monopole operators. These operators are charged under the topological currents $J^{m}_{i} = \frac{1}{2\pi}\ast F_{i}$ \cite{Borokhov:2002ib, Borokhov:2002cg, Borokhov:2003yu}. In radial quantisation, a monopole operator inserted at the origin creates a quantised magnetic flux through the surrounding two-sphere
\begin{align}
    \frac{1}{2\pi}\int_{S^2} F_{i} =m_{i} \ ,  \qquad m_{i}\in \mathbb{Z} \ .
\end{align}
For a general compact gauge group $G$, the magnetic charge belongs to the cocharacter lattice of $G$, modulo the action of the Weyl group \cite{Goddard:1976qe}. 
For the abelian gauge group $G=U(1)^n$, a monopole operator can be specified by $n$ integers giving the first Chern numbers of the corresponding $U(1)$ bundles over $S^2$. 

The geometric branch is associated with diagonal magnetic charges $m_{i} = m$. Indeed, a bifundamental field $X_{ij}$ on $S^2$ can be viewed as a section of the line bundle $L_i\otimes L_j^{-1}$, whose first Chern class in the monopole background is
\begin{align}
    c_1(L_i\otimes L_j^{-1})=m_i-m_j \ .
\end{align}
Consequently, a covariantly constant section of $L_{i}\otimes L_{j}^{-1}$, corresponding to a non-zero constant vacuum expectation value of $X_{ij}$, exists when $m_i=m_j$. I restrict to diagonal monopole operators and denote by $T, \tilde{T}$ the bare monopole operators with magnetic charges $m=\pm1$.

Due to the Chern-Simons couplings, the bare monopole operators $T, \tilde{T}$ generally carry non-zero gauge charges, so gauge-invariant monopole operators must be obtained by dressing them with matter fields. Classically, the gauge current is related to the monopole flux through 
\begin{align}
    \frac{k_{i}}{2\pi}F_{i} +\ast J_{i} = 0 \ , 
\end{align}
and consequently the diagonal monopole operators considered here have the classical gauge charges 
\begin{align}
    Q^{\mathrm{cl}}_{i}[T] =-k_{i}  \ , \qquad Q_{i}^{\mathrm{cl}}[\tilde{T}] = k_{i} \ .
\end{align}
At the quantum level, the gauge charges of the monopole operators acquire one-loop shifts. These follow from the effective Chern-Simons levels obtained after integrating out massive complex fermions \cite{deBoer:1997kr, Aharony:1997bx} (see also \cite{Benini:2011cma, Cremonesi:2016nbo}):
\begin{align}\label{2.15}
    k_{ij}^{\mathrm{eff}} = k_{ij}+\frac{1}{2}\sum_{\psi}Q_{i}[\psi]Q_{j}[\psi]\operatorname{sign}(M_{\psi}) \ ,
\end{align}
where the sum is over complex fermions with non-zero effective mass $M_{\psi}$ and, in the present case, the bare Chern-Simons couplings are $k_{ij} = k_{i}\delta_{ij}$. In a BPS monopole background with magnetic charges $m_i$, and in the absence of bare real masses, the scalars $\sigma_{i}$ acquire expectation values and the sign of the induced fermionic mass satisfies
\begin{align}
    \operatorname{sign}(M_{\psi}) = \operatorname{sign}\big(\sum_{i=1}^{n}m_{i} Q_{i}[\psi]\big ) \ .
\end{align}
Fermions for which this sum vanishes remain massless and are not integrated out. In particular, for diagonal monopole operators, the fermions in bifundamental and adjoint chiral multiplets do not acquire an effective mass and therefore do not contribute to the one-loop monopole charges.  In the presence of flavouring, using~\eqref{2.10}, the one-loop corrected monopole charges can be expressed as
\begin{align}
    &Q_{i}[T] = -\sum_{j=1}^{n}k_{ij}^{\mathrm{eff}} = -k_{i}+\delta Q_{i}[T] \ , \qquad Q_{i}[\tilde T] = \sum_{j=1}^{n}k_{ij}^{\mathrm{eff}} = k_{i}+\delta Q_{i}[\tilde{T}] \ , \label{mongch}\\
    &\delta Q_{i}[T] = \delta Q_{i}[\tilde{T}] = \frac{1}{2}\sum_{\alpha=1}^{N_{f}}Q_{i}[\mathcal{O}_{\alpha}] \ ,
\end{align}
where $k_{ij}^{\mathrm{eff}}$ is evaluated in the corresponding monopole background. Similarly, the R-charges of the bare monopole operators receive one-loop corrections so that \cite{Benini:2009qs, Jafferis:2009th}
\begin{align}
    &R[T] = \Delta_{m}+\delta R[T] \ , \qquad R[\tilde{T}] = -\Delta_m+\delta R[\tilde{T}] \ , \label{monrch} \\
    &\delta R[T] = \delta R[\tilde{T}] = \frac{1}{2}\sum_{\alpha=1}^{N_{f}}(2-R[p_{\alpha}]-R[q_{\alpha}])=\frac{1}{2}\sum_{\alpha=1}^{N_{f}}R[\mathcal{O}_{\alpha}] \ , 
\end{align}
where the sums run over all flavour pairs. I have allowed possible trial R-charges $R^{\mathrm{cl}}[T]=\Delta_m$ and $R^{\mathrm{cl}}[\tilde{T}]=-\Delta_m$, so that $R[T] - R[\tilde{T}] = 2\Delta_{m}$. The one-loop shifts of the gauge charge and R-charge are consistent with the quantum chiral-ring relation proposed in \cite{Benini:2009qs, Jafferis:2009th}
\begin{align}
    T\tilde{T} = \prod_{\alpha=1}^{N_{f}}\mathcal{O}_{\alpha} \ .
\end{align}
As discussed in \cite{Benini:2009qs, Jafferis:2009th}, on the M-theory side flavouring corresponds to the insertion of $N_{f}$ KK-monopoles.\footnote{Despite similar names, the KK-monopoles in M-theory should not be confused with the monopole operators on the gauge theory side. The latter, in particular, have a holographic description in terms of the massless supergravity modes with momentum along the M-theory circle \cite{Aharony:2008ug}, as further used in Section~\ref{m-theory}.} In fact, this identification becomes evident by viewing the KK-monopole solution as an algebraic variety, as reviewed in Appendix~\ref{A}. The loci defined by $\mathcal{O}_{\alpha}=0$ and $T=\tilde{T}=0$ naturally correspond to the KK-monopole insertions. One can further define the M-theory circle action $U(1)_m$ to act trivially on the mesonic coordinates and on the monopole coordinates as
\begin{align}\label{2.22}
    T \mapsto \nu T \ , \qquad
    \tilde{T} \mapsto \nu^{-1}\tilde{T} \ .
\end{align}

As in the well-known ABJM theory \cite{Aharony:2008ug}, in the presence of magnetic fluxes the effective gauge group on the geometric branch is modified. Concretely, the overall diagonal $U(1)$ decouples from the bifundamental chiral fields, while its gauge field enters the Chern-Simons action through the BF coupling \cite{Martelli:2008si}
\begin{align}
    &S_{\mathrm{CS}}(A) =\frac{k}{2\pi n }\int b\wedge da +\dots  \ , \qquad a = \sum_{i=1}^{n}A_{i} \ , \qquad b = \frac{1}{k}\sum_{i=1}^{n} k_{i}A_{i} \ ,  \label{bf}
\end{align}
where $k = \operatorname{gcd}\{k_{i}\}$ and the remaining terms in the action are independent of $a$. The gauge transformations 
\begin{align}
    A_{i} \mapsto A_{i}+d\alpha_{i} \ , \qquad \sum_{i} \alpha_{i}k_{i} = 0 \ ,
\end{align}
act trivially on $b$. At the same time, in the presence of the diagonal magnetic flux, under the gauge transformations
\begin{align}\label{2.24}
    A_{i} \to A_{i}+d\alpha_{i} \ , \qquad \sum_{i}\alpha_{i}k_{i} = k\alpha \ ,
\end{align}
the Chern-Simons action \eqref{bf} transforms as
\begin{align}
    \delta S_{\mathrm{CS}}(A) = \frac{k}{2\pi n}\, \alpha\int_{S^{2}}da = (km)\alpha \ .  
\end{align}
Thus, in the presence of the monopole operators, only a $\mathbb{Z}_{k}$ subgroup of constant gauge transformations in \eqref{2.24} acts on the moduli space. Equivalently, the $\mathbb{Z}_k$ group can be viewed as the residual gauge symmetry left after setting to zero the periodic dual photon associated with the overall diagonal $U(1)$ \cite{Aharony:2008ug, Martelli:2008si}. 

Note that, in the presence of flavouring and in the sector with a fixed diagonal magnetic flux $m$, one should replace the classical Chern-Simons couplings by their quantum-corrected counterparts in \eqref{2.15}. For the theories of interest, the bare couplings are diagonal and the flavoured chiral fields are each charged under only one $U(1)_{i}$, so the off-diagonal effective couplings remain zero. I further assume that the one-loop-corrected diagonal couplings $k_{i}^{\mathrm{eff}}$ satisfy \eqref{cszero}.

\subsubsection*{Geometric branch and monopole loci}

Putting everything together, it is convenient to treat $T, \tilde{T}, X_{ij}$ as generators and to impose the relations arising from the F-term equations and the monopole relation, while gauge invariance is imposed under the full effective gauge group. The geometric branch, viewed as a symplectic quotient, can equivalently be expressed as the affine GIT quotient with a trivial character\footnote{The F-term relations are understood to include any additional equations selecting the geometric branch of the moduli space. See, for instance, Section~\ref{sec24}.}
\begin{align} \label{2.26}
    \mathcal{M}_{\mathrm{geom}} = \operatorname{Spec}\Big[\Big(\frac{\mathbb{C}[X_{ij}, T, \tilde{T}]}{\langle\partial W ,  \ T\tilde{T} - \prod_{\alpha}\mathcal{O}_{\alpha}\rangle}\Big)^{(\mathbb{C}^{\ast})^{n-1}} \Big] \ , 
\end{align}
where the complexified gauge group acts with the bifundamental charges and the quantum-corrected monopole charges given above.

The resulting moduli space $\mathcal{M}_{\mathrm{geom}}$ is required to be a four-dimensional Calabi-Yau cone $\cyf$ so that the gravity dual admits an M-theory description on $\ads \times \ses$, with Sasaki-Einstein link $\ses=\cyf|_{r=1}$. Below, a special role is played by the monopole loci $\mathcal{M}_{\pm}\subset\mathcal{M}_{\mathrm{geom}}$. The monopole loci $\mathcal{M}_{\pm}$ can be defined as subcones whose coordinate rings $\mathbb{C}[\mathcal{M}_{\pm}]$ are generated by gauge-invariant diagonal monopole operators with strictly positive or strictly negative magnetic charges, respectively. In addition, with respect to the R-symmetry grading specified below, one can require that the generators of $\mathbb{C}[\mathcal{M}_{\pm}]$ have a fixed ratio of magnetic charge to R-charge. When such loci exist and the bare monopole operators $T,\tilde{T}$
have non-zero gauge charges, the gauge-invariant monopole operators are obtained by dressing $T,\tilde{T}$ with matter fields.\footnote{Note that, for a fixed sign of magnetic charge, there may be several distinct monopole loci $\mathcal{M}_{\pm,(\alpha)}$, characterised by different ratios of magnetic charge to the superconformal R-charge. For instance, in the one-flavour $\mathcal{N}=2$ ABJM model with gauge group $U(1)_{3/2}\times U(1)_{-3/2}$, introduced in \cite{Benini:2009qs} (see also Section~6.2.2 of \cite{Cremonesi:2016nbo}), the geometric moduli space is the Calabi-Yau cone over $Y^{1,2}(\mathbb{CP}^{2})$. The two gauge-invariant monopole operators $TA_{1}$ and $TA_{2}$ have the same magnetic charge but different R-charges, $R[TA_{1}]\neq R[TA_{2}]$, and therefore belong to different monopole loci in the above sense. In this work, I will mainly consider theories with a unique monopole locus for each sign of magnetic charge and suppress the indices distinguishing the corresponding loci.} I will consider the gauge-invariant diagonal monopole operators explicitly for the examples below.

The R-symmetry action on chiral operators induces a grading on $\mathbb{C}[\mathcal{M}_{\mathrm{geom}}]$. In $\mathcal{N}=2$ theories, however, the constraints of conformal invariance are in general insufficient to fix all
the R-charges of gauge-invariant operators. The exact superconformal R-symmetry may nevertheless be determined by maximising the sphere free energy over the admissible trial R-charges, including their possible mixing with abelian IR symmetries \cite{Jafferis:2010un,Jafferis:2011zi}. In what follows, I grade chiral operators by their exact superconformal R-charge.

With the identification of the R-charges, the chiral ring decomposes into a direct sum of irreducible representations of the R-symmetry $U(1)_r$. For the examples of interest, the monopole loci $\mathcal{M}_{\pm}\subset \mathcal{M}_{\mathrm{geom}}$ are R-invariant subcones, and their coordinate rings $\mathbb{C}[\mathcal{M}_{\pm}]$ inherit the $U(1)_r$-grading. The decomposition into the $U(1)_r$ irreducible representations can be encoded in the $U(1)_r$-character or, in other words, the unrefined Hilbert series. If $M$ is a locus of the geometric branch defined by the vanishing of a homogeneous ideal, the unrefined Hilbert series is defined by
\begin{align} \label{characterdef}
    h(t, M) = \operatorname{Tr}_{\mathbb{C}[M]} t^{2R} = \sum_{r \geq 0}h_{r}t^{2r} \ , 
\end{align}
where $h_{r} = \dim\mathbb{C}[M]_{r}$ denotes the dimension of the $U(1)_r$ component $\mathbb{C}[M]_{r}$ of the coordinate ring with weight $r$. I use the convention in which the grading is by twice the physical R-charge, so that twice the R-charges correspond to the weights of the Reeb vector. I will further denote the Hilbert series on the full geometric branch and monopole loci by
\begin{align}
    H(t) = h(t, \mathcal{M}_{\mathrm{geom}}) \ , \qquad h_{\pm}(t) = h(t, \mathcal{M}_{\pm}) \ .
\end{align}
The Hilbert series $H(t)$ can be used to determine the volume of the dual geometry $\ses$. Similarly, as discussed in Section~\ref{m-theory}, when the monopole loci have complex dimension two, $h_{\pm}(t)$ determine the volumes of the three-dimensional links $\Sigma_{\pm}$ wrapped by the BPS M2-brane instantons. In particular, with the chosen normalisation of the R-symmetry generator:\footnote{More precisely, the character formula applies to any admissible choice of the Reeb vector on a Sasakian link. If $\mathfrak{t}$ denotes the Lie algebra of a maximal torus acting 
on $\cyf$, the Sasaki-Einstein Reeb vector is obtained by minimising the volume over the Reeb cone in $\mathfrak{t}$, subject to the Calabi-Yau normalisation $\mathcal{L}_{\xi}\Omega=4i\Omega$ \cite{Martelli:2005tp,Martelli:2006yb}. In the holographic theories considered here, this geometric volume minimisation, in the large-$N$ limit, corresponds to $F$-maximisation \cite{Jafferis:2010un,Jafferis:2011zi}.}
\begin{align}
    \frac{\mathrm{Vol}(\ses)}{\mathrm{Vol}(S^7)} = \lim_{t\to 1}\, (1-t)^4H(t) \ , \qquad \frac{\mathrm{Vol}(\Sigma_{\pm})}{\mathrm{Vol}(S^3)} = \lim_{t\to 1}\, (1-t)^2 h_{\pm}(t)  \ , \label{2.29}
\end{align}
where $C(\Sigma_{\pm}) = \mathcal{M}_{\pm}$. This relation is a manifestation of the character formula for the volume of the Sasakian link \cite{Martelli:2006yb} (see also \cite{Bergman:2001qi}). Physically, \eqref{2.29} implies that the classical volumes of the corresponding links are controlled by the asymptotic growth of the number of chiral operators in the associated coordinate rings.

It should be stressed, however, that monopole loci need not have complex dimension two. In particular, in Section~\ref{sec2.5} I consider examples of flavoured $\mathcal{N}=8$ SYM theories for which the monopole loci have complex dimension one. The coordinate rings of these loci are generated by bare gauge-invariant monopole operators, and the corresponding links in $\ses$ are circles rather than three-dimensional cycles. As also discussed in Section~\ref{m-theory}, the M2-brane instanton cycles in these cases can still be obtained as links of complex two-dimensional cones containing the monopole loci.

With the identification of the geometric branch
$\mathcal{M}_{\mathrm{geom}}$ with the Calabi-Yau cone of the dual geometry, one can note a further relation between the R-charges of chiral operators and the Reeb flow. Concretely, let $\xi = J(r\partial_{r})$ be the Reeb vector, so it has unit norm on the link $\ses=\cyf|_{r=1}$ and is tangent to the slices of fixed radius. A holomorphic function $f\in\mathbb{C}[\mathcal{M}_{\mathrm{geom}}]$ of physical R-charge $R[f]$ satisfies
\begin{align}
    -i\mathcal{L}_{\xi}f
    =
    2R[f]f\ . 
\end{align}
Equivalently, $\xi$ generates a $U(1)_{r}$-action: 
\begin{align}
    f\bigl(\varphi_{\alpha}(p)\bigr)
    =
    e^{2i\alpha R[f]}f(p)\ , 
\end{align}
where $\varphi_{\alpha}=\exp(\alpha\xi)$ denotes the Reeb flow. I restrict to quasi-regular Sasakian links, so that the \mbox{$U(1)_{r}$-action} on the link is locally free and the space of orbits of $\xi$ is a compact complex orbifold. Since $\xi$ has unit norm on $\ses$, the parameter $\alpha$ is the natural parameter along the orbits of the Reeb flow on the link. The length of an orbit through a point can therefore be determined by looking at the stabiliser of that point.

At the same time, the stabiliser can be conveniently determined from the chiral ring perspective \eqref{2.26}. One can use that the R-symmetry action commutes with the complexified gauge-group action $(\mathbb{C}^{\ast})^{n-1}$, while the ideal generated by the F-term equations and the monopole relation is homogeneous under the R-symmetry. Consequently, the R-symmetry descends to the gauge-invariant coordinate ring. Let $\{f_{a}\}$ be a generating set of $\mathbb{C}[\mathcal{M}_{\mathrm{geom}}]$ with definite R-charges. At a point $p\in\ses$, the flow $\varphi_{\alpha}$ returns to $p$ when $\exp(2i\alpha R[f_{a}])=1$ for every generator $f_{a}$ such that $f_{a}(p)\neq 0$. It follows that, in the quasi-regular case,
\begin{align}\label{reeb-pointwise-length}
    \ell_{\xi}(p)
    =
    \frac{2\pi}{
    \operatorname{gcd}
    \left\{
       2R[f_{a}]
       \,:\,
       f_{a}(p)\neq0
    \right\}}\ .
\end{align}

In particular, suppose that the coordinate ring of a monopole locus
$\mathcal{M}_{\pm}$ is generated by homogeneous gauge-invariant monopole
operators $\mathrm{T}^{\pm}_{a}$, carrying magnetic charges $\pm1$ and
having a fixed ratio of magnetic charge to R-charge. Since their magnetic
charges are equal, the generators have a common R-charge
\mbox{$R_{\pm}=R[\mathrm{T}^{\pm}_{a}]$}. At $\Sigma_{\pm}$, the Reeb-orbit length is therefore
\begin{align}\label{2.33}
\ell_{\xi}\big|_{\Sigma_{\pm}}
    =
    \frac{\pi}{R_{\pm}}\ .
\end{align}
I will further use this fact in Section~\ref{m-theory} and compute the lengths of the Reeb orbits on the monopole loci explicitly for the examples below.

\subsection{$\mathcal{N}=3$ flavoured ABJM theory}\label{sec2.1}

One can consider the $\mathcal{N}=3$ flavoured ABJM theory introduced in \cite{Hohenegger:2009as, Gaiotto:2009tk, Hikida:2009tp}. To analyse the geometric branch, one can take the gauge group $U(1)_{k}\times U(1)_{-k}$ with opposite Chern-Simons levels $k$, under which the bifundamental matter fields $A_{l}$ and $B_{l}$, $l=1,2$, have charges $(1,-1)$ and $(-1,1)$, respectively. The $SU(2)_r$-preserving flavouring is implemented by adding $N_{f}^{(i)}$ flavour pairs $(p_{i},q_{i}), \ i=1,2$, with flavour indices suppressed, to the $i$-th $U(1)$ factor. The pairs at the first and second factors have charges $(\pm1,0)$ and $(0,\pm1)$ and couple to the bifundamental fields through the gauge-invariant operator
\begin{align}
    \mathcal{O} = B_{2}A_{2}-B_{1}A_{1} \ .
\end{align}
I will further denote by $N_{f}=N_{f}^{(1)}+N_{f}^{(2)}$ the total number of flavours. The $SU(2)_r$ symmetry rotates the scalar doublets $(A_{l}, \bar{B}_{l})$ and $(p_i, \bar{q}_i)$, and the corresponding $U(1)_r\subset SU(2)_r$ subgroup is obtained by choosing a maximal torus. With this assignment, the bifundamental chiral fields have the R-charges \mbox{$R[A_{l}] = R[B_{l}] = \frac{1}{2}$}.

Since the operator $\mathcal{O}$ is gauge invariant, the effective Chern-Simons couplings receive no one-loop shifts in the diagonal monopole background. Using \eqref{2.15}, \eqref{mongch}, and \eqref{monrch} with $R[\mathcal{O}]=1$, the gauge and R-charges of the monopole operators $T, \tilde{T}$ become
\begin{align}
   Q_{i}[T]=-Q_{i}[\tilde{T}] = -k_{i} \ , \qquad R[T] = R[\tilde{T}] = \frac{1}{2}N_{f} \ ,
\end{align}
where the $\mathcal{N}=3$ R-symmetry fixes $\Delta_m = 0$. On the geometric branch $p_{i}=q_{i}=0$, the F-term equations impose no additional relations among the bifundamental fields, and the corresponding chiral ring is
\begin{align}\label{2.37}
    \mathbb{C}[\mathcal{M}_{\mathrm{geom}}] = \Big( \frac{\mathbb{C}[A_{l}, B_{l}, T, \tilde{T}]}{\langle T\tilde{T}- (B_{2}A_{2}-B_{1}A_{1})^{N_{f}}\rangle }\Big)^{\mathbb{C}^{\ast}} \ ,
\end{align}
where the $\mathbb{C}^{\ast}$-action for $\lambda \in \mathbb{C}^{\ast}$ is given by 
\begin{align}\label{2.38}
    A_{l} \mapsto \lambda A_{l} \ , \qquad B_{l}\mapsto \lambda^{-1}B_{l} \ , \qquad T \mapsto \lambda^{-k}T \ , \qquad \tilde{T} \mapsto \lambda^{k}\tilde{T} \ .
\end{align}

It is also straightforward to identify the loci associated with positive and negative magnetic charges. For instance, the positive monopole locus is obtained by setting $B_{l}=0$, and the coordinate ring of this locus is generated by invariant monomials of the form $TA_{l_{1}}\dots A_{l_{k}}$ with $l_{j}=1,2$. Similarly, the negative monopole locus is obtained by setting $A_{l} = 0$ and its coordinate ring is generated by invariant monomials of the form $\tilde{T}B_{l_{1}}\dots B_{l_{k}}$, with $l_{j}=1,2$. Using \eqref{2.33}, the Reeb-orbit length on the monopole loci is
\begin{align}
    \ell_{\xi}\big|_{\Sigma_{\pm}} = \frac{2\pi}{N_{f}+k} \ .
\end{align}

The unrefined Hilbert series of the full geometric branch was computed for generic $\mathcal{N}=3$ circular abelian Chern-Simons quivers with flavours in \cite{Cremonesi:2016nbo}. For the ABJM theory with $N_{f}$ flavours, it takes the form \cite{Cremonesi:2016nbo}
\begin{align}
    H(t) = \frac{(1+t^2)(1-t^{2(N_{f}+k)})+2kt^{N_{f}+k}(1-t^2)}{(1-t^2)^3(1-t^{N_{f}+k})^2} \ ,
\end{align}
and the coefficient of the leading pole is
\begin{align}\label{2.41}
    \lim_{t\to 1} \, (1-t)^4H(t) = \frac{N_{f}+2k}{2(N_{f}+k)^2} \ .
\end{align}
The Hilbert series of either monopole locus can be computed directly from the definition~\eqref{characterdef}:
\begin{align}
    h_{\pm}(t) = \sum_{n=0}^{+\infty}(nk+1)t^{n(N_{f}+k)} = \frac{1+(k-1)t^{N_{f}+k}}{(1-t^{N_{f}+k})^2} \ ,
\end{align}
so that
\begin{align}\label{2.43}
    \lim_{t\to 1} \, (1-t)^2h_{\pm}(t) = \frac{k}{(N_{f}+k)^2} \ .
\end{align}

\subsection{Two-node $Q^{1,1,1}/\mathbb{Z}_{N_{f}}$}\label{sec2.2}

In \cite{Benini:2009qs, Jafferis:2009th}, an $\mathcal{N}=2$ flavoured ABJM quiver with geometric branch described by the cone over $Q^{1,1,1}/\mathbb{Z}_{N_{f}}$ was proposed. One takes vanishing bare Chern-Simons couplings $k_{i}=0$ and introduces $N_{f}$ copies of a flavour pair $(p_{l,\alpha},q_{l,\alpha})$, $\alpha=1,\dots,N_{f}$, for each $l=1,2$. The fields $p_{l,\alpha}$ and $q_{l,\alpha}$ have charges $(0,1)$ and $(-1,0)$, respectively, and couple to the operators $\mathcal{O}_{l}=A_{l}$. In other words, the deformed superpotential takes the form
\begin{align}
    \delta W = \sum_{\alpha=1}^{N_{f}}\left(p_{1,\alpha}A_{1}q_{1,\alpha}
    +p_{2,\alpha}A_{2}q_{2,\alpha}\right) \ . 
\end{align}
Note that, in contrast to the previous case, the effective Chern-Simons levels receive one-loop shifts that depend on the sign of the diagonal monopole background. Using \eqref{2.15} in the $T$ and $\tilde{T}$ backgrounds, one can get
\begin{align}
    \bigl(k_{1}^{\mathrm{eff}}(T),k_{2}^{\mathrm{eff}}(T)\bigr)
    =(-N_{f},N_{f}) \ , \qquad
    \bigl(k_{1}^{\mathrm{eff}}(\tilde{T}),k_{2}^{\mathrm{eff}}(\tilde{T})\bigr)
    =(N_{f},-N_{f}) \ .
\end{align}
With the same R-charge assignment for the bifundamental fields $(A_{l},B_{l})$ as in the previous example and applying \eqref{mongch}, \eqref{monrch}, the gauge and R-charges of the monopole operators $T,\tilde{T}$ become
\begin{align}\label{2.46}
    \bigl(Q_{1}[T],Q_{2}[T]\bigr)
    =\bigl(Q_{1}[\tilde{T}],Q_{2}[\tilde{T}]\bigr)
    =(N_{f},-N_{f}) \ , \qquad
    R[T]=R[\tilde{T}]=\frac{1}{2}N_{f} \ ,
\end{align}
where $\Delta_m=0$.\footnote{More precisely, for the $Q^{1,1,1}/\mathbb{Z}_{N_f}$ theory, with the trial R-charge assignment $R[A_l]=\Delta$ and $R[B_l]=1-\Delta$, the large-$N$ free energy is independent of $\Delta$, while $F$-maximisation fixes $\Delta_m=0$ \cite{Jafferis:2011zi}. This is because, for the $U(N)\times U(N)$ theory with vanishing Chern-Simons levels, the shift $\Delta\mapsto\Delta+\delta$ is a flat direction of the free energy and leaves $\Delta_m$ invariant. One can therefore choose $\Delta=1/2$ as a convenient representative.} Thus, the chiral ring can be expressed as
\begin{align}\label{2.47}
    \mathbb{C}[\mathcal{M}_{\mathrm{geom}}]
    =\Big(
    \frac{\mathbb{C}[A_{l},B_{l},T,\tilde{T}]}
    {\langle T\tilde{T}-(A_{1}A_{2})^{N_{f}}\rangle}
    \Big)^{\mathbb{C}^{\ast}} \ ,
\end{align}
where the $\mathbb{C}^{\ast}$-action for $\lambda\in\mathbb{C}^{\ast}$ is given by
\begin{align}\label{2.48}
    A_{l}\mapsto\lambda A_{l} \ , \qquad
    B_{l}\mapsto\lambda^{-1}B_{l} \ , \qquad
    T\mapsto\lambda^{N_{f}}T \ , \qquad
    \tilde{T}\mapsto\lambda^{N_{f}}\tilde{T} \ .
\end{align}
The loci associated with positive and negative magnetic charges are obtained by setting $A_{l}=\tilde{T}=0$ and $A_{l}=T=0$, respectively. Their coordinate rings are generated by the gauge-invariant monomials
\begin{align}
    TB_{l_{1}}\dots B_{l_{N_{f}}} \ , \qquad
    \tilde{T}B_{l_{1}}\dots B_{l_{N_{f}}} \ , \qquad
    l_{j}=1,2 \ .
\end{align}
Each of these generators has R-charge $N_{f}$, and hence \eqref{2.33} gives
\begin{align}\label{2.50}
    \ell_\xi\big|_{\Sigma_{\pm}} = \frac{\pi}{N_{f}} \ .
\end{align}

Specialising the general expression of \cite{Cremonesi:2016nbo} to the present case, one obtains the unrefined Hilbert series of the geometric branch
\begin{align}
    H(t)
    =\frac{(1+t^2)(1-t^{4N_{f}})
    +2N_{f}(1-t^2)t^{2N_{f}}}
    {(1-t^2)^3(1-t^{2N_{f}})^2} \ ,
\end{align}
with the coefficient of the leading pole
\begin{align}\label{2.52}
    \lim_{t\to 1}\,(1-t)^4H(t)
    =\frac{3}{8N_{f}} \ .
\end{align}
The Hilbert series of either monopole locus can also be computed directly from the definition~\eqref{characterdef}:
\begin{align}\label{2.53}
    h_{\pm}(t)
    =\sum_{n=0}^{+\infty}(nN_{f}+1)t^{2nN_{f}}
    =\frac{1+(N_{f}-1)t^{2N_{f}}}
    {(1-t^{2N_{f}})^2} \ ,
\end{align}
so that
\begin{align}\label{2.54}
    \lim_{t\to 1}\,(1-t)^2h_{\pm}(t)
    =\frac{1}{4N_{f}} \ .
\end{align}

\subsection{Four-node chiral $Q^{1,1,1}$-theory}\label{sec2.3}

Another theory whose geometric branch is described by the cone over $Q^{1,1,1}/\mathbb{Z}_k$ is the four-node $\mathcal{N}=2$ gauge theory with a sextic superpotential proposed in \cite{Franco:2008um} and further studied in \cite{Franco:2009sp}. The gauge theory contains three pairs of chiral bifundamental multiplets $A_l$, $B_l$, and $C_l$, with $l=1,2$.\footnote{I consider the UV theory with Chern-Simons levels $(k,k,-k,-k)$  and charges assigned as in \cite{Franco:2009sp}. The matrix model recently studied in \cite{Hosseini:2025jxb} instead uses the Chern-Simons level assignment $(k,-k,0,0)$. For $k=1$, the two assignments were shown in \cite{Franco:2008um} to yield the same geometric branch. Their equivalence for general $k$ is noted in \cite{Hosseini:2025jxb}.}

The superpotential vanishes identically in the abelian theory. With the Chern-Simons levels $(k,k,-k,-k)$, the two non-trivial moment-map relations result from the gauge transformations
\begin{align}
    &\mathbf{v}_{1}=(-1,1,1,-1) \ , \qquad
    \mathbf{v}_{2}=(1,-1,1,-1) \ , \\
    &A_l\mapsto\lambda_{1}\lambda_{2}A_l \ , \qquad
    B_l\mapsto\lambda_{1}^{-1}B_l \ , \qquad
    C_l\mapsto\lambda_{2}^{-1}C_l \ .
    \label{momentq111}
\end{align}
The gauge and R-charges of the monopole operators $T,\tilde{T}$ receive no one-loop corrections. As a GIT quotient, the chiral ring on the geometric branch is given by
\begin{align}
   \mathbb{C}[\mathcal{M}_{\mathrm{geom}}]
   =
   \Big(
    \mathbb{C}[A_l,B_l,C_l,T,\tilde{T}]
    /
    \langle T\tilde{T}-1\rangle
    \Big)^{(\mathbb{C}^{\ast})^3} \ ,
\end{align}
where the third $\mathbb{C}^{\ast}$ factor corresponds to the complexified gauge direction proportional to the Chern-Simons level vector and acts as
\begin{align}
     &A_l\mapsto A_l \ , \qquad
     (B_1,C_1)\mapsto\lambda(B_1,C_1) \ , \qquad
     (B_2,C_2)\mapsto\lambda^{-1}(B_2,C_2) \ , \nonumber\\
     &T\mapsto\lambda^{-2k}T \ , \qquad
     \tilde{T}\mapsto\lambda^{2k}\tilde{T} \ .
\end{align}
The minimally dressed gauge-invariant monopole operators of unit magnetic charge have the form
\begin{align}
   T(A_{l_1}B_1C_1)\dots(A_{l_k}B_1C_1)
   \ , \qquad
   \tilde{T}(A_{l_1}B_2C_2)\dots(A_{l_k}B_2C_2) \ ,
\end{align}
where $l_j=1,2$. The loci associated with positive and negative magnetic charges are obtained by setting $B_2=C_2=0$ and $B_1=C_1=0$, respectively. I consider the symmetric R-charge assignment $R[A_l]=R[B_l]=R[C_l]=\frac{1}{3}$ and vanishing bare monopole R-charges.

The resulting chiral ring on the geometric branch can be identified with that of the two-node flavoured theory considered in Section~\ref{sec2.2} by the reparametrisation
\begin{align}\label{2.60}
    &a_1=B_1C_2 \ , \qquad
    a_2=B_2C_1 \ , \qquad
    b_l=A_l \ , \nonumber\\
    &t=T(B_1C_1)^k \ , \qquad
    \tilde{t}=\tilde{T}(B_2C_2)^k \ .
\end{align}
These variables satisfy the relation
\begin{align}
    t\tilde{t}-(a_1a_2)^k=0 \ ,
\end{align}
and the remaining $\mathbb{C}^{\ast}$ quotient is parametrised by
$(\lambda_1\lambda_2)^{-1}$. Thus, although the UV theories are distinct, their geometric branches and the corresponding monopole loci are isomorphic upon identifying $N_f=k$.

\subsection{Two-node $V^{5,2}/\mathbb{Z}_{k}$}\label{sec24}
One can also consider the theory proposed in \cite{Martelli:2009ga}, whose abelian geometric branch is the cone over $V^{5,2}/\mathbb{Z}_k$. The abelian theory has gauge group $U(1)_k\times U(1)_{-k}$, two pairs of bifundamental chiral multiplets $(A_l, B_l)$, $l=1,2$, and two adjoint chiral multiplets $\Phi_{i}$, $i=1,2$, which are gauge neutral in the abelian theory. The superpotential takes the form
\begin{align}
    W = \frac{1}{3}(\Phi_{1}^3+\Phi_{2}^3)+\Phi_{1}(A_{1}B_{1}+A_{2}B_{2})+\Phi_{2}(B_{1}A_{1}+B_{2}A_{2}) \ , 
\end{align}
so that the F-term equations become
\begin{align}\label{2.63b}
    (\Phi_{i})^2+(A_{1}B_{1}+A_{2}B_{2}) = 0 \ , \qquad (\Phi_{1}+\Phi_{2})A_{l} = 0 \ , \qquad (\Phi_{1}+\Phi_{2})B_{l} = 0 \ .
\end{align}
As discussed in \cite{Cremonesi:2016nbo}, the F-flat moduli space has two branches. To allow non-zero vacuum expectation values for $A_{l}$ and $B_{l}$, one restricts to the geometric branch $\Phi_{1}=-\Phi_{2} \equiv \Phi_{0}$. The gauge and R-charges of the bare monopoles receive no one-loop corrections, and I take vanishing bare monopole R-charge.\footnote{For the two-node MS theory with Chern-Simons levels $(k,-k)$ and the symmetric R-charge assignment for $A_l$ and $B_l$, large-$N$ $F$-maximisation fixes $\Delta_m=0$ \cite{Jafferis:2011zi}.} The geometric branch of the theory can be described as
\begin{align}\label{2.64c}
    \mathbb{C}[\mathcal{M}_{\mathrm{geom}}] = \Big(
    \frac{\mathbb{C}[\Phi_{0}, A_{l},B_{l},T,\tilde{T}]}
    {\langle\Phi_{0}^2+(A_{1}B_{1}+A_{2}B_{2}) , T\tilde{T}-1\rangle}
    \Big)^{\mathbb{C}^{\ast}} \ , 
\end{align}
where $\mathbb{C}^{\ast}$ acts as
\begin{align}\label{2.65a}
    \Phi_{0}\mapsto \Phi_{0} \ , \qquad A_{l}\mapsto \lambda A_{l} \ , \qquad B_{l}\mapsto \lambda^{-1}B_{l} \ , \qquad T \mapsto  \lambda^{-k}T \ , \qquad \tilde{T} \mapsto\lambda^{k}\tilde{T} \ .
\end{align}
The positive and negative monopole loci are obtained by setting $B_{l}=\Phi_{0}=0$ and $A_{l}=\Phi_{0}=0$, respectively, and their coordinate rings are generated by the gauge-invariant dressed monopoles $TA_{l_{1}}\dots A_{l_{k}}$ and $\tilde{T}B_{l_{1}}\dots B_{l_{k}}$. The matter fields have R-charges $R[A_l]=R[B_l]=R[\Phi_{0}]=\frac{2}{3}$, consistent with the enhanced $SO(5)$ symmetry at $k=1$ \cite{Martelli:2009ga, Martelli:2011qj}. Thus, the Reeb-orbit length \eqref{2.33} becomes
\begin{align}\label{2.66a}
    \ell_\xi\big|_{\Sigma_{\pm}} = \frac{3\pi}{2k} \ .
\end{align}

The Hilbert series of the geometric branch was computed in \cite{Cremonesi:2016nbo}:
\begin{align}\label{2.67a}
    H(t)  = \frac{(1+t^{4/3})\big((1+t^{8/3})(1-t^{8k/3})+2kt^{4k/3}(1-t^{8/3})\big)}{(1-t^{8/3})^3(1-t^{4k/3})^2} \ , \qquad \lim_{t\to 1} \, (1-t)^4H(t) = \frac{81}{128k} \ , 
\end{align}
and those of the monopole loci are
\begin{align}\label{2.68a}
    h_{\pm}(t) = \sum_{n=0}^{+\infty}(nk+1)t^{4kn/3} = \frac{1+(k-1)t^{4k/3}}{(1-t^{4k/3})^2} \ , \qquad \lim_{t\to 1} \, (1-t)^2h_{\pm}(t) = \frac{9}{16k} \ .
\end{align}

\subsection{Flavoured $\mathcal{N}=8$ SYM}\label{sec2.5}

Another theory whose abelian geometric branch is the cone over
$V^{5,2}/\mathbb{Z}_{N_f}$ is the flavoured version of
$\mathcal{N}=8$ SYM constructed in \cite{Jafferis:2009th}.
The theory has a single gauge node with vanishing classical
Chern-Simons level, three adjoint chiral multiplets
$\Phi_{i'}$, $i'=1,2,3$, and $N_f$ fundamental-antifundamental
flavour pairs $(p_\alpha,q_\alpha)$.
The abelian superpotential can be written as
\begin{align}
    \delta W = \sum_{\alpha=1}^{N_f}
    p_\alpha\big((\Phi_1)^2+(\Phi_2)^2+(\Phi_3)^2\big)q_\alpha \ .
\end{align}
As above, I restrict to the geometric branch, on which the vacuum expectation values of the flavour fields
$(p_\alpha,q_\alpha)$ vanish.
Using \eqref{2.15} and \eqref{mongch}, the effective gauge
Chern-Simons level and the gauge charges of the diagonal
monopole operators vanish.
With $R[\Phi_{i'}]=\frac{2}{3}$ and $\Delta_m=0$
\cite{Jafferis:2011zi}, \eqref{monrch} gives
\mbox{$R[T]=R[\tilde{T}]=\frac{2}{3}N_f$}. The geometric branch is the cone over $V^{5,2}/\mathbb{Z}_{N_f}$, and the corresponding chiral ring can be expressed as
\begin{align}
    \mathbb{C}[\mathcal{M}_{\mathrm{geom}}]
    =
    \frac{\mathbb{C}[\Phi_{i'},T,\tilde{T}]}
    {\langle T\tilde{T}
    -\big((\Phi_1)^2+(\Phi_2)^2+(\Phi_3)^2\big)^{N_f}\rangle}
    \ .
\end{align}
Note that, in contrast to \eqref{2.64c}, the bare monopole
operators remain gauge invariant, while the $\mathbb{Z}_{N_f}$ quotient is encoded in the monopole OPE.
The positive and negative monopole loci are obtained by setting
$\Phi_{i'}=\tilde{T}=0$ and $\Phi_{i'}=T=0$, respectively. Notably, their coordinate rings are $\mathbb{C}[T]$ and $\mathbb{C}[\tilde{T}]$, so each locus has complex dimension one, and their links in $V^{5,2}/\mathbb{Z}_{N_f}$ are circles. The Hilbert series on the monopole loci and the lengths of the corresponding circles can be expressed as
\begin{align}
    h_\pm(t) = \frac{1}{1-t^{4N_f/3}} \ , \qquad
    \mathrm{Vol}(S^1)
    = 2\pi\lim_{t\to 1}\,(1-t)h_\pm(t)
    = \frac{3\pi}{2N_f} \ .
\end{align}

The same situation occurs in the better-studied
$\mathcal{N}=4$ ADHM theory
\cite{Benini:2009qs,Jafferis:2011zi}.
Namely, the ADHM theory can be constructed from
$\mathcal{N}=8$ SYM by flavouring a single adjoint chiral multiplet. For the abelian theory, the chiral ring on the geometric branch $\mathbb{C}^2\times(\mathbb{C}^2/\mathbb{Z}_{N_f})$
is (see, for instance, \cite{Benini:2009qs}):
\begin{align}
    \mathbb{C}[\mathcal{M}_{\mathrm{geom}}]
    =
    \frac{\mathbb{C}[\Phi_{i'},T,\tilde{T}]}
    {\langle T\tilde{T}-(\Phi_1)^{N_f}\rangle} \ .
\end{align}
With the superconformal R-charge assignment
$R[\Phi_1]=1$, $R[\Phi_2]=R[\Phi_3]=\frac{1}{2}$
and $\Delta_m=0$ \cite{Jafferis:2011zi}, the R-charges
of the diagonal monopole operators are
$R[T]=R[\tilde{T}]=N_f/2$. The Hilbert series of the monopole loci and the lengths
of their circular links are
\begin{align}
    h_\pm(t) = \frac{1}{1-t^{N_f}} \ , \qquad
    \mathrm{Vol}(S^1)
    = 2\pi\lim_{t\to 1}\,(1-t)h_\pm(t)
    = \frac{2\pi}{N_f} \ .
\end{align}

The lower-dimensional monopole loci in these cases suggest a different behaviour of the non-perturbative corrections to the free energy. For instance, in the ADHM theory, the monopole relation does not distinguish between the chiral fields $\Phi_2$ and $\Phi_3$. One can naturally associate an M2-brane instanton with the link of the complex two-dimensional locus $\tilde{T}=\Phi_1=\Phi_2=0$, which contains the positive monopole locus. However, the freedom to replace $\Phi_2=0$ by a linear condition on $(\Phi_2,\Phi_3)$ gives a family of such cycles. As discussed below, this degeneracy gives rise to additional M2-brane zero modes and is reflected in both the matrix model and M2-brane instanton contributions.

\section{M2-branes in M-theory}\label{m-theory}

The $\mathcal{N}=2$ Chern-Simons matter theories discussed in the previous section are expected to be dual to M-theory on the Freund-Rubin background $\ads \times \ses$. The compact space $\ses$ is a seven-dimensional Sasaki-Einstein manifold, admitting at least two real Killing spinors. The Euclidean background can be written as
\begin{align}
    ds^2_{11} = \Le^2 \big(\tfrac{1}{4}ds^2_{\ads}+ds^2_{\ses} \big) \ , \qquad F_{4} = \frac{3i}{8}\Le^3 \vol(\ads) \ ,
\end{align}
where $\Le$ denotes the radius of $\ses$, $\ell_{p}$ is the eleven-dimensional Planck length, and $F_{4}$ denotes the four-form flux. At leading order at large $N$, the flux-quantisation condition \cite{Klebanov:1996un, Fabbri:1999hw} gives
\begin{align}
    N = \frac{1}{(2\pi\ell_{p})^6}\int_{\ses} F_{7}
      = \frac{6\Le^6}{(2\pi \ell_{p})^6}\mathrm{Vol}(\ses) \ . \label{3.2a}
\end{align}
The relation \eqref{3.2a} between $N$ and $\Le$ may receive subleading corrections from higher-derivative curvature terms and, when present, discrete torsion (for $S^7/\mathbb Z_k$ these corrections were analysed in \cite{Bergman:2009zh}).\footnote{See also \cite{Cassia:2025jkr, BenettiGenolini:2026cyc} for related discussions of quantum corrections and changes of ensemble in more general $\ses$ examples.} 
Below, I consider the one-loop approximation to the M2-brane partition function around a BPS saddle. This approximation is controlled by the large effective tension
\begin{equation}
    \T_{2}=\frac{\Le^3}{(2\pi)^2\ell_p^3}\gg1 \ , 
\end{equation}
which can be obtained by taking the large $N$ limit with the effective Chern-Simons couplings and number of flavours held fixed. In this limit, the corrections to \eqref{3.2a} enter the classical probe action beyond one-loop order.

At the same time, it has recently been proposed \cite{Gautason:2025per, Gautason:2025plx} (see also \cite{Bobev:2026gir}) that the M2-brane partition function should be compared with the gauge theory in the grand canonical ensemble, rather than at fixed rank. In the proposed dictionary, for $\ads$ with a round $S^3$ boundary, the chemical potential $\mu$ is identified as $\mu=(\Le/2\ell_p)^3$, and the large effective tension limit corresponds to large chemical potential. In the Fermi gas picture, for theories with grand canonical potential $J(\mu)=\tfrac{1}{3}C\mu^3+\mathcal{O}(\mu)$ at large $\mu$ (free energy $F(N)\sim N^{3/2}$ at large $N$), the large chemical potential limit corresponds to large $N$ with the asymptotic relation \cite{Marino:2011eh, Marino:2012az}:
\begin{align}\label{gce}
    N \approx C\mu^{2} \ , \qquad C = \frac{2}{\pi^2} \frac{\mathrm{Vol}(\ses)}{\mathrm{Vol}(S^7)} \ ,
\end{align}
where $C$ can be determined by matching the leading free energy on the gravitational and gauge theory sides. At the accuracy of the one-loop probe approximation, comparison of the semiclassical M2-brane partition function with the gauge theory does not distinguish between the two
ensembles, and the saddle contribution can be transformed between the two ensembles using \eqref{gce}.\footnote{It should be understood, however, that finite $N$ or finite $\mu$ corrections in the two ensembles generically lead to different predictions for higher-loop contributions to the semiclassical M2-brane partition function in the large effective tension expansion. The two-loop analysis of the M2-brane partition function in the holographic context was recently discussed in \cite{Beccaria:2025ahf, Beccaria:2026zjj, Tseytlin:2026dhy}. In \cite{Beccaria:2026zjj}, it was found that the two-loop correction to the $\mathrm{AdS}_{2}\times S^{1}$ M2-brane partition function, describing the $\tfrac{1}{2}$-BPS Wilson loop in ABJM, vanishes with supersymmetric boundary conditions, supporting the proposed grand canonical interpretation of \cite{Gautason:2025plx}. Similarly, the two-loop correction to the partition function of the $S^3/\mathbb{Z}_{k}$ M2-brane instanton, which reproduces the leading worldsheet-instanton contribution to the ABJM matrix model for $k>2$, was argued to vanish in \cite{Tseytlin:2026dhy}. Analogous higher-loop corrections for the less supersymmetric theories considered here remain to be understood.} For convenience, one can introduce a common variable
\begin{align}
    \mu_{\ast}\equiv \sqrt\frac{N}{C} \ ,
\end{align}
and express the observables in terms of it, so that the effective tension can be equivalently expressed as
\begin{align}\label{tensionm2}
    \T_{2} = \frac{\Le^3}{(2\pi)^2\ell_p^3} = \frac{2\mu_{\ast}}{\pi^2} = \sqrt{\frac{2\pi^2N}{3\mathrm{Vol}(\ses)}} \ .
\end{align}

The M2-brane instantons in the $\ses$ geometry arise from BPS three-cycles $\Sigma\subset \ses$, and the one-loop partition function is determined by the quadratic fluctuations of the BST action \cite{Bergshoeff:1987cm} around $\Sigma$. As demonstrated in \cite{Beccaria:2023ujc}, such a semiclassical M2-brane calculation, for Chern-Simons level $k>2$, successfully reproduces the leading ``worldsheet'' instanton contribution to the free energy of ABJM on $S^3$. In \cite{Kurlyand:2026yke}, M2-brane instantons were considered in generic $\ses$ backgrounds, and a special simplification in the one-loop analysis was observed for M2-branes preserving the two real Killing spinors associated with the Sasaki-Einstein structure. The Sasaki-Einstein manifold has a Calabi-Yau cone $\cyf = C(\ses)$, and these M2-branes wrap three-dimensional Sasakian (invariant) links of holomorphic subcones of $\cyf$. As further discussed in Section~\ref{m2inst}, the one-loop determinant for such branes depends on the Reeb data of $\ses$, while the transverse structure enters through equivariant indices.

Identifying the geometric branch of the moduli space \eqref{2.26}
with the cone $\cyf=C(\ses)$ and twice the superconformal R-charges
with the Reeb weights, one can use the generators of the chiral
ring as convenient algebraic coordinates on $\cyf$. The M2-brane instantons can then be specified by holomorphic subcones of complex dimension two in these coordinates. Interestingly, the ABJM example suggests that the candidate cones can be supplied by the two-dimensional monopole loci of the geometric branch \cite{Kurlyand:2026yke}. Guided by this intuition, one can identify M2-brane instantons in the examples of Section~\ref{2} by examining subcones containing the monopole loci. As demonstrated below, both the classical exponent and the one-loop prefactor of an isolated M2-brane instanton can be computed directly from the Reeb and algebraic data. Before turning to the M2-brane instanton calculation, I first comment on how the monopole loci appear on the gravity side.

\subsection{The monopole loci in $\cyf$ and links in $\ses$}\label{sec3.1}

With the identification of the geometric branch $\mathcal{M}_{\mathrm{geom}}$ on the gauge side with the Calabi-Yau cone $\cyf$ over the Sasaki-Einstein manifold $\ses$ on the M-theory side, the monopole loci $\mathcal{M}_{\pm}\subset\cyf$ are holomorphic subcones. Assuming that they are smooth away from the apex, the corresponding links $\Sigma_{\pm}=\mathcal{M}_{\pm}|_{r=1}$ are Sasakian.

I will further assume that the $U(1)_m$-action is holomorphic, tangent to the slices of fixed radius, and preserves the holomorphic four-form on $\cyf$. Consequently, $U(1)_m$ is an isometry of $\ses$ and commutes with the $U(1)_{r}$-action. I denote by $\zeta$ the Killing vector generating the $U(1)_m$-action, normalised so that its flow has period $2\pi$. For concreteness, and as in the explicit examples of Section~\ref{2}, I consider the case for which the generators of each coordinate ring $\mathbb C[\mathcal M_\pm]$ have the same R-charge and $U(1)_m$-charges $\pm1$ on $\mathcal M_\pm$, respectively. In this case, the $U(1)_m$-action is free on $\Sigma_\pm$, and the $U(1)_m$- and $U(1)_r$-orbits on $\mathcal M_\pm$, and consequently on the links $\Sigma_\pm$, coincide up to orientation.

Alternatively, to identify the links on which the $U(1)_m$- and $U(1)_r$-orbits coincide, one can introduce a Hamiltonian function $h_m$ and consider its connected critical components $\Sigma$ \cite{Farquet:2013cwa}:
\begin{align}
h_m=\eta(\zeta)\ , \qquad
dh_m\big|_{\Sigma}=0\ . \label{mcrit}
\end{align}
Here $\eta$ denotes the contact form on $\ses$, with $\eta(\xi)=1$. Since $\zeta$ is Killing and preserves the contact form, from the Cartan formula and the non-degeneracy of $d\eta|_{\ker\eta}$ it immediately follows
\begin{align}\label{propreeb}
\zeta\big|_{\Sigma}=h_{\ast}\xi\big|_{\Sigma}\ ,\qquad
h_{\ast}=h_m\big|_{\Sigma} \ ,
\end{align}
where $h_\ast$ is constant on each connected critical component. For $h_\ast\neq 0$, it then follows that the $U(1)_m$- and $U(1)_r$-orbits coincide on $\Sigma$, up to orientation. Moreover, since the $U(1)_m$-action is holomorphic and preserves the radial coordinate, the cone over each smooth connected critical component is a holomorphic subcone of $\cyf$, and hence its link is Sasakian. Thus, the subcones corresponding to the non-zero critical values of $h_{m}$ provide natural candidates for the monopole loci $\mathcal M_\pm$. 

The relation between the cones over the critical sets of $h_{m}$ and the monopole loci can be further supported by a simple argument. To model the bulk state dual to a gauge-invariant BPS monopole operator, consider a massless particle moving along the M-theory circle. 
In the Lorentzian $\ads \times \ses$ background, the action for the massless particle with momentum in $\ses$ can be written as
\begin{align}
    S = \mathrm{L}^2\int e^{-1}\big( -\frac{1}{4}\dot{t}^2+\langle \dot{\mathrm{X}}, \dot{\mathrm{X}}\rangle_{\ses} \big)d\tau \ , 
\end{align}
where $e$ is the einbein and the particle is at the centre of $\ads$. A simple calculation shows that the energy $E$ of the massless particle is bounded below by its physical R-charge
\begin{align}
    E \geq R \ , \qquad R = \frac{1}{2}|J_{r}| \ ,  
\end{align}
where $J_{r}$ denotes the momentum component along the Reeb circle. For $\dot{\mathrm{X}}\sim  \zeta$, the bound $E\geq R$ is saturated precisely on the critical sets \eqref{mcrit}.\footnote{The resulting BPS condition is equivalent to the supersymmetric M-theory circle condition discussed in \cite{Farquet:2013cwa}.} As the momentum along the M-theory circle is quantised, on a critical component $\Sigma$, the energy of the particle can be expressed in terms of the critical value $h_{\ast}=h_m|_{\Sigma}$ as
\begin{align}
    E = R = \frac{|n|}{2|h_{\ast}|} \ , \qquad n \in \mathbb{Z} \ . \label{3.6}
\end{align}
On the critical components identified with $\Sigma_\pm$, the $U(1)_m$- and $U(1)_r$-orbits coincide. With the normalisation for $\zeta$ and $\xi$ stated above, it follows that $2\pi |h_{\ast}| = \ell_{\xi}|_{\Sigma_{\pm}}$. Thus, for the minimal non-zero physical R-charge of such a massless particle:
\begin{align}
   R_{\pm}  = \frac{1}{2|h_{\ast}|}=\frac{\pi}{\ell_{\xi}|_{\Sigma_{\pm}}} \ , 
\end{align}
which is just the monopole R-charge formula \eqref{2.33}.

\subsection{M2-brane instantons}\label{m2inst}

When the monopole loci have complex dimension two, it is
straightforward to relate them to M2-brane instantons. In this case, the corresponding Sasakian links $\Sigma_{\pm}=\mathcal{M}_{\pm}|_{r=1}$ have $\dim_{\mathbb{R}}\Sigma_{\pm}=3$ and can be wrapped by Euclidean M2-branes. For the examples of Section~\ref{sec2.5}, where the monopole loci have complex dimension one, candidate M2-brane cycles $\Sigma_{\pm}'$ can instead be defined as the links of complex two-dimensional holomorphic subcones containing these loci, as specified explicitly below. In either case, the corresponding M2-branes wrap Sasakian links of holomorphic subcones of $\cyf$ and thus preserve the two real Killing spinors associated with the Sasaki-Einstein structure on $\ses$ and are BPS \cite{Kurlyand:2026yke}.

With the dictionary \eqref{tensionm2}, the classical exponents for the M2-brane instantons of interest become
\begin{align}\label{m2class}
    S
    =
    \frac{\Le^3}{(2\pi)^2\ell_p^3}
    \mathrm{Vol}(\Sigma)
    =4\mu_{\ast} \,  \frac{\mathrm{Vol}(\Sigma)}{\mathrm{Vol}(S^3)}  = 
    2\pi\sqrt{\frac{N}{6}}\,
    \frac{\mathrm{Vol}(\Sigma)}
         {\sqrt{\mathrm{Vol}(\ses)}} \ ,
\end{align}
where the volumes of both $\Sigma$ and $\ses$ can be computed
from the Reeb/superconformal R-symmetry data using the character
formulae \eqref{2.29}, without requiring an explicit metric on $\ses$.\footnote{For the examples of Section~\ref{sec2.5}, the character
formula for $\mathrm{Vol}(\Sigma_{\pm}')$ is applied to the
coordinate rings of the corresponding complex two-dimensional
subcones containing the monopole loci.}

In \cite{Kurlyand:2026yke}, a general formula for the one-loop
prefactor of an isolated M2-brane instanton wrapping a Sasakian
link $\Sigma$ was derived from the semiclassical analysis of
quadratic fluctuations around the BPS M2-brane. The crucial observation is that, for a Sasakian (invariant) link $\Sigma$, the worldvolume supersymmetry associated with the two preserved real Killing spinors on $\ses$ organises the quadratic fluctuations into a Rozansky-Witten (RW) multiplet and a McLean (ML) multiplet, describing normal fluctuations to $\Sigma$ in $\ads$ and $\ses$, respectively. The supercharge $\mathcal{Q}$ acting on the worldvolume closes
off-shell as
\begin{align}
    \mathcal{Q}^2=R \ , \qquad
    R=-i\mathcal{L}_{\xi}+\mathcal{R} \ ,
\end{align}
where $\mathcal{L}_{\xi}$ is the Lie derivative on the relevant
vector bundle along the induced Reeb vector, while
$\mathcal{R}=2$ for the RW multiplet and $\mathcal{R}=0$ for the
ML multiplet.

The one-loop factor associated with a bundle $E$ can be expressed
as
\begin{align}\label{equivoneloop}
    Z_1^{(E)}
    =
    \frac{\det_{\operatorname{coker}\bar{\partial}_E}R}
         {\det_{\ker\bar{\partial}_E}R} \ ,
\end{align}
where $E$ denotes the relevant $U(1)_r$-equivariant complex vector
bundle over $\Sigma$. For the RW multiplet, $E$ is the direct sum
of two trivial line bundles, while for the ML multiplet,
$E=N\Sigma$ is the normal bundle of $\Sigma$ in $\ses$, endowed
with the transverse complex structure. The Dolbeault-type operators $\bar{\partial}_E$ appearing are transversally elliptic operators relative to the $U(1)_{r}$ action (see \cite{Kurlyand:2026yke} for details). The determinant formula \eqref{equivoneloop} requires separate
treatment when $R$ has zero modes in $\ker\bar{\partial}_E$ or
$\operatorname{coker}\bar{\partial}_E$. Such zero modes are absent for the two-dimensional monopole loci considered below, but appear, for instance, for the flavoured $\mathcal{N}=8$ examples with one-dimensional loci. The treatment of zero modes for these cases is discussed separately. 

For a quasi-regular Sasaki-Einstein manifold, the $U(1)_r$-action
is locally free, and the link $\Sigma$ may contain points with
non-trivial isotropy group $\mathbb{Z}_l\subset U(1)_r$.
The orbit spaces $\ses/U(1)_r$ and $\Sigma/U(1)_r$ are naturally
compact complex orbifolds (see, for instance,
\cite{Martelli:2006yb}), and their isotropy actions must be taken
into account in the $U(1)_r$ Fourier decomposition. If the geometry
is a finite quotient of a regular Sasaki-Einstein manifold, one can
decompose the fluctuations into $U(1)_r$ Fourier modes on the
covering link and project with respect to the finite group action,
as was done in \cite{Kurlyand:2026yke}. I will apply this method directly to the candidate M2-brane instantons appearing in the dual description of the flavoured $\mathcal{N}=8$ examples.

In the case of the links $\Sigma_{\pm}$ resulting directly from the monopole loci, however, the $U(1)_r$- and $U(1)_m$-orbits coincide on $\Sigma_{\pm}$ up to orientation, and the Reeb orbits on each link have the same length. As the $U(1)_m$ action on $\Sigma_{\pm}$ is free,
these links can equivalently be viewed as principal $U(1)_m$-bundles
$p:\Sigma_{\pm}\to \Sigma_{\pm}/U(1)_m$ over smooth compact
Riemann surfaces $\B_{\pm} \equiv \Sigma_{\pm}/U(1)_m$, while the
integer magnetic charges ensure that the $U(1)_m$-action
\eqref{2.22} is $2\pi$-periodic. The $U(1)_m$-action commutes with
$U(1)_r$ and the vector bundles $E$ appearing are
$U(1)_r\times U(1)_m$-equivariant. Moreover, the operators
$\bar{\partial}_E$ relevant here are
$U(1)_r\times U(1)_m$-invariant.\footnote{On $\Sigma_{\pm}$,
$\zeta=h_*\xi$ with constant $h_*$ (see \eqref{propreeb}) and thus $[\mathcal{L}_{\zeta},\bar{\partial}_H]=h_*[\mathcal{L}_{\xi},\bar{\partial}_H]=0$.
For the normal bundle, one can use that $\zeta$ is tangent to $\Sigma_{\pm}$,
is Killing on $\ses$, and preserves the contact structure, so that its flow preserves the normal connection $\mathcal{L}_{\zeta}\nabla^{\perp}=0$ and  $[\mathcal{L}_{\zeta},\bar{\partial}_N]=0$. Note also that, since $U(1)_{r}$- and $U(1)_m$-orbits coincide on $\Sigma_{\pm}$ and $\bar{\partial}_{E}$ are $U(1)_{r}$-transversally elliptic, $\bar{\partial}_{E}$ are $U(1)_m$-transversally elliptic.}
This leads to a simplification for M2-branes wrapping $\Sigma_{\pm}$ as it allows \eqref{equivoneloop} to be computed by decomposing the fluctuations into $U(1)_m$ Fourier modes and determining their Reeb weights.\footnote{Equivalently,
one can use a $U(1)_r$ Fourier decomposition for $\Sigma_{\pm}$, which gives the same final result for the one-loop partition function. A useful local description is obtained by choosing the gauge $T=1$ or $\tilde{T}=1$ on the positive or negative monopole locus, respectively. Note, however, that in this gauge one must also impose invariance under residual finite gauge transformations.}

The reduction to the $U(1)_{r}$-equivariant indices on the base $\B_{\pm}$ through the $U(1)_{m}$ Fourier
decomposition is standard \cite{Atiyah:1974obx} (see also
\cite{Pestun:2016qko}).
Concretely, consider the principal $U(1)_m$-bundle
$p:\Sigma_{\pm}\to\B_{\pm}$, and let $\mathcal{L}$ be the associated
line bundle. Since $E$ is $U(1)_{r}\times U(1)_m$-equivariant and
$U(1)_m$ acts freely on $\Sigma_{\pm}$, there is a
$U(1)_{r}$-equivariant vector bundle $\mathcal{E}$ on $\B_{\pm}$
such that $E = p^{\ast}\mathcal{E}$. The one-loop partition function
\eqref{equivoneloop} can therefore be formally expressed as
\begin{align}\label{oneloopprod}
    Z_1^{(E)}=\prod_{m\in\mathbb{Z}}
    \bigl(\lambda_m^{\xi}+\mathcal{R}\bigr)^{-\chi_m(E)} \ ,
\end{align}
where $m$ is the $U(1)_m$ Fourier label and $\lambda_m^{\xi}$
denotes the eigenvalue of $-i\mathcal{L}_{\xi}$ on the
corresponding sections of $\mathcal{E}\otimes\mathcal{L}^m$,
including its action on the local bundle frame. The multiplicity
$\chi_m(E)$ in the present case is the index of the reduced Dolbeault operator:
\begin{align}\label{indexdolbm}
    \chi_m(E) = \ind(\bar{\partial}_{\mathcal W_m}) \  , \qquad \bar{\partial}_{\mathcal W_m}: \Gamma(\B_{\pm},\mathcal W_m) \to \Omega^{0,1}(\B_{\pm},\mathcal W_m) \ , \qquad \mathcal W_m \equiv \mathcal{E}\otimes \mathcal{L}^m \ ,
\end{align}
where $\mathcal{L}^{m}\equiv\mathcal{L}^{\otimes m}$. For
$E=N\Sigma_{\pm}$, I further denote the descended bundle by
$\mathcal{E}=N_{\B_{\pm}}$. Note that if a fixed $U(1)_m$ Fourier sector contains the set $W_{U(1)_{r}}$ of $U(1)_{r}$ weights, the product \eqref{oneloopprod} is understood separately for each weight. In particular, the exponent corresponding to each $U(1)_{r}$ eigenvalue $\lambda_{\alpha,m}^{\xi}$ becomes $\chi_m(E_{\alpha})$, with $\chi_m(E)=\sum_{\alpha\in W_{U(1)_r}}\chi_m(E_{\alpha})$.

In the examples below, the holomorphic scalar modes in the
$U(1)_m$ Fourier decomposition are represented by gauge-invariant
monopole operators $\T$ on the corresponding monopole loci. The normal bundle $N_{\B}$ can be determined explicitly by introducing gauge-invariant monopole and mesonic coordinates near the monopole locus. The eigenvalues $\lambda_m^{\xi}$ can be determined by the induced Reeb action on sections of $\mathcal{E}\otimes\mathcal{L}^{m}$. In particular, for a holomorphic RW mode represented by a gauge-invariant monopole operator $\T^m$ of magnetic charge $m$, one has $\lambda_m^{\xi}=2R[\T^m]$, while for the ML multiplet, one also includes the Reeb weights of the normal frame.

\subsubsection*{$\mathcal{N}=3$ flavoured ABJM theory}

In the case of the flavoured ABJM theory, using \eqref{2.41} and \eqref{2.43}, one can get for the classical exponent \eqref{m2class} of the M2-brane instantons wrapping $\Sigma_{\pm}$:
\begin{align}\label{3.15}
    S_{\pm} = \frac{4k \mu_{\ast} }{(N_{f}+k)^2} \ , \qquad \mu_{\ast} =\pi (N_{f}+k)\sqrt
    \frac{N}{N_{f}+2k} \ .
\end{align}

For definiteness, let us consider an M2-brane wrapping the link $\Sigma_{+}$ of the positive monopole locus $B_{l}=0$ in the geometric branch \eqref{2.37}. The surface $\B_{+}=\Sigma_{+}/U(1)_{m}\simeq\mathbb{P}^1$ has homogeneous coordinates $[A_{1},A_{2}]$. With the gauge transformations \eqref{2.38}, it is convenient to introduce local gauge-invariant coordinates around the monopole locus on two affine charts of $\mathbb{P}^1$:
\begin{align}
    & A_{1}\neq 0: \qquad
    z_{1}=\frac{A_{2}}{A_{1}} \ , \qquad
    \T_{1}=TA_{1}^{k} \ , \qquad
    \delta x^{(l)}_{1}=A_{1}\delta B_{l} \ , \\
    & A_{2}\neq 0: \qquad
    z_{2}=\frac{A_{1}}{A_{2}} \ , \qquad
    \T_{2}=TA_{2}^{k} \ , \qquad
    \delta x^{(l)}_{2}=A_{2}\delta B_{l} \ ,
\end{align}
while the remaining coordinate $\tilde{T}$ is determined from the monopole OPE on each patch. On the intersection of the affine charts, the transition functions become
\begin{align}
    z_{2}=z_{1}^{-1} \ , \qquad
    \T_{2}=z_{1}^{k}\T_{1} \ , \qquad
    \delta x_{2}^{(l)}=z_{1}\delta x_{1}^{(l)} \ ,
\end{align}
so that the $N_{\B_{+}}$ bundle, parametrised by $\delta x^{(l)}$, is
\mbox{$N_{\B_{+}}\simeq\mathcal{O}(-1)^{\oplus 2}$}.
The gauge-invariant operators $\T$ transform as local frames
of $\mathcal{L}\simeq\mathcal{O}(k)$.

The contribution of the RW multiplet to the one-loop partition function can be determined in a straightforward manner. A section of $\mathcal{L}^{m}\simeq\mathcal{O}(mk)$ can be
written locally as
\begin{align}
    f_{m}=f_{m}(z)\T^{m} \ , \qquad
    f_{m}(z_{2})=z_{1}^{-mk}f_{m}(z_{1}) \ ,
\end{align}
and, using the R-charge assignment in Section~\ref{sec2.1},
$f_{m}$ has a Reeb weight twice the R-charge of the
gauge-invariant monopole operator:
\begin{align}
    -i\mathcal{L}_{\xi}f_{m}
    =m(N_{f}+k)f_{m} \ .
\end{align}
As $\chi_{m}=mk+1$, the contribution of the RW multiplet becomes
\begin{align}
    Z_{1}^{\mathrm{RW}}
    =\prod_{m\in\mathbb{Z}}
    \big(m(N_{f}+k)+2\big)^{-2(mk+1)} \ .
\end{align}

Similarly, for the ML multiplet, one can consider sections of
$N_{\B_{+}}\otimes\mathcal{L}^{m}
\simeq\mathcal{O}(mk-1)^{\oplus 2}$
locally parametrised as
\begin{align}
    s_{m}^{(l)}
    =s_{m}^{(l)}(z)\,
      \partial_{x^{(l)}}\otimes\T^m \ , \qquad
    s_{m}^{(l)}(z_{2})
    =z_{1}^{-mk+1}s_{m}^{(l)}(z_{1}) \ ,
\end{align}
with the Reeb weights
\begin{align}
    -i\mathcal{L}_{\xi}s_{m}^{(l)}
    =\big(m(N_{f}+k)-2\big)s_{m}^{(l)} \ ,
\end{align}
and $\chi_{m}(N\Sigma_{+})=2mk$. Thus, the contribution of the
ML multiplet can be written as
\begin{align}
    Z_{1}^{\mathrm{ML}}
    =\prod_{m\in\mathbb{Z}}
    \big(m(N_{f}+k)-2\big)^{-2mk} \ .
\end{align}

Combining the two multiplets and using Riemann $\zeta$-function
regularisation, one gets for the one-loop partition function
for $k>2$:\footnote{For $N_f=0$ and $k>2$, this reproduces the one-loop determinant computed in \cite{Beccaria:2023ujc} for the leading M2-brane instanton in ABJM theory. The overall factor of $4$ in that case can formally be accounted for by counting $\Sigma_{\pm}$ with both orientations.}
\begin{align}
    Z_{1}
    =\frac{1}{4\sin^2(\frac{2\pi}{N_{f}+k})} \ .
\end{align}
The same contribution is obtained for an M2-brane wrapping
$\Sigma_{-}$.

\subsubsection*{$Q^{1,1,1}/\mathbb{Z}_{N_{f}}$ theory}

For the $Q^{1,1,1}/\mathbb{Z}_{N_{f}}$ theory defined in
Section~\ref{sec2.2}, the classical exponents for the links
$\Sigma_{\pm}$ become
\begin{align}
   S_{\pm} =\frac{\mu_{\ast}}{N_{f}} \ , \qquad \mu_{\ast} = 2\pi \sqrt{\frac{N_{f}N}{3}} \ , 
\end{align}
where I have used \eqref{2.52} and \eqref{2.54}.

To determine the one-loop partition function, let us choose
$\Sigma_{+}$ specified by $A_{l}=\tilde{T}=0$ in the coordinates \eqref{2.47}, so that $\B_{+}=\Sigma_{+}/U(1)_m\simeq\mathbb{P}^{1}$ with homogeneous coordinates $[B_{1},B_{2}]$. In the affine charts, one can introduce local gauge-invariant coordinates
\begin{align}
    & B_{1}\neq 0: \qquad
    z_{1}=\frac{B_{2}}{B_{1}} \ , \qquad
    \T_{1}=TB_{1}^{N_{f}} \ , \qquad
    \delta x_{1}^{(l)}=B_{1}\delta A_{l} \ , \\
    & B_{2}\neq 0: \qquad
    z_{2}=\frac{B_{1}}{B_{2}} \ , \qquad
    \T_{2}=TB_{2}^{N_{f}} \ , \qquad
    \delta x_{2}^{(l)}=B_{2}\delta A_{l} \ ,
\end{align}
and, as in the previous case, $\tilde{T}$ is determined by the
monopole OPE. From the transition functions between the charts,
it also follows that $\mathcal{L}\simeq\mathcal{O}(N_{f})$ and
$N_{\B_{+}}\simeq\mathcal{O}(-1)^{\oplus2}$.

With the R-charge assignment in Section~\ref{sec2.2}, the
contribution of the RW multiplet is
\begin{align}\label{rwq111}
    Z_{1}^{\mathrm{RW}}
    =\prod_{m\in\mathbb{Z}}
    \big(2mN_{f}+2\big)^{-2(mN_{f}+1)} \ .
\end{align}

To determine the contribution of the ML multiplet, one can
consider a local parametrisation of the sections of
$N_{\B_{+}}\otimes\mathcal{L}^{m}
\simeq\mathcal{O}(mN_{f}-1)^{\oplus2}$:
\begin{align}
    s_{m}^{(l)}
    =s_{m}^{(l)}(z)\partial_{x^{(l)}}\otimes\T^{m} \ ,
    \qquad
    s_{m}^{(l)}(z_{2})
    =z_{1}^{-mN_{f}+1}s_{m}^{(l)}(z_{1}) \ ,
\end{align}
with the Reeb weights
\begin{align}
    -i\mathcal{L}_{\xi}s_{m}^{(l)}
    =(2mN_{f}-2)s_{m}^{(l)} \ ,
\end{align}
and $\chi_{m}(N\Sigma_{+})=2mN_{f}$.
Therefore, the ML contribution becomes
\begin{align}\label{mlq111}
    Z_{1}^{\mathrm{ML}}
    =\prod_{m\in\mathbb{Z}}
    \big(2mN_{f}-2\big)^{-2mN_{f}} \ .
\end{align}
Combining \eqref{rwq111} and \eqref{mlq111}, and using Riemann
$\zeta$-function regularisation, one gets for $N_{f}>1$:
\begin{align}
    Z_{1}
    =\frac{1}{4\sin^2(\frac{\pi}{N_{f}})} \ .
\end{align}

One can also note that, with the identification of the geometric
branches \eqref{2.60}, the same M2-brane instanton and its one-loop prefactor appear in the four-node $Q^{1,1,1}/\mathbb{Z}_{k}$ theory with $k\leftrightarrow N_{f}$.\footnote{A similar identification between the one-loop partition functions of M2-branes describing Wilson loops in these theories appears in \cite{Gautason:2025bft}.}

\subsubsection*{$V^{5,2}/\mathbb{Z}_k$ theory}

One can also consider the $V^{5,2}/\mathbb{Z}_{k}$ theory discussed in Section~\ref{sec24}. Using \eqref{2.67a} and \eqref{2.68a}, the classical exponents for both links $\Sigma_{\pm}$ associated with the monopole loci become
\begin{align}
    S_{\pm} = \frac{9\mu_{\ast}}{4k}\ , \qquad \mu_{\ast} = \frac{8\pi}{9}\sqrt{kN} \ .
\end{align}

To compute the one-loop determinant, let us consider the case
of $\Sigma_{+}$, whose cone is specified by setting $B_{l}=\Phi_{0}=0$, while the case of $\Sigma_{-}$ can be treated analogously. The corresponding genus-zero curve is $\B_{+}=\Sigma_{+}/U(1)_{m}\simeq\mathbb{P}^1$ with homogeneous coordinates $[A_{1},A_{2}]$. With the gauge transformations \eqref{2.65a}, the local gauge-invariant coordinates near the monopole locus can be chosen as
\begin{align}
    & A_{1}\neq 0: \qquad
    z_{1}=\frac{A_{2}}{A_{1}} \ , \qquad
    \T_{1}=TA_{1}^{k} \ , \qquad
    \delta x_{1}^{(1)}=\delta\Phi_{0} \ , \qquad
    \delta x_{1}^{(2)}=A_{1}\delta B_{2} \ , \\
    & A_{2}\neq 0: \qquad
    z_{2}=\frac{A_{1}}{A_{2}} \ , \qquad
    \T_{2}=TA_{2}^{k} \ , \qquad
    \delta x_{2}^{(1)}=\delta\Phi_{0} \ , \qquad
    \delta x_{2}^{(2)}=-A_{2}\delta B_{1} \ ,
\end{align}
where I have used that linearising the F-term equation
\eqref{2.63b} around the monopole locus gives
\begin{align}
A_{1}\delta B_{1}+A_{2}\delta B_{2}=0 \ ,
\end{align}
and thus the transverse
directions $\delta B_{l}$ are not independent.
The transition functions become
\begin{align}
    z_{2}=z_{1}^{-1} \ , \qquad
    \T_{2}=z_{1}^{k}\T_{1} \ , \qquad
    \delta x_{2}^{(1)}=\delta x_{1}^{(1)} \ , \qquad
    \delta x_{2}^{(2)}=z_{1}^{2}\delta x_{1}^{(2)} \ ,
\end{align}
and thus $\mathcal{L}\simeq\mathcal{O}(k)$ and
$N_{\mathcal{B}_{+}}\simeq\mathcal{O}\oplus\mathcal{O}(-2)$.

With the R-charges
$R[A_{l}]=R[B_{l}]=R[\Phi_{0}]=2/3$
and zero R-charges of the bare monopole operators, one gets for
the RW contribution
\begin{align}
    Z_{1}^{\mathrm{RW}}
    =\prod_{m\in\mathbb{Z}}
    \Big(\frac{4}{3}mk+2\Big)^{-2(mk+1)} \ .
\end{align}

To determine the contribution of the ML multiplet, one considers \mbox{$N_{\B_{+}}\otimes\mathcal{L}^{m}
\simeq\mathcal{O}(mk)\oplus\mathcal{O}(mk-2)$}
with local parametrisation as above, so that the sections
transform as
\begin{align}
    s_{m}^{(1)}(z_{2})
    =z_{1}^{-mk}s_{m}^{(1)}(z_{1}) \ , \qquad
    s_{m}^{(2)}(z_{2})
    =z_{1}^{-mk+2}s_{m}^{(2)}(z_{1}) \ ,
\end{align}
and have Reeb weights
\begin{align}
    -i\mathcal{L}_{\xi}s_{m}^{(1)}
    =\frac{4}{3}(mk-1)s_{m}^{(1)} \ , \qquad
    -i\mathcal{L}_{\xi}s_{m}^{(2)}
    =\frac{4}{3}(mk-2)s_{m}^{(2)} \ .
\end{align}
Because the two normal components have different Reeb weights,
one gets for the ML contribution
\begin{align}
    Z_{1}^{\mathrm{ML}}
    =\prod_{m\in\mathbb{Z}}
    \Big(\frac{4}{3}(mk-1)\Big)^{-mk-1}
    \Big(\frac{4}{3}(mk-2)\Big)^{-mk+1} \ .
\end{align}

Combining the RW and ML contributions and using $\zeta$-function regularisation for the total product, one obtains for $k>2$:
\begin{align}
    Z_{1}
    =
    \frac{\sin(\frac{2\pi}{k})}
         {4\sin(\frac{\pi}{k})\sin^2(\frac{3\pi}{2k})}
    \,e^{\delta S(k)} \ ,
\end{align}
where $\delta S(k)$ denotes:
\begin{align}
    \delta S(k)
    =k
    \Big(
        \ell\Big(\frac{1}{k}\Big)+\ell\Big(\frac{2}{k}\Big)-2\ell\Big(\frac{3}{2k}\Big)
    \Big) \ .
\end{align}
The function $\ell(z)$ is defined as 
\begin{align}\label{3.48aaaa}
     \ell(z) = - z\log \big(1-e^{2\pi iz} \big)+\frac{i}{2}\big(\pi z^2 +\frac{1}{\pi}\mathrm{Li}_{2}(e^{2\pi iz})  \big)-\frac{i\pi}{12}  \ , 
\end{align}
and also appears in \eqref{loneloop} in the discussion of the localisation matrix model.\footnote{Also note that $\ell(z)$ was argued to appear as one of the building blocks of the one-loop determinant for an M2-brane instanton in \cite{Gautason:2025per}. Its exponential gives the $\mathcal{N}=2$ chiral multiplet contribution to the partition function on $S^3$ \cite{Jafferis:2010un, Hama:2010av}:
$$
 e^{\ell(z)} = \prod_{n=1}^{\infty}\Big(\frac{n+z}{n-z}\Big)^{n} \ .
$$}

\subsubsection*{ADHM theory}

For the ADHM theory, discussed in Section~\ref{sec2.5}, let us consider the cone $\mathcal{M}'_{+} = \{\tilde{T} =\Phi_{1}=\Phi_{2} = 0\}$, for which the coordinate ring becomes
\begin{align}
    \mathbb{C}[\mathcal{M}'_{+}] = \mathbb{C}[\Phi_{3}, T]\ .
\end{align}
A similar analysis can be performed for $\mathcal{M}_{-}'= \{T =\Phi_{1}=\Phi_{2} = 0\}$. With the R-charges $R[\Phi_{3}]=1/2$ and $ R[T] =N_{f}/2$, it is straightforward to compute the Hilbert series of $\mathbb{C}[\mathcal{M}'_{+}]$ and the volume of the corresponding $S^3/\mathbb{Z}_{N_{f}}$ link:
\begin{align}
    &h_{+}'(t) = \sum_{m, l\geq 0}t^{l+mN_{f}} = \frac{1}{(1-t)(1-t^{N_{f}})} \ , \\
    &\mathrm{Vol}(\Sigma_{+}') = 2\pi^2 \lim_{t\to 1}\, (1-t)^2 h_{+}'(t) = \frac{2\pi^2}{N_{f}} \ .
\end{align}
With $\ses=S^7/\mathbb{Z}_{N_{f}}$, \eqref{m2class} gives the classical exponent
\begin{align}\label{3.48}
    S_{+} = \frac{4\mu_{\ast}}{N_{f}} \ , \qquad \mu_{\ast} = \pi\sqrt{\frac{N_{f}N}{2}} \ .
\end{align}
Note that the $U(1)_{m}$ action is not free on the link $\Sigma_{+}'$, and thus the reduction based on the $U(1)_{m}$ Fourier mode expansion, outlined around \eqref{oneloopprod}, does not apply in this case. To compute the one-loop partition function, I apply the procedure discussed in \cite{Kurlyand:2026yke} directly.

Namely, let us introduce covering coordinates
\begin{align}
    T = u^{N_{f}} \ , \qquad \tilde{T} = v^{N_{f}} \ , \qquad \Phi_{1} = uv \ ,
\end{align}
where the choice of the $(u, v)$ coordinates is modulo the $\mathbb{Z}_{N_{f}}$ action $(u, v)\sim (\omega u, \omega^{-1}v)$, with $\omega=e^{2\pi i/N_{f}}$, and the R-charges are $R[u] = R[v] = 1/2$. On the covering link, let us consider the principal $U(1)_{r}$-bundle $\pi : \Sigma_{+}'\to B_{+}'\equiv \Sigma_{+}'/U(1)_{r}$ and let $L$ be the associated line bundle.\footnote{I will slightly abuse the notation and denote by $\Sigma_{+}'$ both its covering link $S^3$ and the projected link $S^3/\mathbb{Z}_{N_{f}}$.} It is convenient to introduce local coordinates and normal fluctuations over the two affine charts of $B_{+}'\simeq \mathbb{P}^{1}$:
\begin{align}
    u \neq 0 &: \qquad z_{1} = \frac{\Phi_{3}}{u} \ , \qquad \mathrm{r}_{1} = u \ , \qquad \delta y_{1}^{(1)} = u^{-1}\delta v \ , \qquad \delta y_{1}^{(2)} = u^{-1}\delta \Phi_{2} \ , \\
    \Phi_{3}\neq 0&: \qquad z_{2} = \frac{u}{\Phi_{3}} \ , \qquad \mathrm{r}_{2} = \Phi_{3} \ , \qquad \delta y_{2}^{(1)} = (\Phi_{3})^{-1}\delta v \ , \qquad \delta y_{2}^{(2)} = (\Phi_{3})^{-1}\delta \Phi_{2} \ .
\end{align}
The transition functions become
\begin{align}
    z_{2} = z_{1}^{-1} \ , \qquad \mathrm{r}_{2} = z_{1}\mathrm{r}_{1} \ , \qquad \delta y_{2}^{(1)} = z_{1}^{-1}\delta y_{1}^{(1)} \ , \qquad \delta y_{2}^{(2)} = z_{1}^{-1}\delta y_{1}^{(2)} \ ,
\end{align}
so that $L \simeq \mathcal{O}(1), \ N_{B_{+}'} \simeq \mathcal{O}(1)\oplus \mathcal{O}(1)$. The sections of $L^n$ and $N_{B_{+}'}\otimes L^{n}$ can be locally parametrised by
\begin{align}
    f_{n} = f_n(z)\mathrm{r}^n \ , \qquad s^{(l)}_{n} = s_{n}^{(l)}(z)\partial_{y^{(l)}} \otimes \mathrm{r}^{n} \ ,
\end{align}
where $l=1,2$ and $n\in \mathbb{Z}$ now parametrises the $U(1)_{r}$ Fourier mode, so that the Reeb weights become
\begin{align}
    2R[f_{n}] = n \ , \qquad 2R[s_{n}^{(l)}] = n \ .
\end{align}
The finite gauge transformations $G = \mathbb{Z}_{N_{f}}$ act on $B_{+}'\simeq \mathbb{P}^1$ with fixed points at the north and south poles, and, as discussed in \cite{Kurlyand:2026yke}, the net multiplicity of each gauge-invariant Fourier mode can be computed using the equivariant Lefschetz formula. In particular, for the RW multiplet one gets for $N_{f}\geq 2$ (with the $p=0$ term treated as in \cite{Kurlyand:2026yke}):
\begin{align}
    \widehat{\chi}_{n}^{G} = \frac{1}{N_{f}}\sum_{p=0}^{N_{f}-1}\Big(\frac{\omega^{-np}}{1-\omega^p} +\frac{1}{1-\omega^{-p}}\Big) = \big\lfloor\frac{n}{N_{f}}\big\rfloor+1 \ ,
\end{align}
where $\lfloor \, \cdot \, \rfloor$ denotes the floor function. Similarly, for the ML multiplet:
\begin{align}
    \widehat{\chi}_{n}^{G}(N\Sigma'_{+}) &= \frac{1}{N_{f}}\sum_{p=0}^{N_{f}-1}\Big(\frac{\omega^{-(n+2)p}+\omega^{-(n+1)p}}{1-\omega^{p}}+\frac{\omega^{-p}+1}{1-\omega^{-p}}\Big) \nonumber \\
    &= 2\big\lfloor\frac{n+2}{N_{f}}\big\rfloor +1-\delta_{n+2\mid N_{f}} \ ,
\end{align}
where $\delta_{a\mid N_{f}}=1$ for $a\in N_{f}\mathbb{Z}$ and vanishes otherwise. The one-loop partition function can be formally written as
\begin{align}
    Z_{1} = \prod_{n\in \mathbb{Z}}(n+2)^{-2\widehat{\chi}_{n}^{G}}\prod_{n\in \mathbb{Z}}n^{-\widehat{\chi}_{n}^{G}(N\Sigma'_{+})} \ .
\end{align}
Crucially, and in contrast to the previous cases, the partition function formally diverges due to zero eigenvalues of $R$. Indeed, although for the RW part $\widehat{\chi}^G_{-2}=0$ for $N_{f}\geq 2$, for the ML part one gets
\begin{align}
    \widehat{\chi}_{0}^{G}(N\Sigma'_{+}) = \begin{cases}
        2 \ , & N_{f}=2 \ , \\
        1 \ , & N_{f}>2
    \end{cases} \ .
\end{align}
As discussed in \cite{Kurlyand:2026yke}, the zero modes in $\ker R$ control the infinitesimal BPS deformations of the cycle and the effective dimension of the BPS moduli space is determined by the corresponding index. In the present case, the positive value of $\widehat{\chi}_{0}^{G}(N\Sigma'_{+})$ suggests a moduli space of complex dimension two for $N_{f}=2$ and one for $N_{f}>2$. To obtain the total one-loop partition function, one should formally separate these zero modes and integrate over the moduli space. Although I will not attempt to perform the integral over the moduli space here, the scaling of its volume can be determined by dimensional analysis. Namely, the expansion around the BPS saddle is done in terms of the effective tension $\T_{2} \sim (\Le/\ell_{p})^3$, so, for a positive index, one expects the volume to scale as
\begin{align}
    \mathrm{Vol}(\mathcal{M}_{\mathrm{BPS}}^{E}) \sim \Big(\frac{\Le^3}{\ell_{p}^3}\Big)^{\widehat{\chi}^{G}(E)} \ ,
\end{align}
where $\widehat{\chi}^{G}(E)$ denotes the index restricted to $\ker R$ and determines the effective complex dimension of the moduli space. Using \eqref{tensionm2}, for the M2-brane instanton of interest one thus obtains
\begin{align}
    \mathrm{Vol}(\mathcal{M}^{N\Sigma'_{+}}_{\mathrm{BPS}})\sim \begin{cases}
        (8\mu_{\ast})^2 \ , & N_{f}=2 \ , \\
        8\mu_{\ast} \ , & N_{f}>2
    \end{cases} \ .
\end{align}
In the present case, reindexing $n\mapsto -n$ for the RW multiplet and $n\mapsto n-2$ for the ML multiplet, and using $\zeta$-function regularisation for the combined product, one obtains the non-zero-mode contribution to the one-loop partition function for $N_{f}\geq 2$:\footnote{It is also used that $\big\lfloor-\frac{n}{N_{f}}\big\rfloor = -\big\lfloor\frac{n}{N_{f}}\big\rfloor-1+\delta_{n\mid N_{f}}$. Note that the value of the partition function itself depends on the phase of the resulting determinant.}
\begin{align}
    |Z_{1}'| = \prod_{n\neq 2}|n-2|^{-1-\delta_{n\mid N_{f}}}=\begin{cases}
        \frac{1}{2\pi^2} \ , & N_{f}=2 \ , \\
        \frac{1}{4\pi\sin(\frac{2\pi}{N_{f}})} \ , & N_{f}>2
    \end{cases} \ .
\end{align}

One thus obtains from the M2-brane analysis the one-loop partition function for $N_{f}>2$:
\begin{align}\label{adhmm2}
    Z_{1} = \gamma\frac{2\mu_{\ast}}{\pi \sin(\frac{2\pi}{N_{f}})} \ ,
\end{align}
where the constant factor $\gamma$ includes the normalisation from integration over the BPS moduli space and the possible determinant phase. 
As discussed in the next section, up to this numerical factor, \eqref{adhmm2} reproduces the prefactor of the leading ``worldsheet'' instanton contribution in the large-$N$/large chemical-potential expansion of the ADHM matrix model.

\subsubsection*{$V^{5,2}/\mathbb{Z}_{N_{f}}$ theory}

For the $V^{5,2}/\mathbb{Z}_{N_{f}}$ theory one can follow the same procedure as in the ADHM case. It is first convenient to introduce coordinates
\begin{align}
X_{1} =\Phi_{1} \ , \qquad X_{2} = \Phi_{2}+i\Phi_{3} \ , \qquad X_{3} = \Phi_{2}-i\Phi_{3} \ ,
\end{align}
in which the monopole relation becomes
\begin{align}
T\tilde{T} -\big(X_{2}X_{3}+X_{1}^2\big)^{N_{f}} = 0 \ .
\end{align}

Let us consider the cone $\mathcal{M}_{+}'=\{\tilde{T} = X_{1} = X_{2} = 0\}$ with coordinate ring
\begin{align}
\mathbb{C}[\mathcal{M}_{+}'] = \mathbb{C}[X_{3}, T] \ .
\end{align}
With the superconformal R-charge assignment $R[X_{3}] = 2/3$ and $R[T]=2N_{f}/3$, the Hilbert series and the volume of the corresponding three-dimensional link become
\begin{align}
h_{+}'(t) = \frac{1}{(1-t^{4/3})(1-t^{4N_{f}/3})} \ , \qquad \mathrm{Vol}(\Sigma_{+}') = 2\pi^2 \frac{9}{16N_{f}} \ ,
\end{align}
while the classical exponent becomes
\begin{align}
S_{+} = \frac{9\mu_{\ast}}{4N_{f}}\ , \qquad \mu_{\ast} = \frac{8\pi}{9}\sqrt{N_{f}N} \ . 
\end{align}
The covering coordinates
\begin{align}\label{flavv52mon}
T = u^{N_{f}} \ , \qquad \tilde{T} = v^{N_{f}} \ , \qquad uv =  X_{2}X_{3}+X_{1}^2 \ ,
\end{align}
are determined up to the $\mathbb{Z}_{N_{f}}$ action $(u, v) \sim (\omega u, \omega^{-1}v), \ \omega = e^{2\pi i /N_{f}}$, and have R-charges \mbox{$R[u]=R[v] =2/3$}. As in the ADHM case, $B_{+}'\simeq \mathbb{P}^1$, while the local coordinates on the two affine charts can be introduced as
\begin{align}
u \neq 0 &: \qquad z_{1} = \frac{X_{3}}{u} \ , \qquad \mathrm{r}_{1} = u \ , \qquad \delta y_{1}^{(1)} =u^{-1}\delta X_{1}\ , \qquad \delta y_{1}^{(2)} = u^{-1}\delta X_{2}  \ , \\
X_{3} \neq 0&: \qquad z_{2} = \frac{u}{X_{3}} \ , \qquad \mathrm{r}_{2} = X_{3} \ , \qquad \delta y_{2}^{(1)} = X_{3}^{-1}\delta X_{1} \ , \qquad \delta y_{2}^{(2)} = X_{3}^{-1}\delta v \ ,
\end{align}
where one can use that, around the locus, the monopole relation in \eqref{flavv52mon} linearises as
\begin{align}
u\delta v = X_{3}\delta X_{2} \ ,
\end{align}
so that $L \simeq \mathcal{O}(1), \ N_{B_{+}'}\simeq \mathcal{O}(1)\oplus \mathcal{O}$ and the transition functions become
\begin{align}
z_{2} = z_{1}^{-1} \ , \qquad \mathrm{r}_{2} = z_{1}\mathrm{r}_{1} \ , \qquad \delta y_{2}^{(1)} = z_{1}^{-1}\delta y_{1}^{(1)} \ , \qquad \delta y_{2}^{(2)} = \delta y_{1}^{(2)} \ .
\end{align}
Using the equivariant Lefschetz formula, under the action of $G=\mathbb{Z}_{N_{f}}$ it is straightforward to obtain the projected indices. For the RW multiplet the calculation is the same as for the ADHM case and gives
\begin{align}
\widehat{\chi}_{n}^{G} = \big\lfloor\frac{n}{N_{f}}\big\rfloor+1 \ ,
\end{align}
while for the ML multiplet one gets
\begin{align}
\widehat{\chi}_{n}^{G}(N\Sigma'_{+}) &= \frac{1}{N_{f}}\sum_{p=0}^{N_{f}-1}\Big(\frac{2\omega^{-(n+1)p}}{1-\omega^{p}}+\frac{1+\omega^{-p}}{1-\omega^{-p}}\Big) = 2\big\lfloor \frac{n+1}{N_{f}}\big\rfloor+1  \ .
\end{align}
The RW multiplet has no zero modes, while for $N_{f}\geq 2$ the zero-mode index of the ML multiplet is
\begin{align}
\widehat\chi^{G}_{0}(N\Sigma'_{+}) = 1 \ .
\end{align}
With the $\zeta$-function regularisation applied separately to the RW and ML multiplets, the partition function for the non-zero modes becomes
\begin{align}
|Z_{1}'| = \prod_{n\in\mathbb{Z}}\Big|\frac{4}{3}n+2\Big|^{-2\lfloor n/N_{f}\rfloor-2} \prod_{n\neq 1}\Big|\frac{4}{3}(n-1)\Big|^{-2\lfloor n/N_{f} \rfloor-1} =2\sin^2(\tfrac{\pi}{2N_{f}} )\cdot \frac{N_{f}}{6\pi \sin^2(\frac{\pi}{N_{f}})} \ ,
\end{align}
and thus, modulo the constant coefficient resulting from integration over the moduli space and the phase of the determinant, one obtains for $N_{f}>1$:
\begin{align}
Z_{1} = \gamma \frac{2N_{f}\mu_{\ast}}{3\pi \cos^2(\frac{\pi}{2N_{f}})} \ .
\end{align}

\section{Matrix model instantons}\label{sec4}

In \cite{Kapustin:2009kz}, the partition function of 3d
$\mathcal{N}=2$ superconformal gauge theories was computed on $S^3$ using localisation, with chiral matter assigned canonical R-charges. In \cite{Jafferis:2010un,Hama:2010av}, the localisation formula was extended to general trial R-charges, taking the schematic form
\begin{align}\label{4.1}
    Z(R)
    =
    \int \prod_{\mathfrak{h}} du \,
    e^{i\pi\,\tr u^2}
    \det{}_{\mathrm{Ad}}\big(\sinh(\pi u)\big)
    \prod_{R_i}
    \det{}_{R_i}\big(e^{\ell(1-\Delta_i+iu)}\big)
    \ ,
\end{align}
where $\mathfrak{h}$ denotes the Cartan subalgebra of the Lie
algebra of the gauge group, $R_i$ denotes the gauge representation
of the $i$-th chiral multiplet, and $\Delta_i$ denotes its trial
R-charge. In \eqref{4.1}, $\tr$ includes the normalisation by Chern-Simons levels and the one-loop determinants for the matter fields involve the function
\begin{align}\label{loneloop}
    \ell(z)
    =
    -z\log\big(1-e^{2\pi iz}\big)
    +\frac{i}{2}
    \big(
        \pi z^2+\frac{1}{\pi}\mathrm{Li}_2(e^{2\pi iz})
    \big)
    -\frac{i\pi}{12}
    \ .
\end{align}
As already discussed in Section~\ref{2}, in the case of
$\mathcal{N}=2$ theories the exact superconformal R-symmetry can
be obtained by locally maximising the free energy
$F=-\log|Z|$ \cite{Jafferis:2010un,Jafferis:2011zi}.

Several approaches have been developed to evaluate the resulting matrix model integral at large $N$. In \cite{Marino:2011eh}, the matrix models of a class of $\mathcal{N}=3$ theories were rewritten as partition functions of ideal Fermi gases. This approach was extended to $\mathcal{N}=2$ theories in
\cite{Marino:2012az}, where the corresponding Fermi gas systems are generally interacting. In this formulation, it is convenient to compute
the grand canonical potential $J(\mu)$, related to the canonical
partition function through
\begin{align}
    Z(N)
    =
    \int \frac{d\mu}{2\pi i}\,
    e^{J(\mu)-\mu N}
    \ .
\end{align}
For an ideal Fermi gas, the grand canonical potential can be expressed in terms of the one-particle density of states as \cite{Marino:2011eh}
\begin{align}
    J(\mu)
    =
    \int_0^\infty dE\,\rho(E)\log(1+e^{\mu-E})
    \ ,
    \qquad
    \rho(E)=\frac{dn(E)}{dE}
    \ ,
\end{align}
while for an interacting Fermi gas, one can instead
consider the Hartree/Thomas-Fermi approximation at large $N$ \cite{Marino:2012az}. For the theories of interest, the grand canonical potential is expected to have the large-$\mu$ expansion \cite{Marino:2011eh,Marino:2012az}
\begin{align}
    J(\mu)
    =
    \frac{C}{3}\mu^3+B\mu+A+J_{\mathrm{np}}(\mu)
    \ ,
\end{align}
where $J_{\mathrm{np}}(\mu)$ denotes the exponentially suppressed
terms to which M2-brane instantons of Section~\ref{m-theory} contribute.

Another approach, developed in \cite{Herzog:2010hf} and further
used in \cite{Jafferis:2011zi}, solves the saddle-point equations
directly in the $N\to\infty$ limit with the Chern-Simons levels
held fixed. Using a continuum approximation in the large $N$
limit, the saddle-point equations for the eigenvalue density
and imaginary parts of the eigenvalues can be solved analytically. Evaluating the resulting free-energy functional
on the solution gives the leading contribution to the free
energy.\footnote{A relation between this large-$N$ free-energy
functional and the Thomas-Fermi functional was noted in
\cite{Marino:2012az}.} In this framework, a further interesting relation between the eigenvalue density and the asymptotic counting of operators in the chiral ring was proposed in \cite{Gulotta:2011si,Gulotta:2011aa}.

Below, I briefly comment on how the M2-brane instantons of Section~\ref{m-theory} may appear in both of these approaches.

\subsection{Fermi gas and quantum mechanical instantons}

A convenient concept in the Fermi gas framework is the ``classical'' Fermi surface. In the case of the ideal Fermi gas, this can be obtained directly from the one-particle Hamiltonian and, in the thermodynamic limit, the area enclosed by the Fermi surface determines the leading free energy of the system \cite{Marino:2011eh}. If $H(p, q)$ denotes the ``classical'' Hamiltonian, the Fermi surface is defined by
\begin{align}
    H(p, q) = E \ ,
\end{align}
and controls the density of the eigenvalues in the thermodynamic limit. As discussed in \cite{Marino:2011eh}, the area $\mathrm{Vol}_{\mathrm{F}}$ enclosed by the Fermi surface determines the distribution of the states at large energies
\begin{align}
    n(E) \approx CE^{s} \approx \frac{\mathrm{Vol}_{\mathrm{F}}}{2\pi \hbar} \ , \qquad E\gg 1 \ ,
\end{align}
where $\hbar$ denotes the effective Planck's constant and $s=2$ for the theories of interest.

More relevant to the non-perturbative corrections is the suggestion in \cite{Marino:2011eh} (see also \cite{Kallen:2013qla}) that ``worldsheet'' matrix model instantons can be associated with quantum mechanical instantons whose actions are periods of the complexified Fermi surface. If $S(E)$ is the action of the corresponding quantum mechanical instanton, the integrated one-particle density of states receives a non-perturbative correction
\begin{align}
    n_{\mathrm{ws}}^{(1)}(E) \sim \exp\big(\frac{i}{\hbar} S(E)\big) \ , \qquad E\gg 1 \ .
\end{align}
For $S(E)=ES_0$, this gives a non-perturbative correction to the grand canonical potential of the form
\begin{align}
   J_{\mathrm{ws}}^{(1)}(\mu) \sim \exp\big(\frac{i\mu}{\hbar}S_{0} \big) \ .
\end{align}
In the present approach, it is natural to associate such ``worldsheet'' matrix model instantons with the BPS M2-branes discussed in Section~\ref{m2inst}.

Let us illustrate this approach for the case of $\mathcal{N}=3$ flavoured ABJM theory. With $N_{f}$ denoting the total number of flavours, the semiclassical Fermi surface is given by \cite{Marino:2011eh}:
\begin{align}\label{4.1aa}
    H(p,q) = \log\big(2\cosh \frac{p}{2}\big)
    +\log\big(2\cosh \frac{q}{2}\big)
    +N_{f}\log\big(2\cosh \frac{p+q}{2k}\big) = E \ .
\end{align}
As discussed in \cite{Marino:2011eh}, for large energy the Fermi surface approaches a polygon
\begin{align}
    |p|+|q|+\frac{N_{f}}{k}|p+q| = 2E \ ,
\end{align}
and, with the effective Planck's constant $\hbar=2\pi k$, the volume of the classical Fermi surface determines
\begin{align}
    \frac{\pi^2}{2E^2}\cdot \frac{\mathrm{Vol}_{\mathrm{F}}}{2\pi \hbar}
    = \frac{\pi^2}{2}C
    = \frac{N_{f}+2k}{2(N_{f}+k)^2} \ ,
\end{align}
matching \eqref{2.41}.

To study the periods of the complexified Fermi surface, it is convenient to analyse the tropical curve associated with the surface, following the approach of \cite{Santamaria:2010dm}.\footnote{See, for instance, \cite{ItenbergMikhalkinShustin2009,brugalle2015brief} for an introduction to tropical geometry.} Namely, one can introduce the coordinates $z=e^{(p+q)/2k}, \ w=e^{q/2}$, together with $t=e^{E}$, in which \eqref{4.1aa} can be viewed as a one-parameter family of curves $\mathcal{C}_{t}\subset(\mathbb{C}^{\ast})^2$, defined by the zero locus of the Laurent polynomial
\begin{align}\label{4.4aa}
    P_{t}(z,w)
    = (z+z^{-1})^{N_{f}}
    (z^{k}+z^{-k}+w^2z^{-k}+w^{-2}z^k)-t \ ,
\end{align}
and the holomorphic symplectic form becomes
\begin{align}
    \omega = 4k \, d\log z\wedge d\log w \ .
\end{align}
To visualise the surface, one can consider the logarithmic map
\begin{align}
    (x,y) = \big(\log_{t}|z|,\log_{t}|w|\big)\in\mathbb{R}^2 \ ,
\end{align}
with the image of $\mathcal{C}_t$ under this map defining its amoeba $\log_{t}\mathcal{C}_{t}$ as a subset of $\mathbb{R}^2$. The corresponding tropical curve can be viewed as a limiting amoeba $\log_{t}\mathcal{C}_{t}$ as $t\to\infty$, describing the large energy approximation of the Fermi surface. It can be conveniently determined by taking the tropical polynomial
\begin{align}
    P_{t}^{\mathrm{trop}}(x,y)
    = \max\big((N_{f}+k)|x|,N_{f}|x|+|2y-kx|,1\big) \ ,
\end{align}
and considering its tropical hypersurface, defined as the set of points at which the maximum among the linear terms is attained at least twice.

\begin{figure}[htbp]
    \centering
    \begin{minipage}[t]{0.4\linewidth}
        \vspace{0pt}
        \centering
        \includegraphics[width=\linewidth]
            {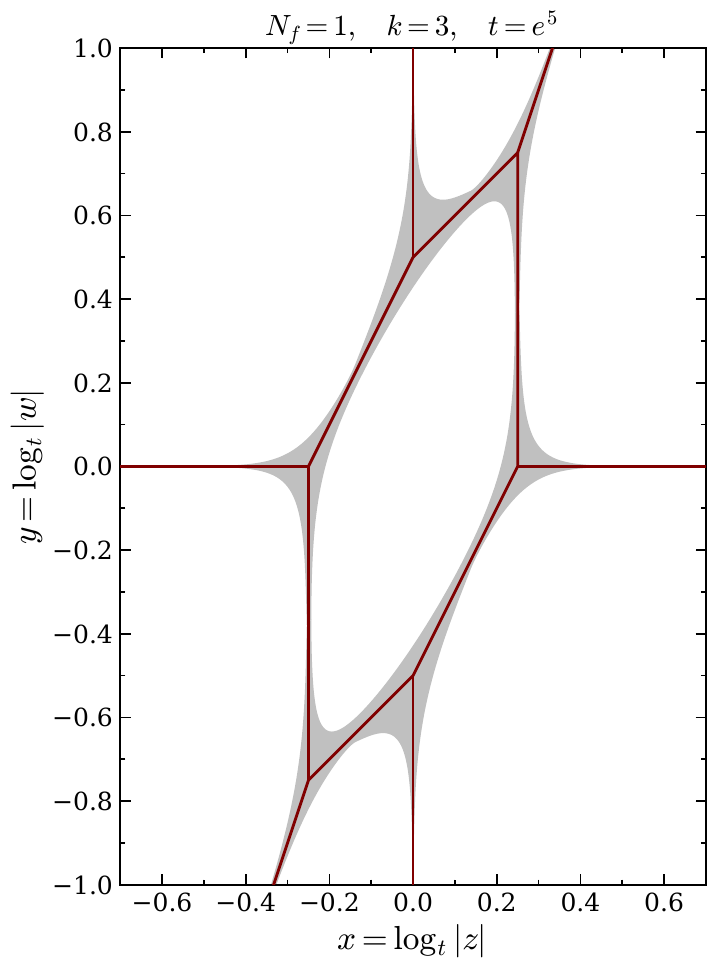}
    \end{minipage} \hspace{50pt}
    \begin{minipage}[t]{0.25\linewidth}
        \vspace{10pt}
        \centering
        \includegraphics[width=\linewidth]
            {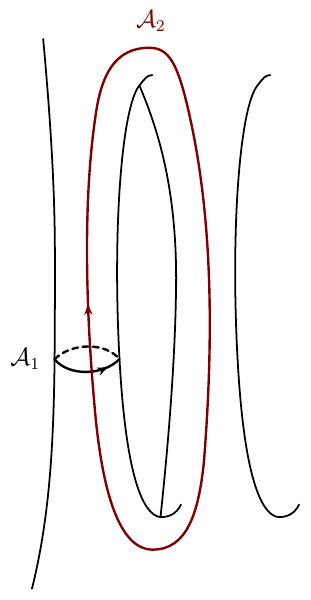}
    \end{minipage}

   \caption{Left: Amoeba (grey) of $\mathcal{C}_{t}$, defined by \eqref{4.4aa} ($N_{f}=1$, $k=3$, $t=e^{5}$), and the corresponding tropical curve (red). Right: Cycles $\mathcal{A}_{1}$ and $\mathcal{A}_{2}$ associated with the edge $x_{\ast}=1/(N_{f}+k)$ of the tropical curve.}
    \label{fig:amoeba}
\end{figure}

To study the periods in more detail, one can ``zoom in'' on the edge $x_{\ast}=1/(N_f+k)$. In the interior of the edge, $|z|=t^{x_{\ast}}\gg 1$ and $1\ll |w|\ll t^{kx_{\ast}}$. The complexified Fermi surface \eqref{4.4aa} can be approximated by a collection of $N_{f}+k$ tubes:
\begin{align}
    z = t^{x_{\ast}}\exp(\tfrac{2\pi i l}{N_{f}+k}) \ ,
    \qquad l=0,\dots,N_{f}+k-1 \ ,
\end{align}
which degenerate to a single edge in the tropical limit (see also similar examples in \cite{Iwao2010}). To compute a period, one can consider a cycle $\mathcal{A}_{1}$ going around one of the tubes. In the large energy approximation, the leading contribution of such a cycle can be written as
\begin{align}\label{a1cycle}
    \exp\Big(\frac{4ik}{\hbar}
    \oint_{\mathcal{A}_{1}}\log z\,d\log w\Big)
    \approx \exp\Big(-\frac{4E}{N_{f}+k}\Big) \ ,
\end{align}
where the $2\pi i$ monodromy of $\log w$ along this cycle is used. Yet, in the present case one can also consider a cycle $\mathcal{A}_{2}$ going around two neighbouring tubes with $\Delta\log z=\frac{2\pi i}{N_{f}+k}$ and $\Delta \log w= \frac{kE}{N_{f}+k}+\mathcal{O}(1)$. In the large energy limit, the contribution of such a cycle can be approximated by an area
\begin{align}
    \exp\Big(\frac{4ik}{\hbar}\Delta\log z\,\Delta\log w\Big)
    \approx \exp\Big(-\frac{4kE}{(N_{f}+k)^2}\Big) \ ,
\end{align}
which corresponds to the leading instanton contribution for $N_{f}>0$, compared to \eqref{a1cycle}. Following the discussion in \cite{Marino:2011eh}, this suggests the non-perturbative contribution to the grand canonical potential
\begin{align}
    J_{\mathrm{ws}}^{(1)}(\mu,N_{f})
    \sim \exp\Big(-\frac{4k\mu}{(N_{f}+k)^2}\Big) \ ,
\end{align}
which corresponds precisely to the classical exponent of the M2-brane instanton \eqref{3.15} identified for this theory.

Another relevant example where matrix model instantons were analysed in the Fermi gas approach is the ADHM theory discussed above. In the Fermi gas formulation, the ``worldsheet'' instantons for the ADHM matrix model were discussed in \cite{Grassi:2014vwa,Hatsuda:2014vsa}, with their actions explicitly identified with periods of the complexified Fermi surface in \cite{Hatsuda:2015lpa}. In \cite{Hatsuda:2014vsa}, the contribution of the leading ``worldsheet'' instanton to the non-perturbative part of the grand canonical potential for $N_{f}>2$ was proposed to take the form
\begin{align}\label{adhmmatrix}
    J_{\mathrm{ws}}^{(1)}(\mu,N_{f})
    = -\frac{4\mu+N_{f}}{2\pi\sin(\frac{2\pi}{N_{f}})}
    e^{-4\mu/N_{f}} \ .
\end{align}
In the large $\mu$ limit, $\mu\approx\pi\sqrt{NN_{f}/2}\gg 1$ and the matrix model instanton contribution \eqref{adhmmatrix} to the partition function matches both the classical exponent \eqref{3.48} and the one-loop prefactor \eqref{adhmm2} from the M2-brane analysis, up to the numerical coefficient $\gamma$ resulting from zero modes and the phase of the determinant (with agreement requiring a total coefficient of $-1$ after integrating over the zero modes and summing over the contributing cycles).\footnote{Note that, in the grand canonical interpretation proposed in \cite{Gautason:2025per, Gautason:2025plx}, the chemical potential is related to the effective M2-brane tension by $\T_{2}=2\mu/\pi^2$ (see also \eqref{tensionm2}). Under this proposal, after factoring out the leading dependence on $\T_{2}$ associated with the zero modes, the prefactor in \eqref{adhmmatrix} would be proportional to $1+N_f/(2\pi^2\T_{2})$. In other words, the $\mu$-independent term in the prefactor of \eqref{adhmmatrix} suggests a non-vanishing two-loop correction to the semiclassical M2-brane partition function. It would be interesting to determine whether the methods developed in \cite{Beccaria:2026zjj,Tseytlin:2026dhy} can be extended to derive this correction.}

\subsection{Operator counting and matrix model}\label{sec4.2}

Building on the saddle point analysis of \cite{Herzog:2010hf, Jafferis:2011zi}, in \cite{Gulotta:2011si, Gulotta:2011aa} it was observed that in the large $N$ limit the eigenvalue density $\rho(x)$ of the matrix model is related to the count of chiral operators on the geometric branch in the abelian theory (see also \cite{Gulotta:2011vp, Kim:2012vza} for further examples). Namely, let $\psi(r, m)$ be the number of operators with $R$-charge at most $r$ and monopole charge at most $m$. Then the large $N$ eigenvalue density can be expressed as \cite{Gulotta:2011si, Gulotta:2011aa}:
\begin{align}
    \frac{\partial^3\psi}{\partial r^2\partial m}\Big|_{m=rx/\kappa} =\frac{r}{\kappa}\, \rho(x)\ , 
\end{align}
where $\kappa$, in the present notation, plays the role of the Lagrange multiplier imposing the normalisation constraint $\int dx \rho (x)=1$ \cite{Herzog:2010hf} in the large $N$ limit. If one introduces $\delta$ as the shift of the R-charge of the bare monopole operator, the volume of the dual $\ses$ can be expressed as
\begin{align}\label{ghp}
    \frac{\mathrm{Vol}(\ses, \delta)}{\mathrm{Vol}(S^{7})} = \frac{1}{8}\int d\hat{x} \frac{\hat{\rho}(\hat{x})}{(1+\hat{x}\delta)^3} \ ,
\end{align}
where $\hat{x}=x/\kappa, \ \hat{\rho}(\hat{x}) = \rho(x)/\kappa$ (with normalisation of the free-energy functional as in \cite{Gulotta:2011aa}). This allows one to reconstruct $\hat{\rho}(\hat{x})$ from $\mathrm{Vol}(\ses, \delta)$ for $\delta$ in a small neighbourhood of zero. For $\hat{\rho}$ with a compact support $[\hat{x}_{-}, \hat{x}_{+}]$, \eqref{ghp} converges for $\Re\delta\in (-1/\hat{x}_{+}, -1/\hat{x}_{-})$. Away from the singularities, one can integrate by parts to obtain
\begin{align}
    \frac{\mathrm{Vol}(\ses, \delta)}{\mathrm{Vol}(S^{7})} = \frac{1}{16\delta^3}\int d\hat{x} \frac{\hat{\rho}''(\hat{x})}{(\hat{x}+1/\delta)} \ .
\end{align}
In particular, if $\hat{\rho}$ is piecewise linear, its second derivative can be expressed, in a distributional sense, by
\begin{align}\label{ghp1}
   \hat{\rho}''(\hat{x}) = \sum_{\alpha}[\hat{\rho}_{\alpha}]\delta'(\hat{x}-\hat{x}_{\alpha})+\sum_{\beta}[\hat{\rho}_{\beta}']\delta(\hat{x}-\hat{x}_{\beta}) \ , 
\end{align}
where $[f_{\alpha}]=f(\hat{x}_{\alpha}+0) - f(\hat{x}_{\alpha}-0)$. Using \eqref{ghp1}, the function $\mathrm{Vol}(\ses, \delta)$ then develops first order and second order poles at the discontinuities \cite{Gulotta:2011aa}:
\begin{align}\label{4.26}
    \frac{\mathrm{Vol}(\ses, \delta)}{\mathrm{Vol}(S^{7})} = \frac{1}{16\delta^3}\Big(\sum_{\alpha}\frac{[\hat{\rho}_{\alpha}]}{(\hat{x}_{\alpha}+1/\delta)^2} +\sum_{\beta}\frac{[\hat{\rho}_{\beta}']}{(\hat{x}_{\beta}+1/\delta)} \Big) \ . 
\end{align}

In \cite{Farquet:2013cwa}, the discontinuities of $\rho(x)$ and its derivative $\rho'(x)$ in the matrix model were given a geometric interpretation. In particular, it was suggested that the interval $[\hat{x}_{-}, \hat{x}_{+}]$ corresponds to the image of the M-theory Hamiltonian $2h_m$, with critical values $2h_{\ast}$ in the interval identified with discontinuities of the eigenvalue density or its derivative. With the interpretation of Section~\ref{sec3.1}, the cones over the critical sets with non-zero $h_{\ast}$ can be identified with the monopole loci and the non-zero critical values can be expressed through the corresponding monopole R-charges as $2h_{\ast} = \pm1/R_{\pm}$. Guided by this, it is further interesting to relate the discontinuity of the eigenvalue density to the volume of the links resulting from the monopole loci, computed from the Hilbert series in Section~\ref{2}.

Concretely, in the case of the shift of the bare monopole R-charge by $\delta$, the modified Hilbert series of the chiral ring becomes
\begin{align}\label{4.27}
    H(t, \delta) = \sum_{r\geq0, m}H_{r, m}t^{2(r+m\delta)} \ , 
\end{align}
where $H_{r, m}$ denotes the number of chiral operators of physical R-charge $r$ and magnetic charge $m$. The volume $\mathrm{Vol}(\ses, \delta)$ can then be expressed as
\begin{align}\label{4.28}
     \frac{\mathrm{Vol}(\ses, \delta)}{\mathrm{Vol}(S^{7})} = \lim_{t\to 1}\, (1-t)^4H(t, \delta) \ .
\end{align}
The Hilbert series of the chiral ring can be decomposed into
sectors of zero, positive, and negative magnetic charge. For theories with gauge-invariant BPS monopoles, the zero-charge sector is generated by mesonic operators and is expected to have at most a third-order pole as $t\to1$. More importantly, the Hilbert series restricted to the purely mesonic sector is independent of $\delta$, and hence only the magnetically
charged sectors contribute to the non-zero poles in $\delta$ of \eqref{4.26}.

For definiteness, let me assume that the monopole loci $\mathcal{M}_{(\alpha)}$ have complex dimension two. On each locus, the R-charge is proportional to
the absolute value of the magnetic charge by definition, so the
corresponding Hilbert series can be written as
\begin{align}
    h_{(\alpha)}(t,\delta)
    =
    \sum_{n=0}^{\infty}
    h_{(\alpha),n}
    t^{2n(R_{(\alpha)}+s_{(\alpha)}\delta)}
    \ , \qquad
    s_{(\alpha)}=\pm1 \ ,
\end{align}
where $h_{(\alpha),n}$ counts operators on
$\mathcal{M}_{(\alpha)}$ with magnetic charge
$m=s_{(\alpha)}n$, and $R_{(\alpha)}$ is the common R-charge
of the monopole generators with magnetic charge $s_{(\alpha)}$. Using \eqref{2.29}, it then follows that the leading singularity of $h_{(\alpha)}(t, \delta)$ can be expressed in terms of the volume of the link $\Sigma_{(\alpha)}$ of the two-dimensional monopole locus (in the sense of Section~\ref{2}) as:
\begin{align}\label{4.30b}
    \lim_{t\to1}(1-t)^2h_{(\alpha)}(t,\delta)
    =
    \frac{R_{(\alpha)}^2}
    {(R_{(\alpha)}+s_{(\alpha)}\delta)^2}
    \frac{\mathrm{Vol}(\Sigma_{(\alpha)})}
    {\mathrm{Vol}(S^3)} \ .
\end{align}

It is natural to expect that the leading growth at large magnetic
charge is obtained by dressing operators on the monopole loci
with two independent local mesonic transverse coordinates. Denoting the corresponding mesonic R-charges by $r_1^{(\alpha)},r_2^{(\alpha)}$, the contribution governing the leading double poles in $\delta$, after taking the limit in \eqref{4.28}, takes the form
\begin{align}\label{4.29}
    H(t,\delta)
    \sim
    \sum_{\alpha}
    h_{(\alpha)}(t,\delta)
    \sum_{n_1,n_2\geq0}
    t^{2n_1r_1^{(\alpha)}+2n_2r_2^{(\alpha)}} \ .
\end{align}
Assuming that each pole is associated with
a single monopole locus, from \eqref{4.30b} it follows that
\begin{align}
    \lim_{\delta\to-s_{(\alpha)}R_{(\alpha)}}
    (\delta+s_{(\alpha)}R_{(\alpha)})^2
    \frac{\mathrm{Vol}(\ses,\delta)}{\mathrm{Vol}(S^7)}
    =
    \frac{R_{(\alpha)}^2}
    {4r_1^{(\alpha)}r_2^{(\alpha)}}
    \frac{\mathrm{Vol}(\Sigma_{(\alpha)})}
    {\mathrm{Vol}(S^3)} \ .
\end{align}
On the other hand, identifying
$\hat{x}_{\alpha}=s_{(\alpha)}/R_{(\alpha)}$, one obtains from
\eqref{4.26}
\begin{align}
    \lim_{\delta\to-s_{(\alpha)}R_{(\alpha)}}
    (\delta+s_{(\alpha)}R_{(\alpha)})^2
    \frac{\mathrm{Vol}(\ses,\delta)}{\mathrm{Vol}(S^7)}
    =
    -\frac{s_{(\alpha)}R_{(\alpha)}}{16}
    [\hat{\rho}_{\alpha}] \ .
\end{align}
Comparing the two expressions gives
\begin{align}\label{4.34a}
    \frac{\mathrm{Vol}(\Sigma_{(\alpha)})}
    {\mathrm{Vol}(S^3)}
    =
    -\frac{r_1^{(\alpha)}r_2^{(\alpha)}}
    {4s_{(\alpha)}R_{(\alpha)}}
    [\hat{\rho}_{\alpha}]
    =
    -\frac{r_1^{(\alpha)}r_2^{(\alpha)}}{4}
    \hat{x}_{\alpha}[\hat{\rho}_{\alpha}] \ .
\end{align}
Thus, for two-dimensional monopole loci, \eqref{4.34a} relates the volumes of the M2-brane instanton links, and hence their classical actions, to
discontinuities of the eigenvalue density of the large-$N$ matrix model.

A non-trivial check of this relation can be done in the case of $\mathcal{N}=3$ flavoured ABJM theory. In this case, the rescaled eigenvalue density and its compact support become \cite{Herzog:2010hf, Santamaria:2010dm}:
\begin{align}
    \hat\rho(\hat{x}) = 2-N_{f}|\hat{x}| \ , \qquad \hat{x}\in [-\hat{x}_{\ast}, \hat{x}_{\ast}] \ , \quad \hat{x}_{\ast} = \frac{2}{N_{f}+k} \ ,
\end{align}
with the edges of the cut corresponding to the inverse R-charges of the gauge-invariant monopole operators of Section~\ref{sec2.1}. In this case, the mesonic operators of the abelian theory are of the form $M = AB$ and have unit R-charge. The eigenvalue density is discontinuous at the edges of the cut, so that the leading pole contribution there gives
\begin{align}
    \frac{\mathrm{Vol}(\Sigma_{\pm})}{\mathrm{Vol}(S^3)}=-\frac{1}{4}\, \hat{x}_{\ast}[\hat{\rho}(\hat{x}_{\ast})] = \frac{k}{(N_{f}+k)^2} \ , 
\end{align}
which agrees precisely with \eqref{2.43}.

Similarly, in the case of the $Q^{1,1,1}/\mathbb{Z}_{N_{f}}$ theory, the rescaled eigenvalue density becomes \cite{Jafferis:2011zi, Gulotta:2011aa}:
\begin{align}
   \hat\rho(\hat{x}) = 2-N_{f}|\hat{x}| \ , \qquad \hat{x}\in [-\hat{x}_{\ast}, \hat{x}_{\ast}] \ , \quad \hat{x}_{\ast} = \frac{1}{N_{f}} \ , 
\end{align}
so that $R_{\pm} = 1/|\hat{x}_{\ast}|$ coincide with the monopole R-charges discussed in Section~\ref{sec2.2}. The transverse mesonic coordinates to the monopole loci are of the form $M = AB$. The eigenvalue density discontinuity of the matrix model through \eqref{4.34a} gives
\begin{align}
    \frac{\mathrm{Vol}(\Sigma_{\pm})}{\mathrm{Vol}(S^3)}=-\frac{1}{4}\, \hat{x}_{\ast}[\hat{\rho}(\hat{x}_{\ast})] = \frac{1}{4N_{f}} \ , 
\end{align}
which agrees with \eqref{2.54}.

Another interesting example is the $V^{5,2}/\mathbb{Z}_{k}$ theory. In this case, the rescaled eigenvalue density and its compact support become \cite{Martelli:2011qj}:
\begin{align}
    \hat{\rho}(\hat{x}) = \frac{27}{16} \ , \qquad \hat{x}\in[-\hat{x}_{\ast}, \hat{x}_{\ast}] \ , \quad \hat{x}_{\ast} = \frac{3}{2k} \ ,
\end{align}
so that $\hat{x}_{\ast}$ can again be expressed in terms of the R-charges of the monopoles of Section~\ref{sec24} as $\hat{x}_{\ast} = 1/R_{\pm}$. The gauge-invariant monopole operators on the loci of positive or negative magnetic charge can be dressed by mesonic operators $\Phi_{0}$ or $AB$ with R-charges $\frac{2}{3}$ and $\frac{4}{3}$, respectively. The discontinuity of the eigenvalue density at the edge of the cut gives
\begin{align}
     \frac{\mathrm{Vol}(\Sigma_{\pm})}{\mathrm{Vol}(S^3)}=-\frac{8}{4\cdot 9}\, \hat{x}_{\ast}[\hat{\rho}(\hat{x}_{\ast})] = \frac{9}{16k} \ ,
\end{align}
again reproducing the volume of the M2-brane instanton link \eqref{2.68a}.

\section{Concluding remarks}\label{sec5}
In this paper I have identified a class of M2-brane instantons in $\ads\times\ses$ through the monopole loci on the geometric branch of the abelian moduli space. The one-loop calculation considered in Section~\ref{m-theory} relies on the semiclassical derivation of \cite{Kurlyand:2026yke}. It would be interesting to explore whether this construction can be extended beyond quadratic order, and whether the integrals over the bosonic zero modes, such as those appearing in the ADHM and $V^{5,2}/\mathbb{Z}_{N_f}$ theories considered above, can be evaluated explicitly.

The relation to the abelian chiral ring also provides an
interesting perspective on matrix model instantons. In
particular, under the assumptions of Section~\ref{sec4.2},
the classical actions of M2-brane instantons associated with
complex two-dimensional monopole cones are related to
discontinuities of the large-$N$ eigenvalue density.
It would be interesting to derive a direct relation between
these discontinuities and non-perturbative corrections to
the matrix model, as well as to understand matrix model
instantons in more general three-dimensional SCFTs.

To conclude, let me also comment on the simplification
appearing in the $\mathcal{N}=3$ flavoured ABJM and
$Q^{1,1,1}/\mathbb{Z}_{N_f}$ theories. In these theories,
the monopole loci have complex dimension two, while the local transverse mesonic coordinates are of the form $AB$ and have
superconformal R-charge $R[AB]=1$. Using the reduction along
the M-theory circle in \eqref{oneloopprod}, and in the absence of zero modes, the contributions from the RW and ML
multiplets can be combined as
\begin{align}\label{5.1}
    Z_{1}
    = Z_{1}^{\mathrm{RW}}Z_{1}^{\mathrm{ML}}
    = \prod_{m\in\mathbb{Z}}
    \big(2R_{\pm}m-2\big)^{-2\chi_{-m}-\chi_m(N\Sigma)}
    \ ,
\end{align}
where $R_{\pm}$ is the R-charge of a gauge-invariant BPS
monopole operator of magnetic charge $\pm1$ on the
corresponding locus. Through the $U(1)_m$ reduction described
in Section~\ref{m-theory}, the multiplicities $\chi_m$ and
$\chi_m(N\Sigma)$ can be expressed as indices of Dolbeault operators on the curve $\B_{\pm}$.

In the examples considered, the descended normal bundle
satisfies $\det N_{\B_{\pm}}\simeq K_{\B_{\pm}}$, and thus the total space of $\pi:N_{\B_{\pm}}\to\B_{\pm}$ has a trivial canonical bundle:\footnote{In the examples explicitly
considered for $\mathcal{N}=3$ flavoured ABJM and
$Q^{1,1,1}/\mathbb{Z}_{N_f}$,
$\B_{\pm}\simeq\mathbb{P}^{1}$ and
$N_{\B_{\pm}}\simeq\mathcal{O}(-1)^{\oplus2}$, so that the
total space of $\pi:N_{\B_{\pm}}\to\B_{\pm}$ is the resolved
conifold.}
\begin{align}\label{5.2}
    K_{N_{\B_{\pm}}}
    \simeq
    \pi^{\ast}\big(
        K_{\B_{\pm}}\otimes(\det N_{\B_{\pm}})^{-1}
    \big)
    \simeq \mathcal{O}
    \ .
\end{align}
Using \eqref{indexdolbm} and the Riemann-Roch theorem for
holomorphic vector bundles, it follows that
\begin{align}
    2\chi_{-m}+\chi_m(N\Sigma)=2-2g
    \ ,
\end{align}
where $g$ is the genus of $\B_{\pm}$ and $g=0$ for the curves considered in Section~\ref{m-theory}.

This simplification is reminiscent of the construction considered in \cite{Nekrasov:2014nea}, and a formal extension of \eqref{5.1}, \eqref{5.2} to curves of arbitrary genus, assuming the same eigenvalues for fluctuations, would result in the Gopakumar-Vafa-type expression
\begin{align}
    Z_{1} = \Big(2\sin\big(\frac{\pi}{R_{\pm}}\big)\Big)^{2g-2} \ .
\end{align}
From the gauge-theory perspective, $R_{\pm}$ is the R-charge of a unit BPS monopole operator, while from the gravity perspective, \eqref{2.33} identifies $\pi/R_{\pm}$ with the effective M-theory circle length $\ell_{\xi}|_{\Sigma_{\pm}}$ on the M2-brane.

It would be interesting to investigate this structure
further, in particular its possible extension to multiple
covers and multi-brane configurations, and its relation
to topological strings.

\section*{Acknowledgements}

I thank Sam Bennett for useful discussions and collaboration at the initial stages. I also thank Jesse van Muiden and Arkady Tseytlin for useful discussions and comments on the draft. This work was supported by the President's PhD Scholarship at Imperial College London.

\newpage

\appendix 
\section{Kaluza-Klein monopoles in M-theory} \label{A}

Kaluza-Klein monopole solutions \cite{Sorkin:1983ns, Gross:1983hb} in M-theory take the form $\mathbb{R}^{1,6}\times \mathrm{TN}$, where the metric on the Taub-NUT space can be written as \cite{Hawking:1976jb, Gibbons:1979xm}:
\begin{align}
    & ds^2 = \mathrm{Re}(e\otimes \bar{e})  \ , \qquad e = e^{0}+i_{r}e^{r} \ ,  \nonumber \\ 
    &e^{0} = U^{-1/2}(d\tau +A) \ , \qquad e^{r} = U^{1/2}dx^{r} \ ,   \qquad r=1,2, 3\ , \label{A1}
\end{align}
where $A(x)$ and $U(x)$ are, respectively, a one-form and a scalar function on $\mathbb{R}^3$, satisfying\footnote{The relative sign depends on the orientation conventions and the choice of left/right hyperkähler structure, as will become evident below.}
\begin{align}\label{sdcond}
     dA = -(\ast_{3}dU)\ .
\end{align}
For convenience, I take the orthonormal coframe $e$ to be valued in $\mathbb{H}$ and $i_{r}=\{i, j,k\}$. In four Euclidean dimensions, the spin group is $\mathrm{Spin}(4)\cong \mathrm{Sp}(1)_{L}\times \mathrm{Sp}(1)_{R}$ and acts on an $\mathbb{H}$-valued tangent vector as
\begin{align}
    q \mapsto aqb^{-1} \ , \qquad a\in \mathrm{Sp}(1)_{L}, \ b\in \mathrm{Sp}(1)_{R} \ .
\end{align}
Consequently, the covariant derivative and curvature split into $\mathfrak{sp}(1)_{L, R}$-valued forms
\begin{align}
    &\nabla q = dq+\omega_{L} q - q \omega_{R} \ ,  \qquad \mathcal{R} = \nabla^2 = \mathcal{R}_{L}-\mathcal{R}_{R} \ , \nonumber \\
    & \mathcal{R}_{L, R} = d\omega_{L, R} +\omega_{L, R} \wedge \omega_{L, R} \ .
\end{align}
It is straightforward to check that the condition \eqref{sdcond} ensures the vanishing of the $\mathfrak{sp}(1)_{L}$ part of the connection $\omega_{L}$. Thus, the holonomy of the manifold is contained within $\mathrm{Sp}(1)_{R}$ and the manifold preserves the hyperkähler structure $I_{r}$, corresponding to the left multiplication by $i_{r}$, so that $\nabla I_{r} = 0$. Equivalently, for the hyperkähler form one has
\begin{align}
    &\Omega = -\mathrm{Im}(e\otimes \bar{e})  = i_{r}\Omega_{r} \ , \quad \nabla\Omega = 0 \ , \quad d\Omega = 0 \ , \\
    &\Omega_{r}(q, p) =\langle I_{r}(q), p\rangle  \ , \qquad \Omega_{r} = e^{0}\wedge e^{r}+\frac{1}{2}\epsilon_{rst}e^{s}\wedge e^{t} \ .
\end{align}

A particular choice of $U(x)$ corresponds to the multi-KK-monopole solution
\begin{align}
    U(x) = \frac{1}{R^2}+\frac{1}{2}\sum_{a=1}^{m}\frac{1}{|x-x_{a}|} \ .
\end{align}
As pointed out in \cite{Townsend:1995kk}, for non-coincident KK-monopoles the solution has no singularity, provided the $S^1$ coordinate $\tau$ has periodicity $\tau \sim \tau +2\pi $. The parameter $R$ of the ALF solution controls the asymptotic length of $S^1$. Identifying the $S^{1}$ coordinate $\tau$ with the M-theory circle and using the standard metric ansatz for dimensional reduction to type IIA, the KK-monopole solution becomes
\begin{align}
    e^{2\varphi} = U^{-3/2}(x) \ , \qquad ds^2_{10} = U^{-1/2}(x)ds^2_{\mathbb{R}^{1,6}}+U^{1/2}(x)dx^{r}dx^{r} \ , 
\end{align}
which is the BPS D6-brane solution of type IIA supergravity with the RR 2-form flux $F_{2}=dA$, whose integral over the transverse two-sphere satisfies
\begin{align}
   \frac{1}{2\pi}\int_{S^{2}_{\infty}} F_{2} = m \ . \label{A.9}
\end{align}
Upon dimensional reduction, the points where the norm of the tangent vector $\partial_{\tau}$ to the M-theory circle vanishes map to the D6-brane insertions.

As shown explicitly in \cite{Gibbons:1996nt}, the multi-Taub-NUT space can be obtained as a hyperkähler quotient \cite{Hitchin:1986ea, Kronheimer:1989zs}. Namely, taking the parameter $R=1$ for simplicity, one can consider an action of the abelian group $G$, locally isomorphic to $\mathbb{R}^{m}$, on $\mathbb{H}^{m}\times \mathbb{H}$. In coordinates $(p_{a}, q), \ a=1, \dots, m$, the group acts as
\begin{align}
    p_{a} \mapsto p_{a}e^{it_{a}} \ , \qquad q\mapsto q+\sum_{a=1}^{m}t_{a} \ , \qquad t_{a}\in \mathfrak{g} \simeq \mathbb{R}^{m} \ .  \label{A10}
\end{align}
The corresponding hyperkähler moment-map components, valued in $\operatorname{Im}(\mathbb{H})\simeq \mathbb{R}^3$, are
\begin{align}
    \mu_{a} = \langle \mu, t_{a}\rangle = -\frac{1}{2}p_{a}i\bar{p}_{a}-\frac{1}{2}(\bar{q}-q) \ . \label{A11}
\end{align}
The moment relations then define the pre-quotient space
\begin{align}
    \mu_{a}^{-1}(x_{a}) = \Big\{i_{r}(x^r-x_{a}^r) = \frac{1}{2}p_{a}i\bar{p}_{a}  \Big\} \ , \qquad q =x^{0} +i_{r}x^{r} \ , \qquad x_{a} = i_{r}x_{a}^{r} \ , \label{A12}
\end{align}
and, as explicitly shown in \cite{Gibbons:1996nt}, the flat metric on $\mathbb{H}^{m}\times \mathbb{H}$ induces the line element \eqref{A1} on the quotient $\mu^{-1}_{a}(x_{a})/G$. More relevant to the main part of the paper is the structure of the quotient space $\mu^{-1}_{a}(x_{a})/G$ as a complex manifold. As discussed in \cite{Kronheimer:1989zs, Witten:1997sc}, 
by choosing one of the complex structures $I_{r}$, the action \eqref{A10} in complex coordinates becomes
\begin{align}
    y_{a} \mapsto e^{it_{a}}y_{a} \ , \quad \tilde{y}_{a} \mapsto e^{-it_{a}}\tilde{y}_{a} \ , \quad w \mapsto w \ , \quad z\mapsto z+\sum_{a=1}^{m}t_{a} \ ,  
\end{align}
and the moment maps \eqref{A11} can be written as
\begin{align}
    \mu_{\mathbb{R}, a} =-\frac{1}{2}|y_{a}|^2+\frac{1}{2}|\tilde{y}_{a}|^2 +\operatorname{Im}(z)  \ , \qquad \mu_{\mathbb{C}, a} = -y_{a}\tilde{y}_{a}+w \ .
\end{align}
This allows one to identify the multi-Taub-NUT space with a symplectic quotient
\begin{align}
    \big(\mu_{\mathbb{R}, a}^{-1}(\xi_{a})\cap\mu^{-1}_{\mathbb{C}, a}(e_{a}) \big)/G \ ,
\end{align}
where $\xi_{a} = x_{a}^{1}$, $e_{a} = x_{a}^{2}+ix_{a}^3$. Algebraically, by the Kempf-Ness theorem, this can be written as the GIT quotient\footnote{For a pedagogical exposition of the GIT quotient and its wider applications, see, for instance, \cite{ginzburg2009lectures} and references therein.}
\begin{align}
    \big(\mu_{\mathbb{R}, a}^{-1}(\xi_{a})\cap\mu^{-1}_{\mathbb{C}, a}(e_{a}) \big)/G \simeq \mu^{-1}_{\mathbb{C},a}(e_{a})//_{\chi_{\xi}} \,G^{\mathbb{C}} = \operatorname{Proj}\Big( \bigoplus_{n\geq 0}\mathbb{C}[\mu^{-1}_{\mathbb{C}, a}(e_{a})]^{\chi_{\xi}^n}\Big) \ .
\end{align}
The effective complexified group is $(\mathbb{C}^{\ast})^m$, and the character $\chi_{\xi}:(\mathbb{C}^{\ast})^m \to \mathbb{C}^{\ast}$, corresponding to the real moment-map levels, defines the semi-invariant sections\footnote{In the GIT description the weights $\xi=(\xi_{1},\dots,\xi_{m})\in\mathbb{Z}^{m}$ must be integral in order to define an algebraic character.}
\begin{align}
    &f_{n} \mapsto \chi_{\xi}^{n}(\lambda)f_{n} \ ,  \qquad f_{n}\in \mathbb{C}[\mu_{\mathbb{C},a}^{-1}(e_{a})]^{\chi_{\xi}^n} \ , \\
    &\chi_{\xi}(\lambda) = \lambda_{1}^{\xi_{1}}\dots \lambda_{m}^{\xi_m} \ , \qquad \lambda_{a}=e^{it'_{a}}\in \mathbb{C}^{\ast} \ ,
\end{align} 
where $t'_{a}\in \mathbb{C}$ are parameters of the complexified group $G^{\mathbb{C}}$. Physically, the parameters $\xi_{a}, e_{a}$ specify the positions of the KK-monopoles. When no two KK-monopoles coincide, the GIT quotient provides a smooth resolution of the quotient. For the trivial character $\chi_{0} = 1$, the GIT quotient simply gives
\begin{align}
    \mu_{\mathbb{C}, a}^{-1}(e_{a}) // G^{\mathbb{C}} = \operatorname{Spec}\mathbb{C}[\mu_{\mathbb{C},a}^{-1}(e_{a})]^{G^{\mathbb{C}}} = \operatorname{Spec} \big(\mathbb{C}[T, \tilde{T}, w]/I\big) \ ,
\end{align}
where the generators $T, \tilde{T}, w$ are $G^{\mathbb{C}}$-invariant coordinates and the resulting variety $V(I)$ is determined by the ideal $I$ generated by the relations among these coordinates
\begin{align}
    T \equiv e^{-iz}\prod_{a=1}^{m}y_{a} \ , \qquad \tilde{T} \equiv e^{iz}\prod_{a=1}^{m}\tilde{y}_{a} \ , \qquad  T\tilde{T} = \prod_{a=1}^{m}(w-e_{a}) \ . \label{A.20}
\end{align}
When all KK-monopole positions coincide, the affine GIT quotient is $\mu^{-1}_{\mathbb{C}, a}(0)//G^{\mathbb{C}}\simeq \mathbb{C}^2/\mathbb{Z}_{m}$, giving an $A_{m-1}$ singularity.\footnote{Note that when $e_{a}=e_{b}$ a minimal resolution of singularities is provided by $\xi_{a}\neq \xi_{b}$ \cite{Kronheimer:1989zs}. 
} As also observed in \cite{Witten:1997sc}, in coordinates \eqref{A.20} the symmetry corresponding to the complexified translations along the M-theory circle can be written as
\begin{align}
    T \mapsto \nu T \ , \qquad \tilde{T} \mapsto \nu^{-1}\tilde{T} \ , \qquad w\mapsto w \ .
\end{align}
The points $w=e_{a}$, $T=\tilde{T} = 0$ are the fixed points of the M-theory circle action and correspond to the KK-monopole insertions. The total magnetic charge $m$ in \eqref{A.9} is the sum of the magnetic charges of the individual KK-monopoles.\footnote{More precisely, assuming $e_{a}\neq e_{b}$, locally, around each KK-monopole insertion $w = e_{a}+\varepsilon'$, one can consider a three-sphere $S^3_{a}: |T|^2+|\tilde{T}|^2 = \epsilon^2$. On each such three-sphere, the M-theory circle acts with no fixed points and defines the Hopf fibration. Since $dF_{2} = 0$ away from the KK-monopole loci, Gauss's law for magnetic charges gives $\frac{1}{2\pi}\int_{S^2_{\infty}} F_{2} = m$, where each insertion contributes via the first Chern number of the Hopf fibration.}

\newpage
\small
\bibliographystyle{JHEP-v2.9}
\bibliography{M2-ext-bib}

@article{Hama:2010av,
    author = "Hama, Naofumi and Hosomichi, Kazuo and Lee, Sungjay",
    title = "{Notes on SUSY Gauge Theories on Three-Sphere}",
    eprint = "1012.3512",
    archivePrefix = "arXiv",
    primaryClass = "hep-th",
    reportNumber = "DAMTP-2010-129, YITP-10-100",
    doi = "10.1007/JHEP03(2011)127",
    journal = "JHEP",
    volume = "03",
    pages = "127",
    year = "2011"
}

@article{Franco:2009sp,
    author = "Franco, Sebastian and Klebanov, Igor R. and Rodriguez-Gomez, Diego",
    title = "{M2-branes on Orbifolds of the Cone over $Q^{1,1,1}$}",
    eprint = "0903.3231",
    archivePrefix = "arXiv",
    primaryClass = "hep-th",
    reportNumber = "NSF-KITP-09-22, PUPT-2294",
    doi = "10.1088/1126-6708/2009/08/033",
    journal = "JHEP",
    volume = "08",
    pages = "033",
    year = "2009"
}

@article{Martelli:2008si,
    author = "Martelli, Dario and Sparks, James",
    title = "{Moduli spaces of Chern-Simons quiver gauge theories and AdS$_4$/CFT$_3$}",
    eprint = "0808.0912",
    archivePrefix = "arXiv",
    primaryClass = "hep-th",
    doi = "10.1103/PhysRevD.78.126005",
    journal = "Phys. Rev. D",
    volume = "78",
    pages = "126005",
    year = "2008"
}

@article{Fabbri:1999hw,
    author = "Fabbri, Davide and Fre', Pietro and Gualtieri, Leonardo and Reina, Cesare and Tomasiello, Alessandro and Zaffaroni, Alberto and Zampa, Alessandro",
    title = "{3-D superconformal theories from Sasakian seven manifolds: New nontrivial evidences for AdS$_4$ / CFT$_3$}",
    eprint = "hep-th/9907219",
    archivePrefix = "arXiv",
    reportNumber = "CERN-TH-99-231",
    doi = "10.1016/S0550-3213(00)00098-5",
    journal = "Nucl. Phys. B",
    volume = "577",
    pages = "547--608",
    year = "2000"
}

@article{Kim:2012vza,
    author = "Kim, Hyojoong and Kim, Nakwoo",
    title = "{Operator Counting for N=2 Chern-Simons Gauge Theories with Chiral-like Matter Fields}",
    eprint = "1202.6637",
    archivePrefix = "arXiv",
    primaryClass = "hep-th",
    doi = "10.1007/JHEP05(2012)152",
    journal = "JHEP",
    volume = "05",
    pages = "152",
    year = "2012"
}

@article{Franco:2008um,
    author = "Franco, Sebastian and Hanany, Amihay and Park, Jaemo and Rodriguez-Gomez, Diego",
    title = "{Towards M2-brane Theories for Generic Toric Singularities}",
    eprint = "0809.3237",
    archivePrefix = "arXiv",
    primaryClass = "hep-th",
    reportNumber = "PUPT-2279, NSF-KITP-08-125, IMPERIAL-TP-08-AH-08, SU-ITP-08-22",
    doi = "10.1088/1126-6708/2008/12/110",
    journal = "JHEP",
    volume = "12",
    pages = "110",
    year = "2008"
}

@article{Hanany:2008cd,
    author = "Hanany, Amihay and Zaffaroni, Alberto",
    title = "{Tilings, Chern-Simons Theories and M2 Branes}",
    eprint = "0808.1244",
    archivePrefix = "arXiv",
    primaryClass = "hep-th",
    doi = "10.1088/1126-6708/2008/10/111",
    journal = "JHEP",
    volume = "10",
    pages = "111",
    year = "2008"
}

@article{Hanany:2008fj,
    author = "Hanany, Amihay and Vegh, David and Zaffaroni, Alberto",
    title = "{Brane Tilings and M2 Branes}",
    eprint = "0809.1440",
    archivePrefix = "arXiv",
    primaryClass = "hep-th",
    reportNumber = "IMPERIAL-TP-08-AH-07, BICOCCA-FT-09-15, MIT-CTP-3977",
    doi = "10.1088/1126-6708/2009/03/012",
    journal = "JHEP",
    volume = "03",
    pages = "012",
    year = "2009"
}

@article{Goddard:1976qe,
    author = "Goddard, P. and Nuyts, J. and Olive, David I.",
    title = "{Gauge Theories and Magnetic Charge}",
    reportNumber = "CERN-TH-2255",
    doi = "10.1016/0550-3213(77)90221-8",
    journal = "Nucl. Phys. B",
    volume = "125",
    pages = "1--28",
    year = "1977"
}

@article{Cremonesi:2016nbo,
    author = "Cremonesi, Stefano and Mekareeya, Noppadol and Zaffaroni, Alberto",
    title = "{The moduli spaces of 3d $ \mathcal{N}\ge 2 $ Chern-Simons gauge theories and their Hilbert series}",
    eprint = "1607.05728",
    archivePrefix = "arXiv",
    primaryClass = "hep-th",
    reportNumber = "CERN-TH-2016-167",
    doi = "10.1007/JHEP10(2016)046",
    journal = "JHEP",
    volume = "10",
    pages = "046",
    year = "2016"
}

@article{Jafferis:2011zi,
    author = "Jafferis, Daniel L. and Klebanov, Igor R. and Pufu, Silviu S. and Safdi, Benjamin R.",
    title = "{Towards the F-Theorem: N=2 Field Theories on the Three-Sphere}",
    eprint = "1103.1181",
    archivePrefix = "arXiv",
    primaryClass = "hep-th",
    reportNumber = "PUPT-2366",
    doi = "10.1007/JHEP06(2011)102",
    journal = "JHEP",
    volume = "06",
    pages = "102",
    year = "2011"
}

@article{Benini:2009qs,
    author = "Benini, Francesco and Closset, Cyril and Cremonesi, Stefano",
    title = "{Chiral flavors and M2-branes at toric CY$_4$ singularities}",
    eprint = "0911.4127",
    archivePrefix = "arXiv",
    primaryClass = "hep-th",
    reportNumber = "PUTP-2324, TAUP-2906-09",
    doi = "10.1007/JHEP02(2010)036",
    journal = "JHEP",
    volume = "02",
    pages = "036",
    year = "2010"
}

@article{Jafferis:2009th,
    author = "Jafferis, Daniel Louis",
    title = "{Quantum corrections to $\mathcal{N} = 2$ Chern-Simons theories with flavor and their AdS$_{4}$ duals}",
    eprint = "0911.4324",
    archivePrefix = "arXiv",
    primaryClass = "hep-th",
    doi = "10.1007/JHEP08(2013)046",
    journal = "JHEP",
    volume = "08",
    pages = "046",
    year = "2013"
}

@article{Schwarz:2004yj,
    author = "Schwarz, John H.",
    title = "{Superconformal Chern-Simons theories}",
    eprint = "hep-th/0411077",
    archivePrefix = "arXiv",
    reportNumber = "CALT-68-2525",
    doi = "10.1088/1126-6708/2004/11/078",
    journal = "JHEP",
    volume = "11",
    pages = "078",
    year = "2004"
}

@article{Gaiotto:2009mv,
    author = "Gaiotto, Davide and Tomasiello, Alessandro",
    title = "{The gauge dual of Romans mass}",
    eprint = "0901.0969",
    archivePrefix = "arXiv",
    primaryClass = "hep-th",
    doi = "10.1007/JHEP01(2010)015",
    journal = "JHEP",
    volume = "01",
    pages = "015",
    year = "2010"
}

@article{Sorkin:1983ns,
    author = "Sorkin, R. D.",
    title = "{Kaluza-Klein Monopole}",
    doi = "10.1103/PhysRevLett.51.87",
    journal = "Phys. Rev. Lett.",
    volume = "51",
    pages = "87--90",
    year = "1983"
}

@article{Gross:1983hb,
    author = "Gross, David J. and Perry, Malcolm J.",
    title = "{Magnetic Monopoles in Kaluza-Klein Theories}",
    reportNumber = "PRINT-83-0305 (PRINCETON)",
    doi = "10.1016/0550-3213(83)90462-5",
    journal = "Nucl. Phys. B",
    volume = "226",
    pages = "29--48",
    year = "1983"
}

@article{Gibbons:1979xm,
    author = "Gibbons, G. W. and Hawking, S. W.",
    title = "{Classification of Gravitational Instanton Symmetries}",
    doi = "10.1007/BF01197189",
    journal = "Commun. Math. Phys.",
    volume = "66",
    pages = "291--310",
    year = "1979"
}

@article{Hawking:1976jb,
    author = "Hawking, S. W.",
    title = "{Gravitational Instantons}",
    reportNumber = "Print-77-0294 (CAMBRIDGE)",
    doi = "10.1016/0375-9601(77)90386-3",
    journal = "Phys. Lett. A",
    volume = "60",
    pages = "81",
    year = "1977"
}

@article{Townsend:1995kk,
    author = "Townsend, P. K.",
    title = "{The eleven-dimensional supermembrane revisited}",
    eprint = "hep-th/9501068",
    archivePrefix = "arXiv",
    reportNumber = "DAMTP-R-95-2",
    doi = "10.1016/0370-2693(95)00397-4",
    journal = "Phys. Lett. B",
    volume = "350",
    pages = "184--187",
    year = "1995"
}

@article{Gibbons:1996nt,
    author = "Gibbons, G. W. and Rychenkova, P. and Goto, R.",
    title = "{HyperKahler quotient construction of BPS monopole moduli spaces}",
    eprint = "hep-th/9608085",
    archivePrefix = "arXiv",
    reportNumber = "DAMTP-R-38-96",
    doi = "10.1007/s002200050121",
    journal = "Commun. Math. Phys.",
    volume = "186",
    pages = "585--599",
    year = "1997"
}

@article{Aharony:1997bx,
    author = "Aharony, Ofer and Hanany, Amihay and Intriligator, Kenneth A. and Seiberg, N. and Strassler, M. J.",
    title = "{Aspects of N=2 supersymmetric gauge theories in three-dimensions}",
    eprint = "hep-th/9703110",
    archivePrefix = "arXiv",
    reportNumber = "RU-97-10A, IASSNS-HEP-97-18",
    doi = "10.1016/S0550-3213(97)00323-4",
    journal = "Nucl. Phys. B",
    volume = "499",
    pages = "67--99",
    year = "1997"
}

@article{Borokhov:2002cg,
    author = "Borokhov, Vadim and Kapustin, Anton and Wu, Xin-kai",
    title = "{Monopole operators and mirror symmetry in three-dimensions}",
    eprint = "hep-th/0207074",
    archivePrefix = "arXiv",
    reportNumber = "CALT-68-2397",
    doi = "10.1088/1126-6708/2002/12/044",
    journal = "JHEP",
    volume = "12",
    pages = "044",
    year = "2002"
}

@article{Hitchin:1986ea,
    author = "Hitchin, Nigel J. and Karlhede, A. and Lindstrom, U. and Rocek, M.",
    title = "{Hyperkahler Metrics and Supersymmetry}",
    reportNumber = "STOCKHOLM-ITP-86-8",
    doi = "10.1007/BF01214418",
    journal = "Commun. Math. Phys.",
    volume = "108",
    pages = "535",
    year = "1987"
}

@article{Witten:1997sc,
    author = "Witten, Edward",
    title = "{Solutions of four-dimensional field theories via M-theory}",
    eprint = "hep-th/9703166",
    archivePrefix = "arXiv",
    reportNumber = "IASSNS-HEP-97-19",
    doi = "10.1016/S0550-3213(97)00416-1",
    journal = "Nucl. Phys. B",
    volume = "500",
    pages = "3--42",
    year = "1997"
}

@article{ginzburg2009lectures,
  title="{Lectures on Nakajima's quiver varieties}",
  author={Ginzburg, Victor},
  journal={arXiv preprint arXiv:0905.0686},
  year={2009}
}

@article{Drukker:2011zy,
    author = "Drukker, Nadav and Marino, Marcos and Putrov, Pavel",
    title = "{Nonperturbative aspects of ABJM theory}",
    eprint = "1103.4844",
    archivePrefix = "arXiv",
    primaryClass = "hep-th",
    reportNumber = "IMPERIAL-TP-2011-ND-01",
    doi = "10.1007/JHEP11(2011)141",
    journal = "JHEP",
    volume = "11",
    pages = "141",
    year = "2011"
}

@article{Drukker:2010nc,
    author = "Drukker, Nadav and Marino, Marcos and Putrov, Pavel",
    title = "{From weak to strong coupling in ABJM theory}",
    eprint = "1007.3837",
    archivePrefix = "arXiv",
    primaryClass = "hep-th",
    reportNumber = "HU-EP-10-39",
    doi = "10.1007/s00220-011-1253-6",
    journal = "Commun. Math. Phys.",
    volume = "306",
    pages = "511--563",
    year = "2011"
}

@article{Lin:2026uzp,
    author = "Lin, Jieming and van Muiden, Jesse and Wang, Zihan",
    title = "{M5 Instantons in ABJM}",
    eprint = "2609.09041",
    archivePrefix = "arXiv",
    primaryClass = "hep-th",
    month = "9",
    year = "2026"
}

@article{Distler:1986wm,
    author = "Distler, Jacques",
    title = "{Resurrecting (2,0) Compactifications}",
    doi = "10.1016/0370-2693(87)91643-1",
    journal = "Phys. Lett. B",
    volume = "188",
    pages = "431--436",
    year = "1987"
}

@article{Distler:1987ee,
    author = "Distler, Jacques and Greene, Brian R.",
    title = "{Aspects of (2,0) String Compactifications}",
    reportNumber = "HUTP-87/A065",
    doi = "10.1016/0550-3213(88)90619-0",
    journal = "Nucl. Phys. B",
    volume = "304",
    pages = "1--62",
    year = "1988"
}

@article{Cagnazzo:2009zh,
    author = "Cagnazzo, Alessandra and Sorokin, Dmitri and Wulff, Linus",
    title = "{String instanton in $AdS_4\times CP^3$}",
    eprint = "0911.5228",
    archivePrefix = "arXiv",
    primaryClass = "hep-th",
    doi = "10.1007/JHEP05(2010)009",
    journal = "JHEP",
    volume = "05",
    pages = "009",
    year = "2010"
}

@article{Gautason:2023igo,
    author = "Gautason, Fridrik Freyr and Puletti, Valentina Giangreco M. and van Muiden, Jesse",
    title = "{Quantized strings and instantons in holography}",
    eprint = "2304.12340",
    archivePrefix = "arXiv",
    primaryClass = "hep-th",
    doi = "10.1007/JHEP08(2023)218",
    journal = "JHEP",
    volume = "08",
    pages = "218",
    year = "2023"
}

@article{Kronheimer:1989zs,
    author = "Kronheimer, P. B.",
    title = {{The construction of ALE spaces as hyper-K{\"a}hler quotients}},
    journal = "J. Diff. Geom.",
    volume = "29",
    number = "3",
    pages = "665--683",
    year = "1989"
}

@article{Farquet:2013cwa,
    author = "Farquet, Daniel and Sparks, James",
    title = "{Wilson loops and the geometry of matrix models in AdS$_4$/CFT$_3$}",
    eprint = "1304.0784",
    archivePrefix = "arXiv",
    primaryClass = "hep-th",
    doi = "10.1007/JHEP01(2014)083",
    journal = "JHEP",
    volume = "01",
    pages = "083",
    year = "2014"
}

@article{Gaiotto:2009tk,
    author = "Gaiotto, Davide and Jafferis, Daniel Louis",
    title = "{Notes on adding D6 branes wrapping $RP^3$ in $AdS_4\times CP^3$}",
    eprint = "0903.2175",
    archivePrefix = "arXiv",
    primaryClass = "hep-th",
    doi = "10.1007/JHEP11(2012)015",
    journal = "JHEP",
    volume = "11",
    pages = "015",
    year = "2012"
}

@article{Borokhov:2002ib,
    author = "Borokhov, Vadim and Kapustin, Anton and Wu, Xin-kai",
    title = "{Topological disorder operators in three-dimensional conformal field theory}",
    eprint = "hep-th/0206054",
    archivePrefix = "arXiv",
    reportNumber = "CALT-68-2391",
    doi = "10.1088/1126-6708/2002/11/049",
    journal = "JHEP",
    volume = "11",
    pages = "049",
    year = "2002"
}

@article{Borokhov:2003yu,
    author = "Borokhov, Vadim",
    title = "{Monopole operators in three-dimensional N=4 SYM and mirror symmetry}",
    eprint = "hep-th/0310254",
    archivePrefix = "arXiv",
    reportNumber = "CALT-68-2459",
    doi = "10.1088/1126-6708/2004/03/008",
    journal = "JHEP",
    volume = "03",
    pages = "008",
    year = "2004"
}

@article{deBoer:1997kr,
    author = "de Boer, Jan and Hori, Kentaro and Oz, Yaron",
    title = "{Dynamics of N=2 supersymmetric gauge theories in three-dimensions}",
    eprint = "hep-th/9703100",
    archivePrefix = "arXiv",
    reportNumber = "LBL-40109, LBNL-40109, UCB-PTH-97-13",
    doi = "10.1016/S0550-3213(97)00328-3",
    journal = "Nucl. Phys. B",
    volume = "500",
    pages = "163--191",
    year = "1997"
}

@article{Hohenegger:2009as,
    author = "Hohenegger, Stefan and Kirsch, Ingo",
    title = "{A Note on the holography of Chern-Simons matter theories with flavour}",
    eprint = "0903.1730",
    archivePrefix = "arXiv",
    primaryClass = "hep-th",
    doi = "10.1088/1126-6708/2009/04/129",
    journal = "JHEP",
    volume = "04",
    pages = "129",
    year = "2009"
}

@article{Hatsuda:2013gj,
    author = "Hatsuda, Yasuyuki and Moriyama, Sanefumi and Okuyama, Kazumi",
    title = "{Instanton Bound States in ABJM Theory}",
    eprint = "1301.5184",
    archivePrefix = "arXiv",
    primaryClass = "hep-th",
    reportNumber = "DESY-13-010, TIT-HEP-626",
    doi = "10.1007/JHEP05(2013)054",
    journal = "JHEP",
    volume = "05",
    pages = "054",
    year = "2013"
}

@article{Hatsuda:2013oxa,
    author = "Hatsuda, Yasuyuki and Marino, Marcos and Moriyama, Sanefumi and Okuyama, Kazumi",
    title = "{Non-perturbative effects and the refined topological string}",
    eprint = "1306.1734",
    archivePrefix = "arXiv",
    primaryClass = "hep-th",
    reportNumber = "DESY-13-096, TIT-HEP-627",
    doi = "10.1007/JHEP09(2014)168",
    journal = "JHEP",
    volume = "09",
    pages = "168",
    year = "2014"
}

@article{Moriyama:2014gxa,
    author = "Moriyama, Sanefumi and Nosaka, Tomoki",
    title = "{Partition Functions of Superconformal Chern-Simons Theories from Fermi Gas Approach}",
    eprint = "1407.4268",
    archivePrefix = "arXiv",
    primaryClass = "hep-th",
    reportNumber = "YITP-14-56",
    doi = "10.1007/JHEP11(2014)164",
    journal = "JHEP",
    volume = "11",
    pages = "164",
    year = "2014"
}

@article{Moriyama:2015jsa,
    author = "Moriyama, Sanefumi and Nosaka, Tomoki",
    title = "{Superconformal Chern-Simons Partition Functions of Affine D-type Quiver from Fermi Gas}",
    eprint = "1504.07710",
    archivePrefix = "arXiv",
    primaryClass = "hep-th",
    reportNumber = "YITP-15-18, OCU-PHYS-425",
    doi = "10.1007/JHEP09(2015)054",
    journal = "JHEP",
    volume = "09",
    pages = "054",
    year = "2015"
}

@article{Moriyama:2014nca,
    author = "Moriyama, Sanefumi and Nosaka, Tomoki",
    title = "{Exact Instanton Expansion of Superconformal Chern-Simons Theories from Topological Strings}",
    eprint = "1412.6243",
    archivePrefix = "arXiv",
    primaryClass = "hep-th",
    reportNumber = "YITP-14-98",
    doi = "10.1007/JHEP05(2015)022",
    journal = "JHEP",
    volume = "05",
    pages = "022",
    year = "2015"
}

@article{Calvo:2012du,
    author = "Calvo, Flavio and Marino, Marcos",
    title = "{Membrane instantons from a semiclassical TBA}",
    eprint = "1212.5118",
    archivePrefix = "arXiv",
    primaryClass = "hep-th",
    doi = "10.1007/JHEP05(2013)006",
    journal = "JHEP",
    volume = "05",
    pages = "006",
    year = "2013"
}

@article{Honda:2014ica,
    author = "Honda, Masazumi and Moriyama, Sanefumi",
    title = "{Instanton Effects in Orbifold ABJM Theory}",
    eprint = "1404.0676",
    archivePrefix = "arXiv",
    primaryClass = "hep-th",
    reportNumber = "HRI-ST1403, KEK-TH-1714",
    doi = "10.1007/JHEP08(2014)091",
    journal = "JHEP",
    volume = "08",
    pages = "091",
    year = "2014"
}

@article{Martelli:2009ga,
    author = "Martelli, Dario and Sparks, James",
    title = "{AdS$_4$ / CFT$_3$ duals from M2-branes at hypersurface singularities and their deformations}",
    eprint = "0909.2036",
    archivePrefix = "arXiv",
    primaryClass = "hep-th",
    doi = "10.1088/1126-6708/2009/12/017",
    journal = "JHEP",
    volume = "12",
    pages = "017",
    year = "2009"
}

@article{Hosseini:2025jxb,
    author = "Hosseini, Seyed Morteza",
    title = "{Large-N Free Energy of Chiral N=2 Chern-Simons-Matter Theories}",
    eprint = "2510.24837",
    archivePrefix = "arXiv",
    primaryClass = "hep-th",
    doi = "10.1103/2bpj-jf7x",
    journal = "Phys. Rev. Lett.",
    volume = "136",
    number = "9",
    pages = "091601",
    year = "2026"
}

@article{Hikida:2009tp,
    author = "Hikida, Yasuaki and Li, Wei and Takayanagi, Tadashi",
    title = "{ABJM with Flavors and FQHE}",
    eprint = "0903.2194",
    archivePrefix = "arXiv",
    primaryClass = "hep-th",
    reportNumber = "KEK-TH-1307, IPMU09-0032",
    doi = "10.1088/1126-6708/2009/07/065",
    journal = "JHEP",
    volume = "07",
    pages = "065",
    year = "2009"
}

@article{Nekrasov:2014nea,
    author = "Nekrasov, Nikita and Okounkov, Andrei",
    title = "{Membranes and sheaves}",
    eprint = "1404.2323",
    archivePrefix = "arXiv",
    primaryClass = "math.AG",
    doi = "10.14231/AG-2016-015",
    journal = "Algebr.  Geom.",
    volume = "3",
    number = "3",
    pages = "320--369",
    year = "2016"
}

@article{Jafferis:2010un,
    author = "Jafferis, Daniel L.",
    title = "{The Exact Superconformal R-Symmetry Extremizes Z}",
    eprint = "1012.3210",
    archivePrefix = "arXiv",
    primaryClass = "hep-th",
    doi = "10.1007/JHEP05(2012)159",
    journal = "JHEP",
    volume = "05",
    pages = "159",
    year = "2012"
}

@article{Martelli:2006yb,
    author = "Martelli, Dario and Sparks, James and Yau, Shing-Tung",
    title = "{Sasaki-Einstein manifolds and volume minimisation}",
    eprint = "hep-th/0603021",
    archivePrefix = "arXiv",
    reportNumber = "CERN-PH-TH-2006-039, HUTP-06-A0002",
    doi = "10.1007/s00220-008-0479-4",
    journal = "Commun. Math. Phys.",
    volume = "280",
    pages = "611--673",
    year = "2008"
}

@article{Bergman:2001qi,
    author = "Bergman, Aaron and Herzog, Christopher P.",
    title = "{The Volume of some nonspherical horizons and the AdS / CFT correspondence}",
    eprint = "hep-th/0108020",
    archivePrefix = "arXiv",
    reportNumber = "PUPT-2002",
    doi = "10.1088/1126-6708/2002/01/030",
    journal = "JHEP",
    volume = "01",
    pages = "030",
    year = "2002"
}

@article{Martelli:2005tp,
    author = "Martelli, D. and Sparks, J. and Yau, Shing-Tung",
    title = "{The Geometric dual of a-maximisation for Toric Sasaki-Einstein manifolds}",
    eprint = "hep-th/0503183",
    archivePrefix = "arXiv",
    reportNumber = "CERN-PH-TH-2005-047, HUTP-05-A0012",
    doi = "10.1007/s00220-006-0087-0",
    journal = "Commun. Math. Phys.",
    volume = "268",
    pages = "39--65",
    year = "2006"
}

@article{Kurlyand:2026yke,
    author = "Kurlyand, Stefan A.",
    title = "{Membrane instantons and non-perturbative effects in AdS$_{4}$/CFT$_{3}$}",
    eprint = "2606.19467",
    archivePrefix = "arXiv",
    primaryClass = "hep-th",
    doi = "10.1007/JHEP09(2026)222",
    journal = "JHEP",
    volume = "09",
    pages = "222",
    year = "2026"
}

@article{Bobev:2026gir,
    author = "Bobev, Nikolay and Gautason, Fridrik Freyr and van Muiden, Jesse",
    title = "{Holographic Tests of the $\mu$ Ensemble}",
    eprint = "2607.06493",
    archivePrefix = "arXiv",
    primaryClass = "hep-th",
    month = "7",
    year = "2026"
}

@article{Cassia:2025jkr,
    author = "Cassia, Luca and Hristov, Kiril",
    title = "{M2-brane partition functions and HD supergravity from equivariant volumes}",
    eprint = "2508.21619",
    archivePrefix = "arXiv",
    primaryClass = "hep-th",
    doi = "10.1007/JHEP03(2026)100",
    journal = "JHEP",
    volume = "03",
    pages = "100",
    year = "2026"
}

@article{Beccaria:2025ahf,
    author = "Beccaria, Matteo and Kurlyand, Stefan A. and Tseytlin, Arkady A.",
    title = "{2-loop free energy of M2 brane in AdS$_{7}$ {\texttimes} S$^{4}$ and surface defect anomaly in (2,0) theory}",
    eprint = "2511.22306",
    archivePrefix = "arXiv",
    primaryClass = "hep-th",
    reportNumber = "Imperial-TP-2025-SK-03",
    doi = "10.1007/JHEP03(2026)257",
    journal = "JHEP",
    volume = "03",
    pages = "257",
    year = "2026"
}

@article{Beccaria:2026zjj,
    author = "Beccaria, M. and Kurlyand, S. A. and Tseytlin, A. A. and van Muiden, J.",
    title = "{M2 brane in AdS$_4\times S^7/\mathbb Z_k$: 2-loop correction to $\frac{1}{2}$-BPS Wilson loop vs localization in ABJM theory}",
    eprint = "2609.14497",
    archivePrefix = "arXiv",
    primaryClass = "hep-th",
    month = "9",
    year = "2026"
}

@article{Kallen:2013qla,
    author = "Kallen, Johan and Marino, Marcos",
    title = "{Instanton Effects and Quantum Spectral Curves}",
    eprint = "1308.6485",
    archivePrefix = "arXiv",
    primaryClass = "hep-th",
    doi = "10.1007/s00023-015-0421-1",
    journal = "Annales Henri Poincare",
    volume = "17",
    number = "5",
    pages = "1037--1074",
    year = "2016"
}

@article{Grassi:2014vwa,
    author = "Grassi, Alba and Marino, Marcos",
    title = "{M-theoretic matrix models}",
    eprint = "1403.4276",
    archivePrefix = "arXiv",
    primaryClass = "hep-th",
    doi = "10.1007/JHEP02(2015)115",
    journal = "JHEP",
    volume = "02",
    pages = "115",
    year = "2015"
}

@article{Santamaria:2010dm,
    author = "Santamaria, Ricardo Couso and Marino, Marcos and Putrov, Pavel",
    title = "{Unquenched flavor and tropical geometry in strongly coupled Chern-Simons-matter theories}",
    eprint = "1011.6281",
    archivePrefix = "arXiv",
    primaryClass = "hep-th",
    doi = "10.1007/JHEP10(2011)139",
    journal = "JHEP",
    volume = "10",
    pages = "139",
    year = "2011"
}

@article{Tseytlin:2026dhy,
    author = "Tseytlin, Arkady A.",
    title = "{M2 brane $S^3/\mathbb Z_k$ instanton partition function in ${\rm AdS}_4\times S^7/\mathbb Z_k$: 2-loop correction}",
    eprint = "2609.34777",
    archivePrefix = "arXiv",
    primaryClass = "hep-th",
    month = "9",
    year = "2026"
}

@article{Hatsuda:2014vsa,
    author = "Hatsuda, Yasuyuki and Okuyama, Kazumi",
    title = "{Probing non-perturbative effects in M-theory}",
    eprint = "1407.3786",
    archivePrefix = "arXiv",
    primaryClass = "hep-th",
    reportNumber = "DESY-14-126",
    doi = "10.1007/JHEP10(2014)158",
    journal = "JHEP",
    volume = "10",
    pages = "158",
    year = "2014"
}

@article{Hatsuda:2015lpa,
    author = "Hatsuda, Yasuyuki and Honda, Masazumi and Okuyama, Kazumi",
    title = "{Large N non-perturbative effects in $\mathcal{N}=4$ superconformal Chern-Simons theories}",
    eprint = "1505.07120",
    archivePrefix = "arXiv",
    primaryClass = "hep-th",
    reportNumber = "DESY-15-078, HRI-ST-1505",
    doi = "10.1007/JHEP09(2015)046",
    journal = "JHEP",
    volume = "09",
    pages = "046",
    year = "2015"
}

@article{brugalle2015brief,
  title={Brief introduction to tropical geometry},
  author={Brugall{\'e}, Erwan and Itenberg, Ilia and Mikhalkin, Grigory and Shaw, Kristin},
  journal={arXiv preprint arXiv:1502.05950},
  year={2015}
}

@article{Iwao2010,
    author = {Iwao, Shinsuke},
    title = {Integration over Tropical Plane Curves and Ultradiscretization},
    journal = {International Mathematics Research Notices},
    volume = {2010},
    number = {1},
    pages = {112-148},
    year = {2010},
    month = {01},
    issn = {1073-7928},
    doi = {10.1093/imrn/rnp129},
    url = {https://doi.org/10.1093/imrn/rnp129},
}

@book{ItenbergMikhalkinShustin2009,
  author    = {Itenberg, Illia and Mikhalkin, Grigory and Shustin, Eugenii},
  title     = {Tropical Algebraic Geometry},
  series    = {Oberwolfach Seminars},
  edition   = {2},
  publisher = {Birkh{\"a}user Basel},
  year      = {2009},
  isbn      = {978-3-0346-0047-7},
  doi       = {10.1007/978-3-0346-0048-4}
}

@article{Herzog:2010hf,
    author = "Herzog, Christopher P. and Klebanov, Igor R. and Pufu, Silviu S. and Tesileanu, Tiberiu",
    title = "{Multi-Matrix Models and Tri-Sasaki Einstein Spaces}",
    eprint = "1011.5487",
    archivePrefix = "arXiv",
    primaryClass = "hep-th",
    reportNumber = "PUPT-2359",
    doi = "10.1103/PhysRevD.83.046001",
    journal = "Phys. Rev. D",
    volume = "83",
    pages = "046001",
    year = "2011"
}

@article{Marino:2012az,
    author = "Mari{\~n}o, Marcos and Putrov, Pavel",
    title = "{Interacting fermions and N=2 Chern-Simons-matter theories}",
    eprint = "1206.6346",
    archivePrefix = "arXiv",
    primaryClass = "hep-th",
    doi = "10.1007/JHEP11(2013)199",
    journal = "JHEP",
    volume = "11",
    pages = "199",
    year = "2013"
}

@article{BenettiGenolini:2026cyc,
    author = "Benetti Genolini, Pietro and Gaar, Florian and Gauntlett, Jerome P. and Park, Jaeha and Sparks, James",
    title = "{Airy functions from quantum M-theory}",
    eprint = "2607.07255",
    archivePrefix = "arXiv",
    primaryClass = "hep-th",
    month = "7",
    year = "2026"
}

@article{Benini:2011cma,
    author = "Benini, Francesco and Closset, Cyril and Cremonesi, Stefano",
    title = "{Quantum moduli space of Chern-Simons quivers, wrapped D6-branes and AdS4/CFT3}",
    eprint = "1105.2299",
    archivePrefix = "arXiv",
    primaryClass = "hep-th",
    reportNumber = "PUTP-2373, WIS-04-11-MAY-DPPA, TAUP-2928-11",
    doi = "10.1007/JHEP09(2011)005",
    journal = "JHEP",
    volume = "09",
    pages = "005",
    year = "2011"
}

@article{Gautason:2025plx,
    author = "Gautason, Fridrik Freyr and van Muiden, Jesse",
    title = "{Ensembles in M-theory and holography}",
    eprint = "2505.21633",
    archivePrefix = "arXiv",
    primaryClass = "hep-th",
    doi = "10.1007/JHEP11(2025)078",
    journal = "JHEP",
    volume = "11",
    pages = "078",
    year = "2025"
}

@article{Martelli:2011qj,
    author = "Martelli, Dario and Sparks, James",
    title = "{The large N limit of quiver matrix models and Sasaki-Einstein manifolds}",
    eprint = "1102.5289",
    archivePrefix = "arXiv",
    primaryClass = "hep-th",
    doi = "10.1103/PhysRevD.84.046008",
    journal = "Phys. Rev. D",
    volume = "84",
    pages = "046008",
    year = "2011"
}

@article{Gautason:2025bft,
    author = "Gautason, Fridrik Freyr and Nix, Alexia",
    title = "{Universal holographic Wilson loops in 3d SCFTs}",
    eprint = "2511.04596",
    archivePrefix = "arXiv",
    primaryClass = "hep-th",
    doi = "10.1007/JHEP06(2026)061",
    journal = "JHEP",
    volume = "06",
    pages = "061",
    year = "2026"
}

@article{Hatsuda:2012dt,
    author = "Hatsuda, Yasuyuki and Moriyama, Sanefumi and Okuyama, Kazumi",
    title = "{Instanton Effects in ABJM Theory from Fermi Gas Approach}",
    eprint = "1211.1251",
    archivePrefix = "arXiv",
    primaryClass = "hep-th",
    reportNumber = "DESY-12-196, TIT-HEP-624",
    doi = "10.1007/JHEP01(2013)158",
    journal = "JHEP",
    volume = "01",
    pages = "158",
    year = "2013"
}

@article{Farquet:2014bda,
    author = "Farquet, Daniel and Sparks, James",
    title = "{Wilson loops on three-manifolds and their M2-brane duals}",
    eprint = "1406.2493",
    archivePrefix = "arXiv",
    primaryClass = "hep-th",
    doi = "10.1007/JHEP12(2014)173",
    journal = "JHEP",
    volume = "12",
    pages = "173",
    year = "2014"
}

@article{Gautason:2025per,
    author = "Gautason, Fridrik Freyr and van Muiden, Jesse",
    title = "{Localization of the M2-Brane}",
    eprint = "2503.16597",
    archivePrefix = "arXiv",
    primaryClass = "hep-th",
    doi = "10.1103/67bh-xd42",
    journal = "Phys. Rev. Lett.",
    volume = "135",
    number = "10",
    pages = "101601",
    year = "2025"
}

@book{Atiyah:1974obx,
    author = "Atiyah, M. F.",
    title = "{Elliptic Operators and Compact Groups}",
    doi = "10.1007/BFb0057821",
    isbn = "978-3-540-37811-2, 978-3-540-37811-2",
    publisher = "Springer-Verlag",
    address = "Berlin, Germany",
    volume = "401",
    year = "1974"
}

@article{Pestun:2016qko,
    author = "Pestun, Vasily",
    title = "{Review of localization in geometry}",
    eprint = "1608.02954",
    archivePrefix = "arXiv",
    primaryClass = "hep-th",
    doi = "10.1088/1751-8121/aa6161",
    journal = "J. Phys. A",
    volume = "50",
    number = "44",
    pages = "443002",
    year = "2017"
}

@article{Klebanov:1996un,
    author = "Klebanov, Igor R. and Tseytlin, Arkady A.",
    title = "{Entropy of near extremal black p-branes}",
    eprint = "hep-th/9604089",
    archivePrefix = "arXiv",
    reportNumber = "PUPT-1613, IMPERIAL-TP-95-96-39",
    doi = "10.1016/0550-3213(96)00295-7",
    journal = "Nucl. Phys. B",
    volume = "475",
    pages = "164--178",
    year = "1996"
}

@article{Gulotta:2011si,
    author = "Gulotta, Daniel R. and Herzog, Christopher P. and Pufu, Silviu S.",
    title = "{From Necklace Quivers to the F-theorem, Operator Counting, and T(U(N))}",
    eprint = "1105.2817",
    archivePrefix = "arXiv",
    primaryClass = "hep-th",
    reportNumber = "PUPT-2374",
    doi = "10.1007/JHEP12(2011)077",
    journal = "JHEP",
    volume = "12",
    pages = "077",
    year = "2011"
}

@article{Marino:2011eh,
    author = "Marino, Marcos and Putrov, Pavel",
    title = "{ABJM theory as a Fermi gas}",
    eprint = "1110.4066",
    archivePrefix = "arXiv",
    primaryClass = "hep-th",
    doi = "10.1088/1742-5468/2012/03/P03001",
    journal = "J. Stat. Mech.",
    volume = "1203",
    pages = "P03001",
    year = "2012"
}

@article{Gulotta:2011vp,
    author = "Gulotta, Daniel R. and Ang, J. P. and Herzog, Christopher P.",
    title = "{Matrix Models for Supersymmetric Chern-Simons Theories with an ADE Classification}",
    eprint = "1111.1744",
    archivePrefix = "arXiv",
    primaryClass = "hep-th",
    doi = "10.1007/JHEP01(2012)132",
    journal = "JHEP",
    volume = "01",
    pages = "132",
    year = "2012"
}

@article{Gulotta:2011aa,
    author = "Gulotta, Daniel R. and Herzog, Christopher P. and Pufu, Silviu S.",
    title = "{Operator Counting and Eigenvalue Distributions for 3D Supersymmetric Gauge Theories}",
    eprint = "1106.5484",
    archivePrefix = "arXiv",
    primaryClass = "hep-th",
    reportNumber = "PUPT-2384",
    doi = "10.1007/JHEP11(2011)149",
    journal = "JHEP",
    volume = "11",
    pages = "149",
    year = "2011"
}

@article{Kapustin:2009kz,
    author = "Kapustin, Anton and Willett, Brian and Yaakov, Itamar",
    title = "{Exact Results for Wilson Loops in Superconformal Chern-Simons Theories with Matter}",
    eprint = "0909.4559",
    archivePrefix = "arXiv",
    primaryClass = "hep-th",
    reportNumber = "CALT-68-2750",
    doi = "10.1007/JHEP03(2010)089",
    journal = "JHEP",
    volume = "03",
    pages = "089",
    year = "2010"
}

@article{Aharony:2008ug,
    author = "Aharony, Ofer and Bergman, Oren and Jafferis, Daniel Louis and Maldacena, Juan",
    title = "{${\mathcal{N}}\!=6$ Superconformal Chern-Simons-Matter Theories, M2-Branes and Their Gravity Duals}",
    eprint = "0806.1218",
    archivePrefix = "arXiv",
    primaryClass = "hep-th",
    reportNumber = "WIS-12-08-JUN-DPP",
    doi = "10.1088/1126-6708/2008/10/091",
    journal = "JHEP",
    volume = "10",
    pages = "091",
    year = "2008"
}

@article{Bergman:2009zh,
    author = "Bergman, Oren and Hirano, Shinji",
    title = "{Anomalous Radius Shift in Ad$S^4$/CFT(3)}",
    eprint = "0902.1743",
    archivePrefix = "arXiv",
    primaryClass = "hep-th",
    doi = "10.1088/1126-6708/2009/07/016",
    journal = "JHEP",
    volume = "07",
    pages = "016",
    year = "2009"
}

@article{Bergshoeff:1987cm,
    author = "Bergshoeff, E. and Sezgin, E. and Townsend, P. K.",
    title = "{Supermembranes and Eleven-Dimensional Supergravity}",
    reportNumber = "IC-87-10",
    doi = "10.1201/9781482268737-10",
    journal = "Phys. Lett. B",
    volume = "189",
    pages = "75--78",
    year = "1987"
}

@article{Giombi:2023vzu,
    author = "Giombi, Simone and Tseytlin, Arkady A.",
    title = "{Wilson Loops at Large $N$ and the Quantum M2-Brane}",
    eprint = "2303.15207",
    archivePrefix = "arXiv",
    primaryClass = "hep-th",
    reportNumber = "PUPT-2639, Imperial-TP-AT-2023-02",
    doi = "10.1103/PhysRevLett.130.201601",
    journal = "Phys. Rev. Lett.",
    volume = "130",
    number = "20",
    pages = "201601",
    year = "2023"
}

@article{Beccaria:2023ujc,
    author = "Beccaria, Matteo and Giombi, Simone and Tseytlin, Arkady A.",
    title = "{Instanton Contributions to the ABJM Free Energy from Quantum M2 Branes}",
    eprint = "2307.14112",
    archivePrefix = "arXiv",
    primaryClass = "hep-th",
    reportNumber = "PUPT-2645, Imperial-TP-AT-2023-04",
    doi = "10.1007/JHEP10(2023)029",
    journal = "JHEP",
    volume = "10",
    pages = "029",
    year = "2023"
}
\end{document}